\documentclass[aps,prb,longbibliography,showpacs,floatfix,superscriptaddress,12pt]{revtex4-2}
\usepackage[utf8]{inputenc}
\usepackage{geometry}
\usepackage[english]{babel}
\usepackage{blindtext}
\usepackage{amssymb}
\usepackage{graphicx}

\usepackage{amsmath}
\usepackage{latexsym}
\usepackage{float}
\usepackage{times}
\usepackage{color}
\usepackage{subfigure}
\usepackage{indentfirst}
\usepackage{setspace}
\usepackage{bm}

\usepackage[colorlinks,linkcolor=blue,anchorcolor=blue,citecolor=green]{hyperref}
\usepackage{CJKutf8}  

\graphicspath{{pdf}}

\allowdisplaybreaks

\begin{document}	
	
\title{
	Emergence of spatial patterns and synchronization in superconducting time crystals}
\author{Bo Fan(\begin{CJK*}{UTF8}{gbsn}范波\end{CJK*})
}
\email{bo.fan@sjtu.edu.cn}
\affiliation{Shanghai Center for Complex Physics, School of Physics and Astronomy, 
	\\Shanghai Jiao Tong University, Shanghai 200240, China}
\author{Zi Cai(\begin{CJK*}{UTF8}{gbsn}蔡子\end{CJK*})
}
\email{zcai@sjtu.edu.cn}
\affiliation{Wilczek Quantum Center, School of Physics and Astronomy, 
	\\Shanghai Jiao Tong University, Shanghai 200240, China}
\author{Antonio M. Garc\'ia-Garc\'ia}
\email{amgg@sjtu.edu.cn}
\affiliation{Shanghai Center for Complex Physics, School of Physics and Astronomy, 
	\\Shanghai Jiao Tong University, Shanghai 200240, China}
\vspace{0.2cm}
\date{\today}
\vspace{0.3cm}

\begin{abstract}
We identify a time crystal phase characterized by a frequency half of the driving frequency in disordered superconductors by employing the time dependent Bogoliubov-de Gennes formalism at zero temperature with a periodically driven coupling constant. 
After a period of exponential increase of spatial inhomogeneities and exponential suppression of the order parameter amplitude, the time crystal develops islands of different sizes.  
Each of these islands is a time crystal with the same frequency albeit with a phase shift $\pi$ with respect to the homogeneous time crystal. After its emergence, the island gradually becomes smaller, though the phase shift persists, until it is abruptly synchronized at a time that it depends on its initial size. 
We find a critical disorder strength, still deep in the metallic phase, at which the time crystal phase terminates. 
For even stronger disorder, the order parameter oscillates with the driving frequency in regions where localization effects are not important.
\end{abstract}
\maketitle

\newpage

\tableofcontents
\section{Introduction}
Quantum time crystals \cite{wilczek2012quantum} is a novel form of quantum matter resulting from the  
spontaneous breaking of continuous time translation symmetry in quantum many-body systems. Originally, it was proposed that they could exist \cite{wilczek2012quantum} even in quantum systems at equilibrium. However, it was soon realized that an explicit time dependent Hamiltonian \cite{bruno2013,watanabe2015} was a necessary condition for its existence. Quantum many-body periodically driven systems, usually referred to as Floquet systems, were quickly realized \cite{sacha2015,khemani2016,else2016,yao2017discrete} to be a fertile ground to discover and characterize quantum time crystals. In this context, a signature of the time crystal phase is the observation of oscillations at frequencies other than the driving frequency though related to it by an integer or a simple rational number.    

The recent experimental observation of time crystals in a variety of quantum many body systems \cite{choi2017,zhang2017observation,zhang2017observation,smits2018observation,kleiner2021space,trager2021real,kessler2021observation,autti2022} together with the proposal of quantum computing applications, ranging from the role of time crystals in the enhancement of the qubit coherence time \cite{bao2024} to its relevance \cite{mi2021} in quantum processors, have further boosted the interest in this problem.

A time crystal phase has also been identified theoretically  \cite{collado2021emergent,collado2023dynamical} in a Bardeen-Cooper-Schrieffer (BCS) superconductor \cite{bardeen1957} with a periodically driven coupling constant by the observation of an oscillating order parameter with a frequency half of the driving frequency. Similar time-dependent mean-field equations have also been employed to study the properties of discrete time crystals, {\it e.g.} its excitation \cite{Yang2021} and stability against thermal fluctuations \cite{Yue2022}. Although an experimental confirmation of these findings is still missing, various dynamical phases of BCS superconductors have been observed recently in a cavity quantum electrodynamics simulator \cite{young2024observing} which is an important step towards the discovery of time crystals in driven superconductors.

We note that the mentioned BCS approach cannot account for spatial inhomogeneities that result from either the presence of impurities and dislocations or emerge spontaneously in spatially homogeneous system as a consequence of the out of equilibrium dynamics \cite{dzero2009cooper,garcia2014b,chesler2015defect,kibble1980some,zurek1985cosmological,lamporesi2013spontaneous,cordoba2019spontaneous,yanchuk2014pattern,wang2020pattern,chern2019nonequilibrium,kleiner2021space,trager2021real,yue2023prethermal,fan2024quenched,fan2023exploring}.

Therefore, a natural question to ask is whether the time crystal phase \cite{collado2021emergent,collado2023dynamical} found in a periodically driven BCS superconductor is robust to the expected formation of spatial patterns as a consequence of either the out of equilibrium dynamics or the explicit addition of disorder. Do those spatial patterns coexist with the time crystal phase or do they destroy it? If the time crystal phase is robust, how is it modified by these perturbations? 

In this paper, we address those questions by investigating the time-dependent Bogoliubov de-Gennes (BdG) equations with a periodically driven pairing interaction.  Although the time-dependent self-consistent BdG formalism is still a mean field theory, it allows us to study explicitly the space-time evolution of the time crystal. We show that the time crystal phase is robust to both the presence of a sufficiently weak disorder and spatial inhomogeneities \cite{dzero2009cooper} induced by the driven dynamics. However, the time crystal is qualitatively modified with respect to the one observed in the spatially homogeneous limit.  

We shall see that, even in the presence of a weak disorder, which do not cause any effect to the spatial distribution of the order parameter at equilibrium, the order parameter develops spatial inhomogeneities in the form of isolated metastable islands. Each of these islands becomes an independent time crystal with a phase shift with respect to the time crystal corresponding to the spatial homogeneous region that surrounds the islands. For longer times, we observe the size reduction and eventual synchronization of the coexistent time crystals. Indeed, the time crystal phase survives even in the presence of a relatively strong disorder strength $V \lesssim 0.7$. However, for $V \gtrsim 0.7$, still deep in the metallic region at equilibrium, we observe the sharp termination of the time crystal phase. 

The paper is organized as follows: in Section \ref{sec:method} we introduce the model and describe the numerical procedure to compute the space-time dependence of the order parameter. Section \ref{sec:strictclean} is devoted to the identification of the time crystal phase in the clean limit and the comparison with previous BCS results. The impact of the dynamics and the emergence of spatial inhomogeneities due to a weak disordered potential is discussed in Section \ref{sec:weakdis}. In order to illustrate that, for sufficiently weak disorder, the resulting spatial pattern is not related to disorder, we study in Section \ref{sec:turningoff} the dynamics after removing the random potential once the time crystal oscillations start to develop. The breaking of the time crystal phase at stronger disorder is investigated in Section \ref{sec:strongdis}. Conclusions can be found in Section \ref{sec:con}.

\section{The model and the numerical method} \label{sec:method}

We start with the BdG equations \cite{ghosal2001inhomogeneous} of a two dimensional square lattice defined as follows:

\begin{equation}
	\left(\begin{matrix}
		\hat{K} 		& \hat{\Delta}  \\
		\hat{\Delta}^* 	& -\hat{K}^* 	\\
	\end{matrix}\right)
	\left(\begin{matrix}
		u_n({\bf r}_i)   \\
		v_n({\bf r}_i) \\
	\end{matrix}\right)
	= E_n
	\left(\begin{matrix}
		u_n({\bf r}_i)   \\
		v_n({\bf r}_i) \\
	\end{matrix}\right)
	\label{eq.BdG_mat}
\end{equation} 
where
\begin{equation}
	\hat{K}u_n({\bf r}_i)=-t_{ij}\sum_{\delta}u_n({\bf r}_i+\delta)+(V_i-\mu_i)u_n({\bf r}_i),
\end{equation}

$\delta$ stands for the nearest neighboring sites, $t_{ij}$ is the hopping strength, $V_i$ are random variables from an uniform distribution between $[-V,V]$, $ \mu $ is the chemical potential, and $\hat{\Delta}u_n({\bf r}_i) =\Delta({\bf r}_i)u_n({\bf r}_i)$. The same definition applies to $v_n({\bf r}_i)$. The BdG equations are completed by the self-consistency conditions for the site dependent order parameter 
\begin{equation}
	\Delta({\bf r}_i) = |U_0|\sum_{n}u_n({\bf r}_i)v_n^*({\bf r}_i)
	\label{eq.gap}
\end{equation}
and the local density
\begin{equation}
	n({\bf r}_i) =2\sum_{n}|v_n({\bf r}_i)|^2.
	\label{eq.n}
\end{equation}

We first solve the Bogoliubov-de Gennes Eq.~\eqref{eq.BdG_mat} for a square lattice of $N = L\times L$ sites, where $L$ is in units of the lattice constant, to get the mean field solution ${\Delta({\bf r}_i)}$ and ${n({\bf r}_i)}$. This will be the initial state for the dynamical evolution of the order dependent amplitude using the time dependent BdG formalism. In order to proceed, we need to introduce the time dependent order parameter ${\Delta({\bf r}_i,t)}$ and wavefunctions $\{u_n({\bf r}_i,t),v_n({\bf r}_i,t)\}$ leading to the following time-dependent BdG equations \cite{challis2007bragg,liao2011traveling,scott2012rapid,tokimoto2022josephson}

\begin{equation}
	\left(\begin{matrix}
		\hat{K} 		& \hat{\Delta}({\bf r}_i,t)  \\
		\hat{\Delta}^*({\bf r}_i,t) 	& -\hat{K}^* 	\\
	\end{matrix}\right)
	\left(\begin{matrix}
		u_n({\bf r}_i,t)   \\
		v_n({\bf r}_i,t) \\
	\end{matrix}\right)
	= i \hbar {\partial \over \partial t}
	\left(\begin{matrix}
		u_n({\bf r}_i,t)   \\
		v_n({\bf r}_i,t) \\
	\end{matrix}\right)
	\label{eq.TDBdG_mat}.
\end{equation} 
In order to study the dynamics of the BdG superconductor, we solve  Eq.~\eqref{eq.TDBdG_mat} by using the 4th order Runge Kutta algorithm \cite{butcher2016numerical} where the time dependent order parameter is given by, 
\begin{equation}
	\Delta({\bf r}_i,t) = |U(t)|\sum_{n}u_n({\bf r}_i,t)v_n^*({\bf r}_i,t),
	\label{eq.TDgap}
\end{equation}
and the time dependent local density by
\begin{equation}
	n({\bf r}_i,t) =2\sum_{n}|v_n({\bf r}_i,t)|^2.
	\label{eq.TDn}
\end{equation}
The out of equilibrium dynamics inducing the time crystal phase is generated by a periodic in time coupling constant,  
\begin{align}
	U(t) = U_0 [1 + \alpha \sin(\omega_d t)] \label{eq.Ut},
\end{align}
with $|\alpha| < 1$ and $U_0 < 0$.
We note that this driving protocol has already been employed in Refs.~\cite{collado2021emergent,collado2023dynamical} to investigate the time crystal phase in the simpler spatially homogeneous Bardeen-Cooper-Schrieffer superconductors. 

Due to technical reasons related to the slow convergence of the solutions in the weak electron-phonon coupling region, especially in the disordered case, we will focus on the strong coupling limit $U_0 = -6$. For simplicity, we use a constant hopping energy $t_{ij} = 1$, which is also the unit of energy. The chemical potential is fixed in the main text at $\mu = 0$. Results for another value of the chemical potential, see Appendix \ref{app:difmu}, also shows the same three phases, Rabi-Higgs, Syncronized Higgs and Time crystal, though the order parameter becomes complex and the time crystal frequency is different and less sharply defined than for $\mu = 0$.

The maximum system size that we can reach by solving the time-dependent BdG equations Eq.~\eqref{eq.TDBdG_mat} in the presence of a disordered potential is $N=120\times 120$. The primary results are based on a system size of $N=100\times 100$. In the clean limit, we solve the system in momentum space by using the quasiparticle excitations energy $\xi_k =-2t( \cos(k_x) + \cos(k_y)) - \mu$, where $k_x = 2\pi n_x/L$ and $k_y = 2\pi n_y/L$. A natural question to ask is whether this maximum size $120\times 120$ is large enough to neglect finite size effects at least in the range of times that we explore in the paper. In Appendix \ref{app:size}, we answer this question affirmatively.

In this clean limit, we compare real and momentum space solutions and, as expected, we find excellent agreement. However, in momentum space it is possible to find solutions for much larger system sizes though only in the clean limit. As a consequence, in this limit, we find the full phase diagram of the system with great precision which is instrumental in the identification of the parameters in which the time crystal phase occurs. We employ a small step-size $dt = 1 \times 10^{-3}/\Delta_0$, where $\Delta_0=2.48$, to minimize numerical errors and thus ensure the full convergence of the solutions. More specifically, we have checked, see Figure~\ref{Fig:identimecrystal}, that even for the longest times investigated, the numerical error is small by comparing the time evolution for different $dt$.

\section{Identification of the time crystal phase in the clean limit}\label{sec:strictclean}

\begin{figure}[!htbp]
	\begin{center}	
		{\label{fig.2D_BdG_time_fourier_V0_a0p25_wd0p8}
			\includegraphics[width=5.5cm]{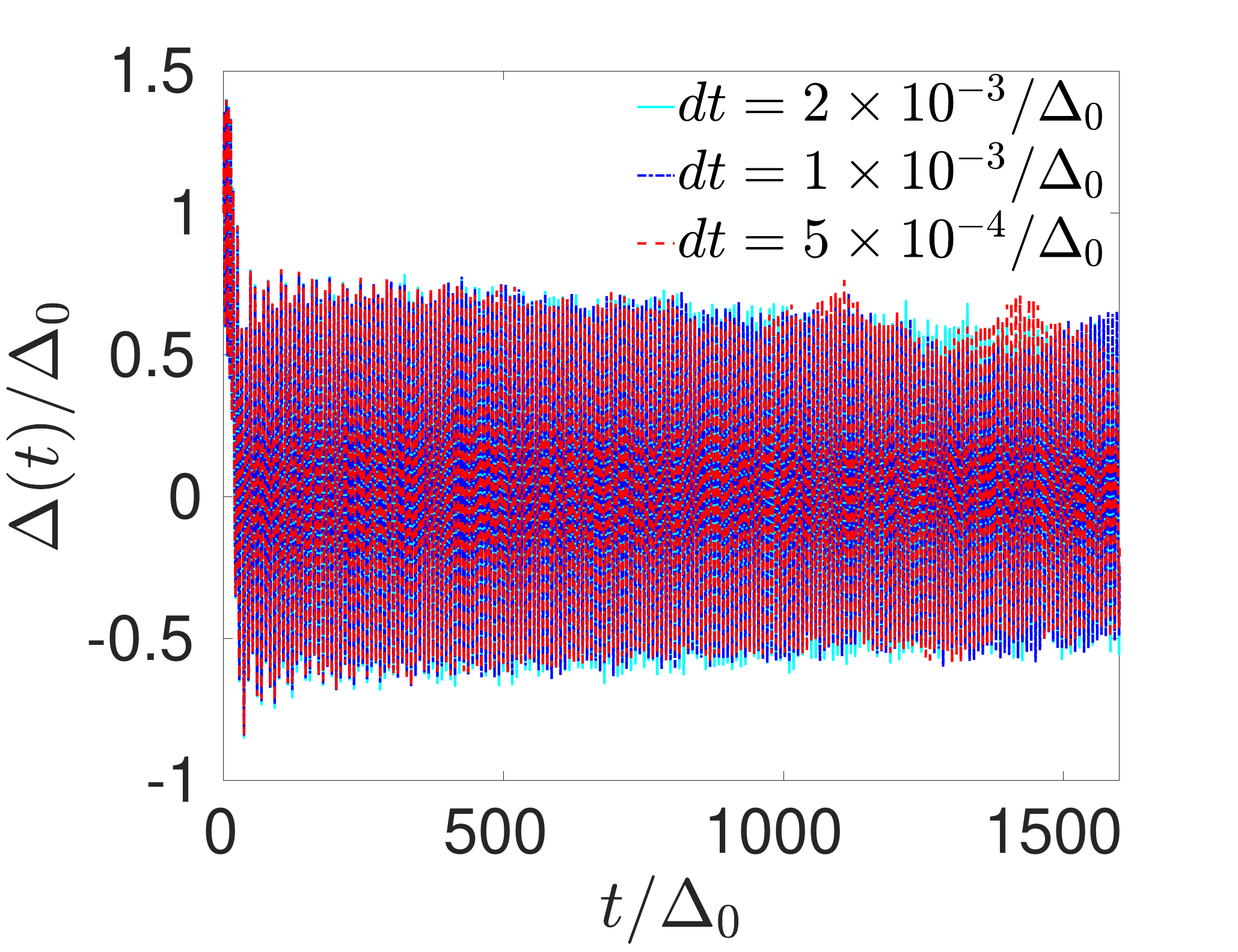}
			\includegraphics[width=5.5cm]{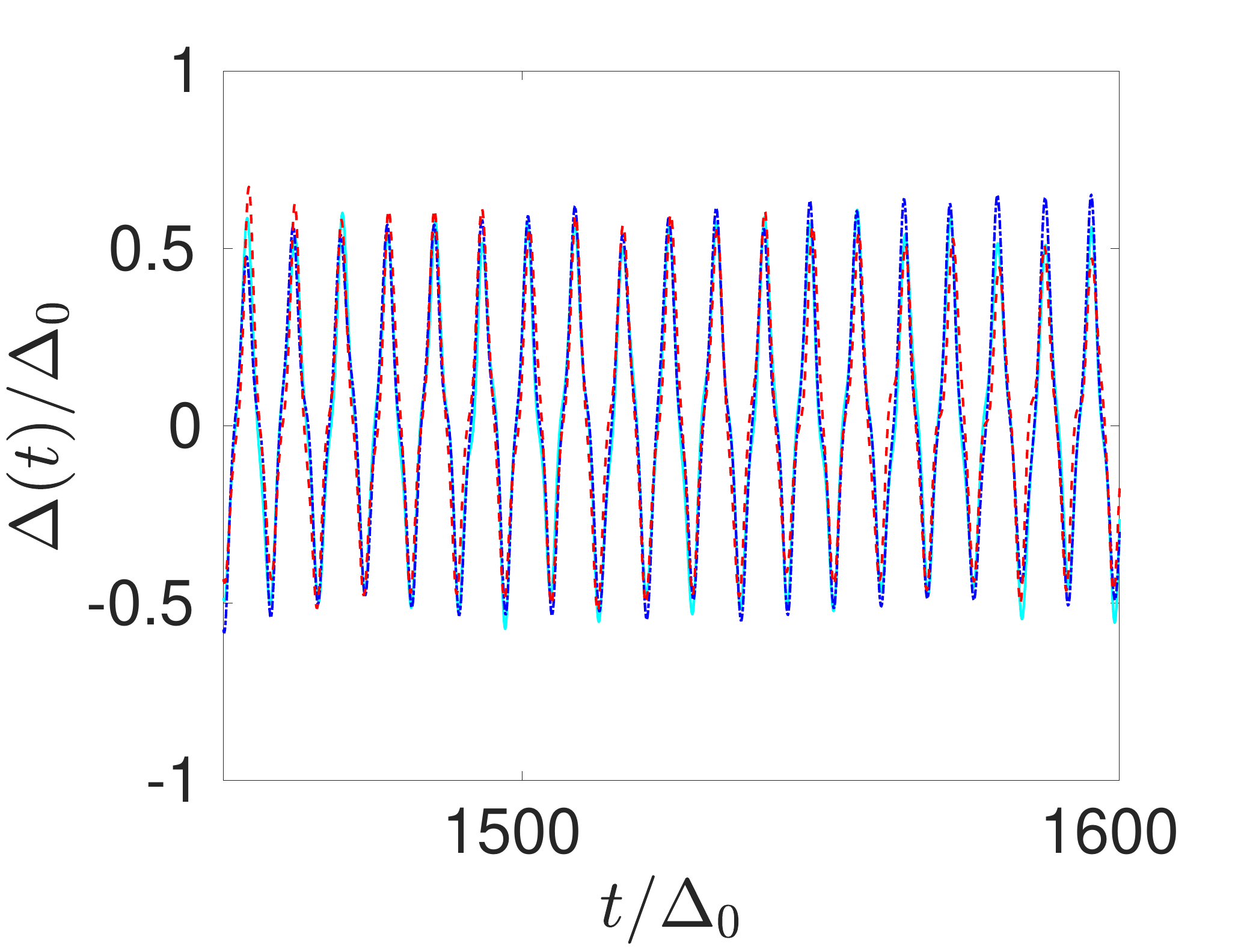}
			\includegraphics[width=5.5cm]{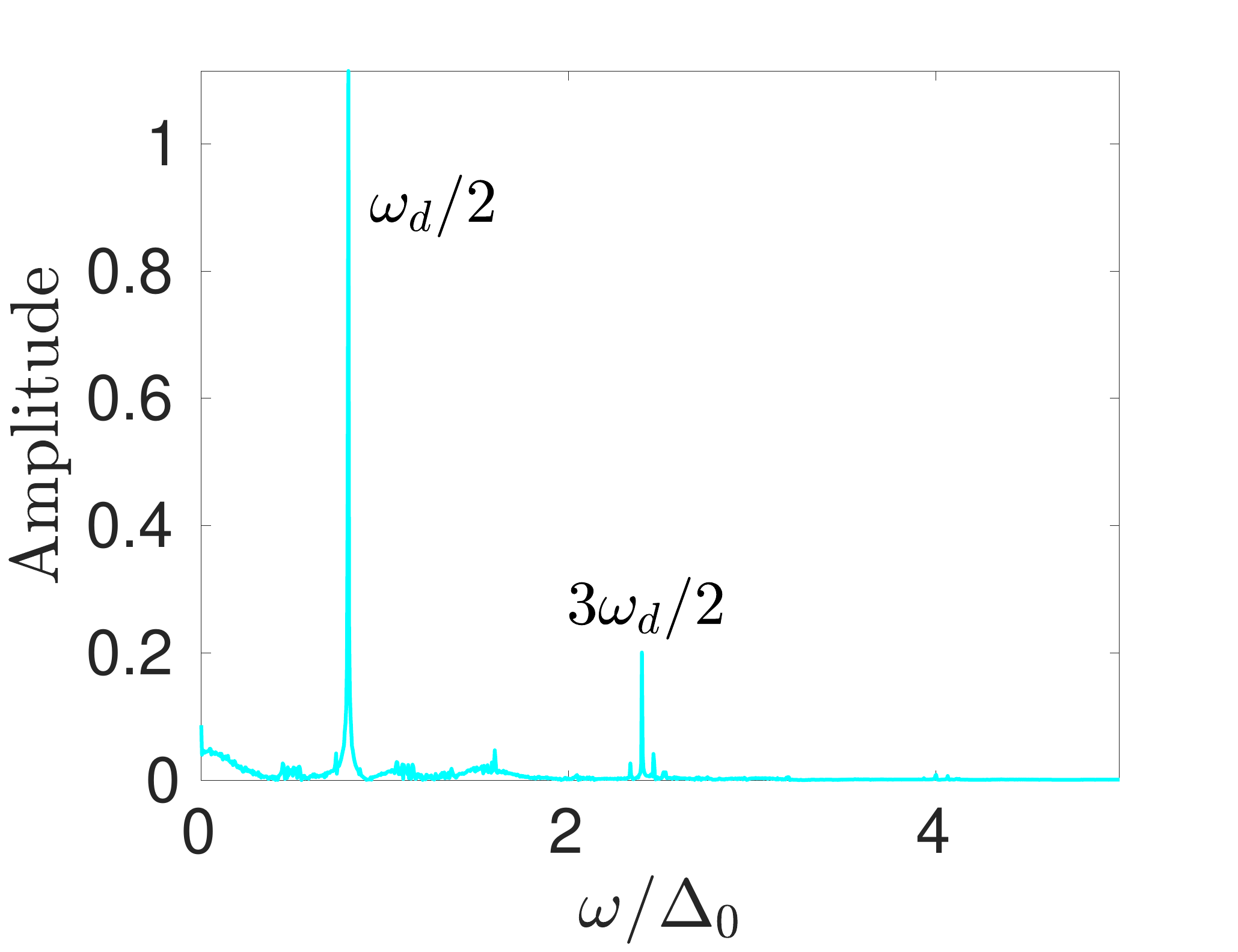}
		}\\
		{Time crystal phase: {$\alpha=0.25,\omega_d = 0.8\times2\Delta_0$} }
		\caption{The time evolution of the order parameter in the clean, homogeneous limit for driving frequencies close to those $\sim 0.8 \times 2\Delta_0/\hbar$ required to break a Cooper pair (left and central) and its Fourier transform (right). The strong peak in the Fourier transform at half the driving frequency is a signature of the time crystal phase. The system size is $N=200\times200$, the equilibrium coupling constant $U_0 = -6$ and the chemical potential is fixed at $\mu = 0$. Different time step-sizes $dt$ are employed in order to confirm the robustness of our results. The accumulated numerical error grows with time. However, our results indicate that even for $t~\Delta_0>1000$ this error is still negligible. 
		}\label{Fig:identimecrystal}
	\end{center}
\end{figure}

We initiate our study with the identification of the range of driving amplitude and frequencies in which the time crystal phase, characterized by an oscillating order parameter with a frequency which is half of the driving frequency. For that purpose, we solve the time dependent BdG equations in the two-dimensional square lattice introduced in the previous section. For this identification, we do not introduce any disorder so we can work in momentum space which makes possible to reach larger sizes.
We have found the time crystal phase, see Figure \ref{Fig:identimecrystal}, by the presence of sharp Fourier peaks at half the driving frequency, in a relatively narrow region near the driving frequency $\omega_d \sim 0.8\times2\Delta_0$. This is because the fluctuation is weak in the strong coupling $U_0 = -6$ limit. For example, if the driving amplitude is fixed at $\alpha = 0.25$, the time crystal phase is restricted to driving frequencies $0.79 \le \omega_d/2\Delta_0 \leq 0.83$. In the weak coupling limit, the time crystal phase occurs for a much broader range of parameters, see Appendix~\ref{app:BCS_result} for more details. This is in good agreement with previous results Ref.~\cite{collado2023dynamical} that employ the simpler BCS approach predicting a spatially homogeneous order parameter in all cases. 
From now on, we fix the driving amplitude to $\alpha = 0.25$ and the driving frequency to $\omega_d = 0.8\times 2\langle \Delta(r)\rangle$, where $\langle\Delta(r)\rangle$ is the initial spatially averaged order parameter before the quench dynamics starts. 

Although the main focus of this paper is the time crystal phase, we have also investigated other driving frequencies and amplitudes which led us to also identify the Synchronized Higgs phase and Rabi-Higgs phase which were reported in Ref.~\cite{collado2023dynamical}. We note that, although the resulting phase diagram shows strong similarities, see Appendix \ref{app:BCS_result}, with that of Ref.\cite{collado2023dynamical}, we could not find a gapless phase without oscillations predicted in Ref.~\cite{collado2023dynamical}. Instead, in this region we observe that despite the fact that the order parameter amplitude decays to zero, it still oscillates between zero and a maximum peak around $0.6\Delta_0$.
The reason of not observing a gapless phase in our study may be that while we employ the realistic energy levels and quantum degeneracies of a square lattice, previous studies \cite{collado2023dynamical, barankov2006synchronization,zhoubcs,yuzbashyan2006dynamical} have used equally spaced eigenvalues with no eigenenergy degeneracy. Since this is not the main interest of the paper, more details of these phases, for instance providing evidence that the mentioned oscillations in the gapless phase is not a finite size effect, are left to Appendix \ref{app:BCS_result}.  
Having identified the time crystal, we now proceed to study its space-time evolution in the presence of disorder.

\section{The time crystal phase in the weak disorder limit}\label{sec:weakdis}

\begin{figure}[!htbp]
   	\begin{center}
   		\subfigure[]{ \label{fig.L100_meangap_vs_V_tall}
   			\includegraphics[height=4.cm]{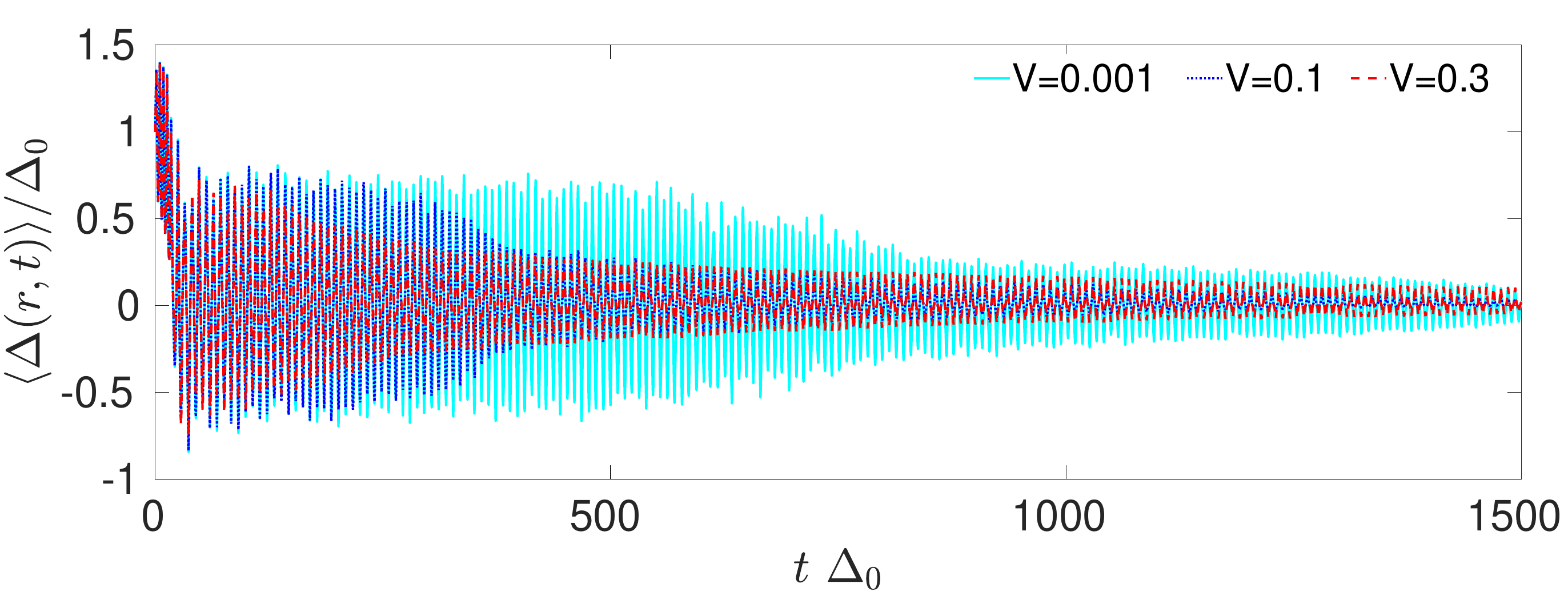} }
   		\subfigure[]{ \label{fig.L100_meangap_vs_V_fourier}
   			\includegraphics[height=4.cm]{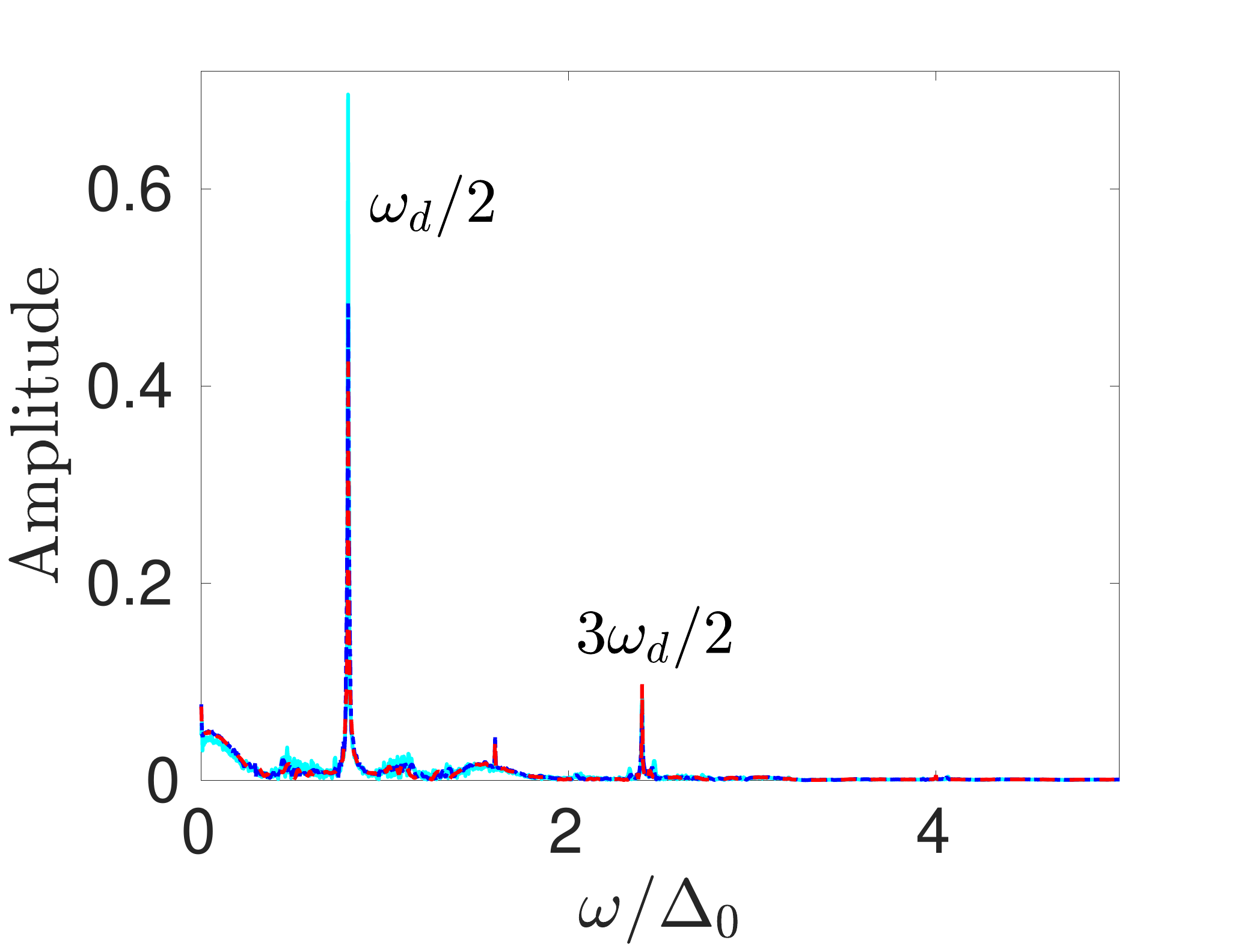} }
   		\subfigure[]{ \label{fig.L100_meangap_vs_V_t100}
   			\includegraphics[height=4.cm]{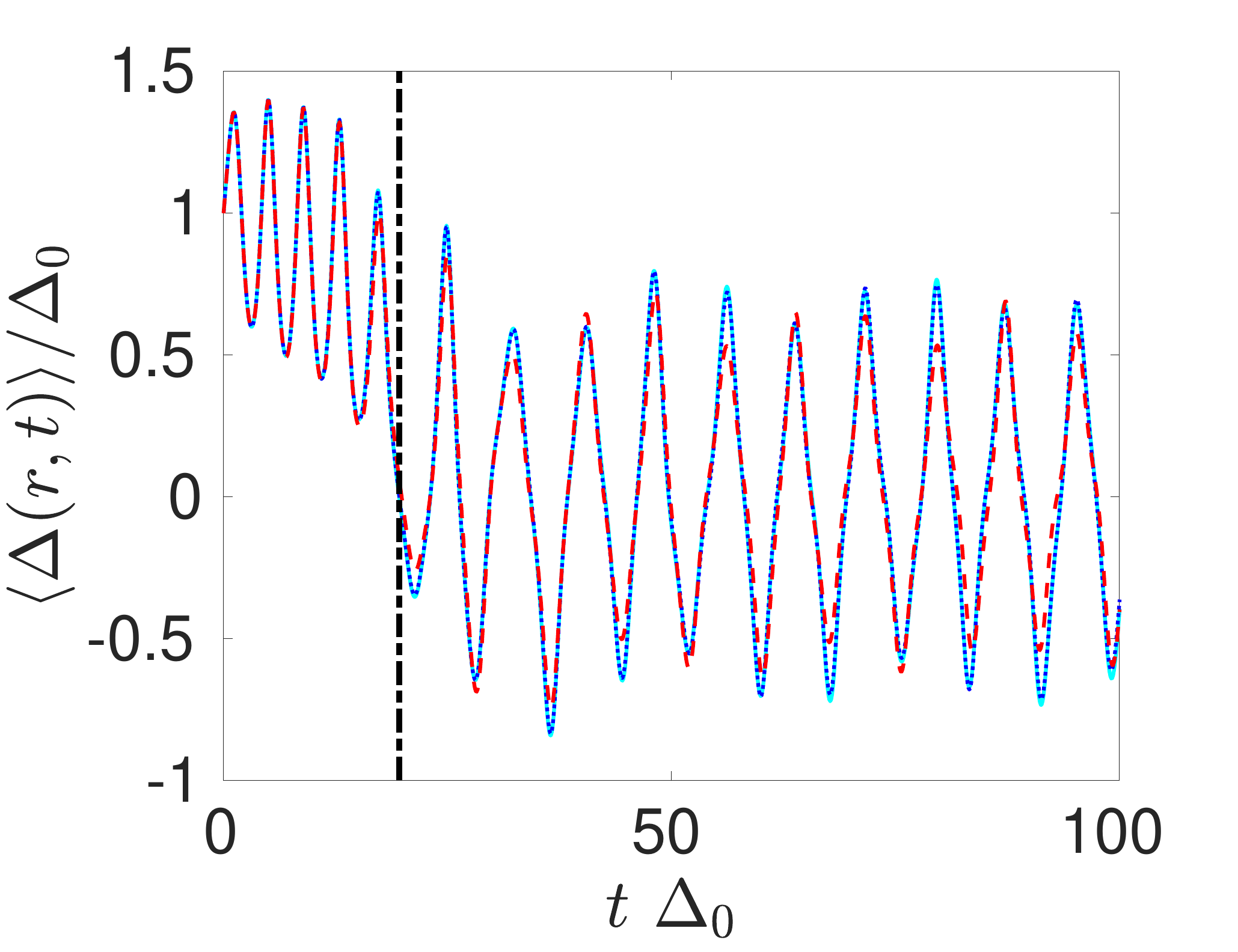} }
   		\subfigure[]{ \label{fig.L100_meangap_vs_V_t600}
   			\includegraphics[height=4.cm]{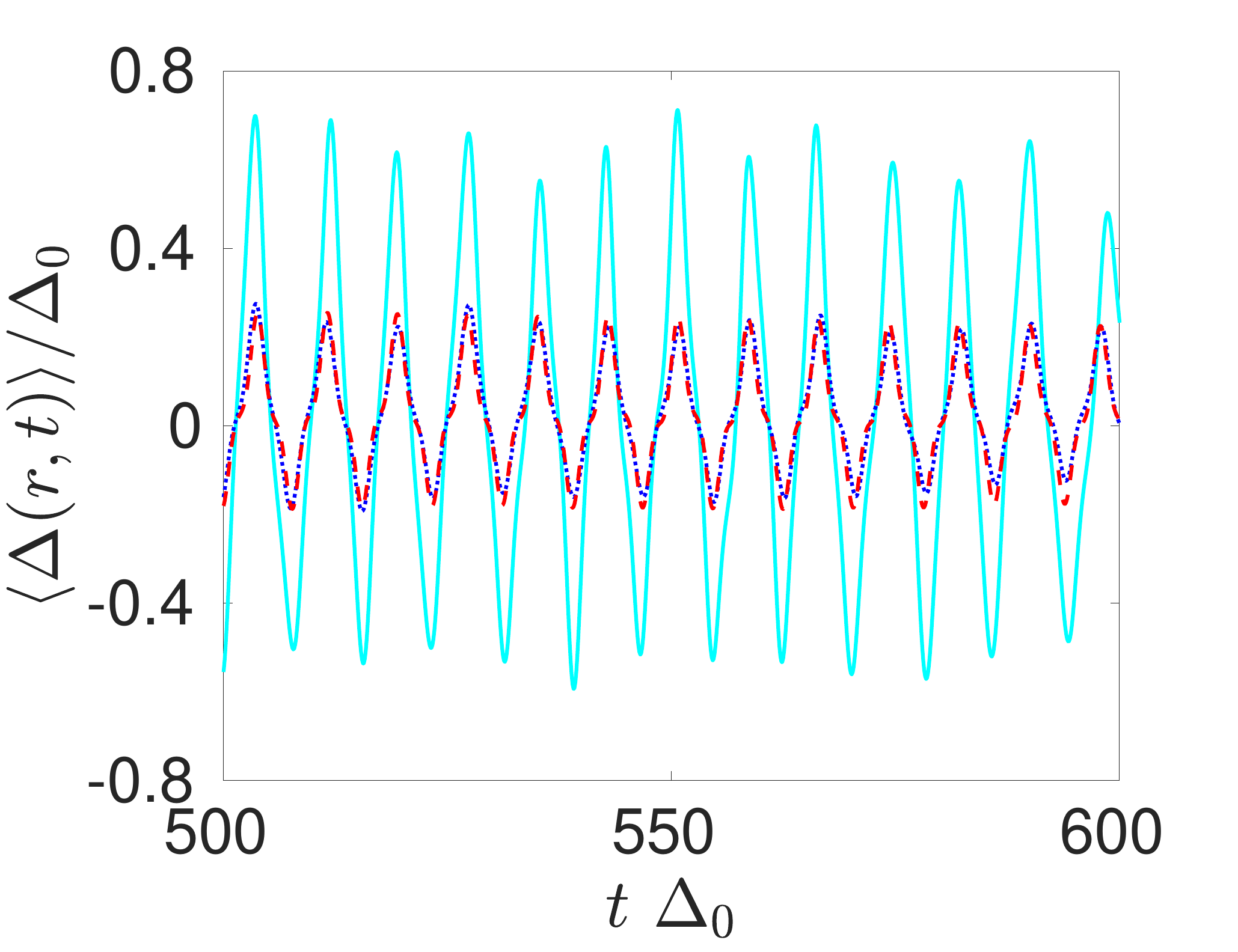} }
   		\subfigure[]{ \label{fig.L100_meangap_vs_V_t1200}
   			\includegraphics[height=4.cm]{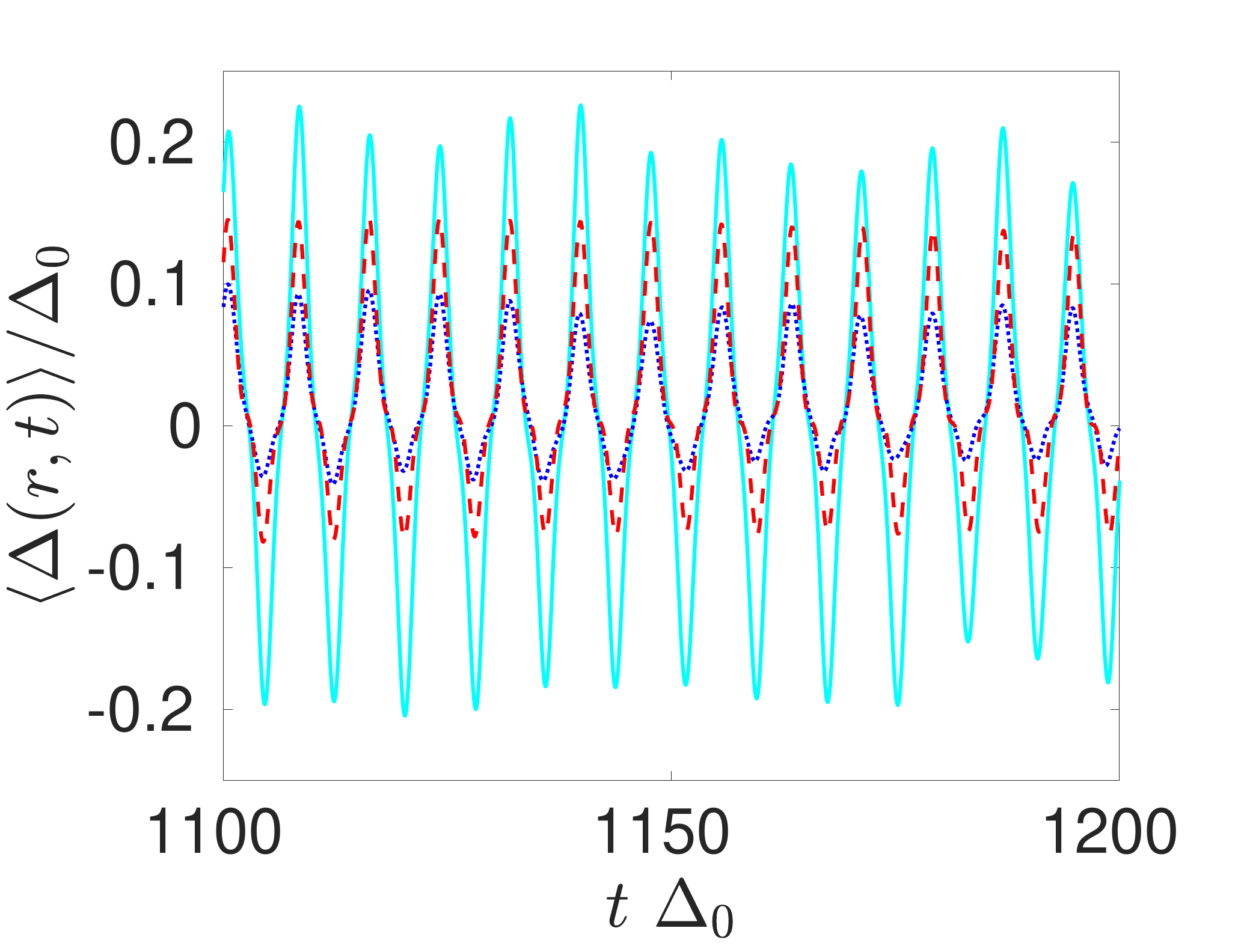} }
   		\caption{The time evolution of the spatial averaged order parameter $\langle\Delta(r,t)\rangle$, where the brackets stand for spatial average, for different disorder strengths $V = 0.001, 0.1, 0.3$ and the corresponding Fourier transformation. \subref{fig.L100_meangap_vs_V_t100}$\sim$ \subref{fig.L100_meangap_vs_V_t1200} are blown up's of \subref{fig.L100_meangap_vs_V_tall} for a narrower time range in order to show the time crystal oscillations in more detail. The vertical black dashed line at $t\Delta_0 \sim 19.6$ represents the time when the system enters the time crystal phase. The rest of parameters are the equilibrium coupling constant $U_0 = -6$ and the chemical potential is fixed at $\mu = 0$. The driving amplitude $\alpha = 0.25$ and the driving frequency $\omega_d = 0.8\times 2\langle\Delta(r)\rangle$.}\label{Fig:L100_meangap_vs_Vweak}
   	\end{center}
 \end{figure}

In this section, we study the impact of disorder by solving the time dependent BdG equations in the presence of a random potential $V_i$ at each site. As mentioned earlier, we have chosen an uniform distribution $V_i \in [-V, V]$ where $V$ is referred to as the strength of the disordered potential. In order to have a glimpse of the role of disorder, we depict in Figure~\ref{Fig:L100_meangap_vs_Vweak} the time evolution of the order parameter after spatial average for three different disorder strengths $V = 0.001, 0.1, 0.3$ corresponding to the weak disorder limit deep in the metallic phase. Indeed, in the static limit, the spatial inhomogeneities for $V = 0.001$ are completely negligible. The sharp peaks in the Fourier transform at $\omega_d/2$ and $3\omega_d/2$, the latter much weaker and likely a subharmonic of the former, see Figure~\ref{fig.L100_meangap_vs_V_fourier}, reveal that the time crystal phase is robust to the addition of a weak disorder. The frequency of oscillations is still half the driving frequency even for quite long times. However, the amplitude of the oscillations are sensitive to disorder. Eventually, the amplitude decreases rather abruptly at a time scale that becomes shorter as the disorder strength increases.
Before this decay, we shall see the order parameter is spatially homogeneous.
We proceed now to the description of the emergence of spatial patterns as a result of the driving. 
   
\subsection{The emergency of the spatial pattern in the presence of weak disorder}

\begin{figure}[!htbp]
	\begin{center}
		\subfigure[]{ \label{fig.BdG_V0p001_spatial}
			\includegraphics[width=15cm]{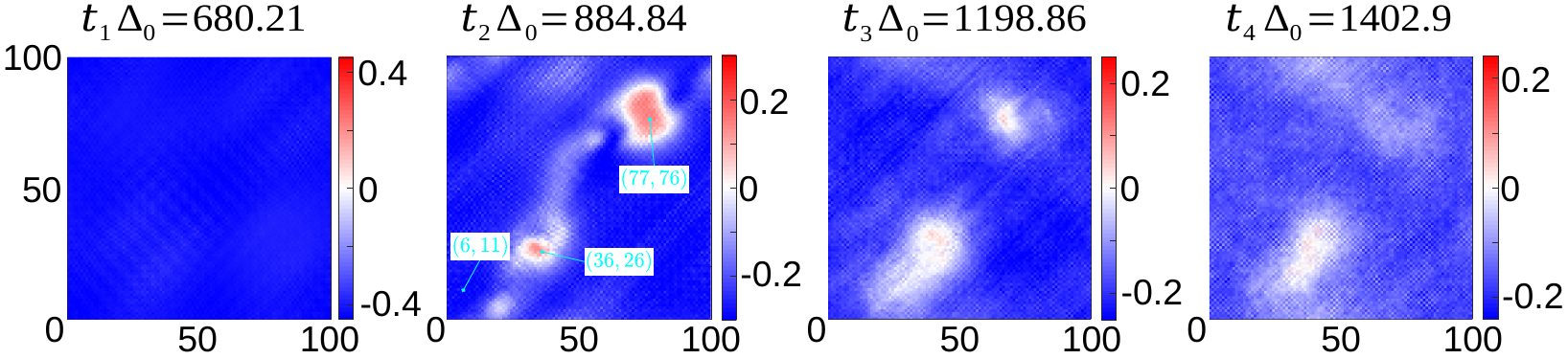}  }
		\subfigure[]{ \label{fig.BdG_V0p001_t1800}
			\includegraphics[width=7.5cm]{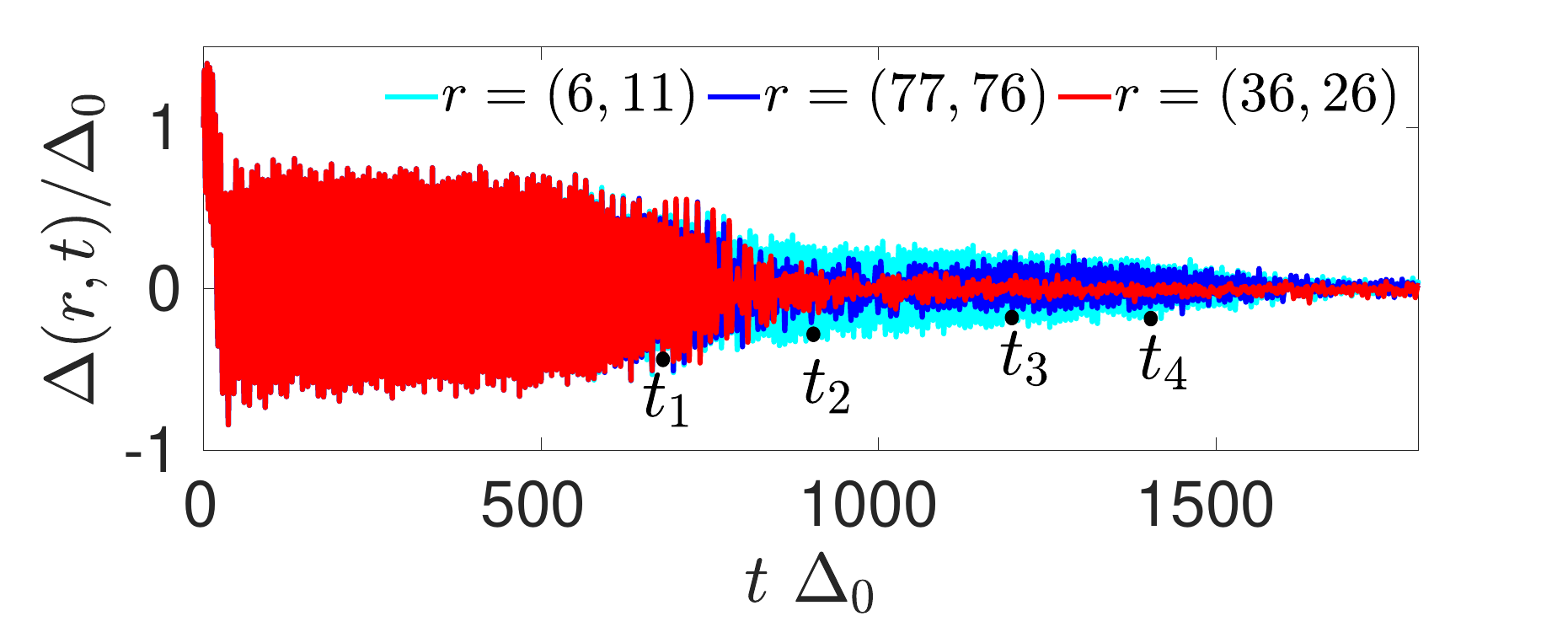}}
		\subfigure[]{ \label{fig.BdG_V0p001_t710}
			\includegraphics[width=7.5cm]{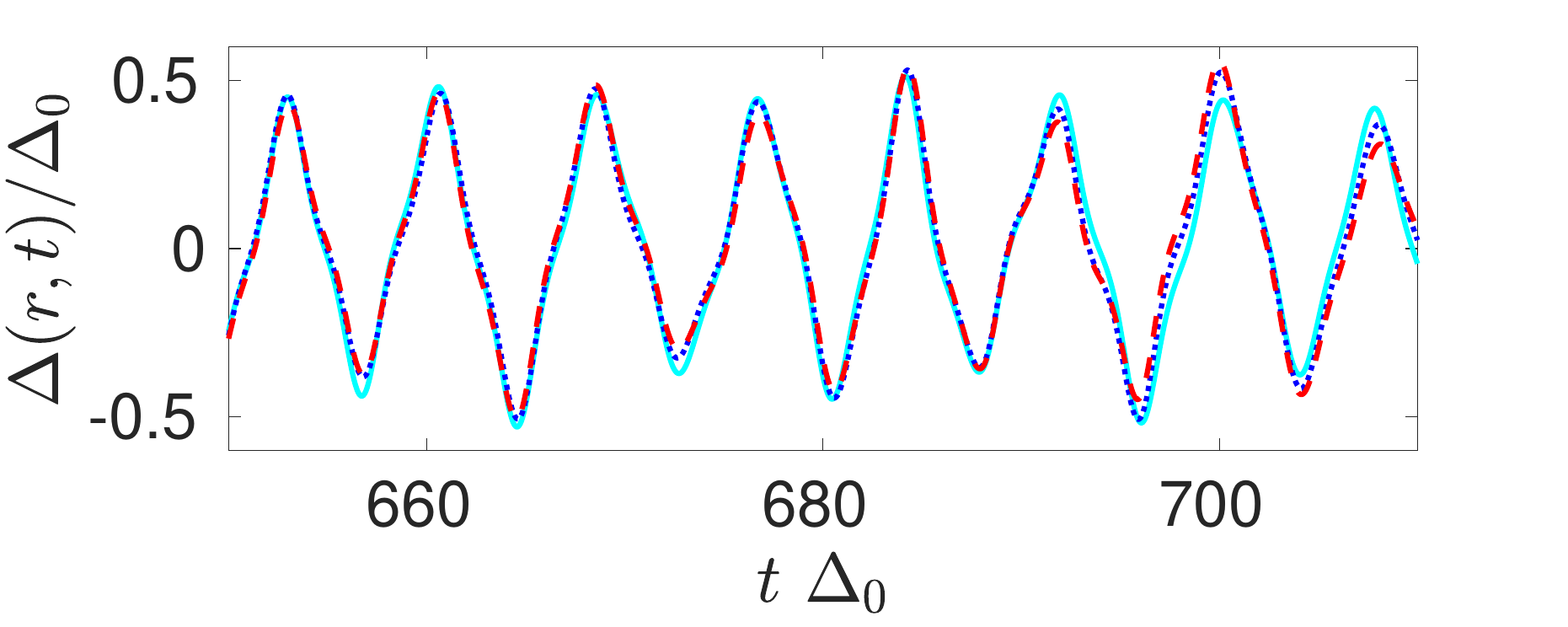}}
		\subfigure[]{ \label{fig.BdG_V0p001_t1000}
			\includegraphics[width=7.5cm]{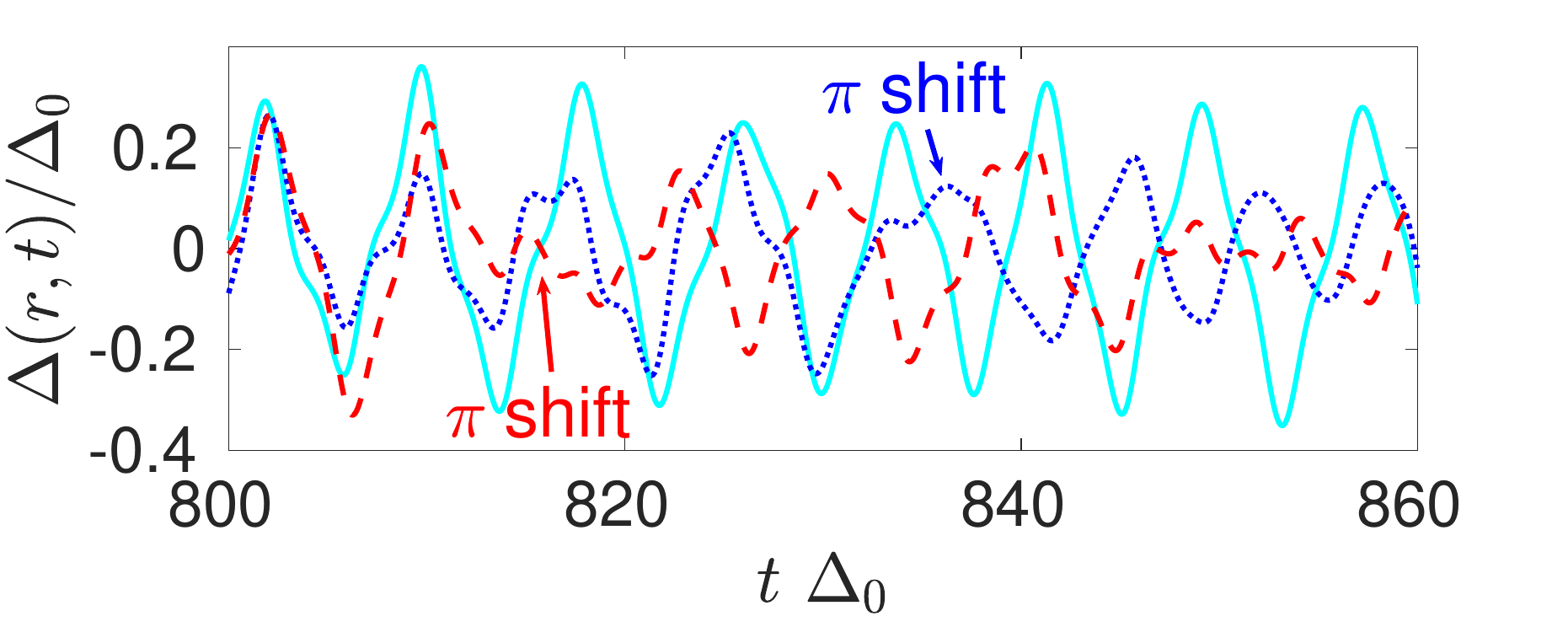}}
		\subfigure[]{ \label{fig.BdG_V0p001_t1460}
			\includegraphics[width=7.5cm]{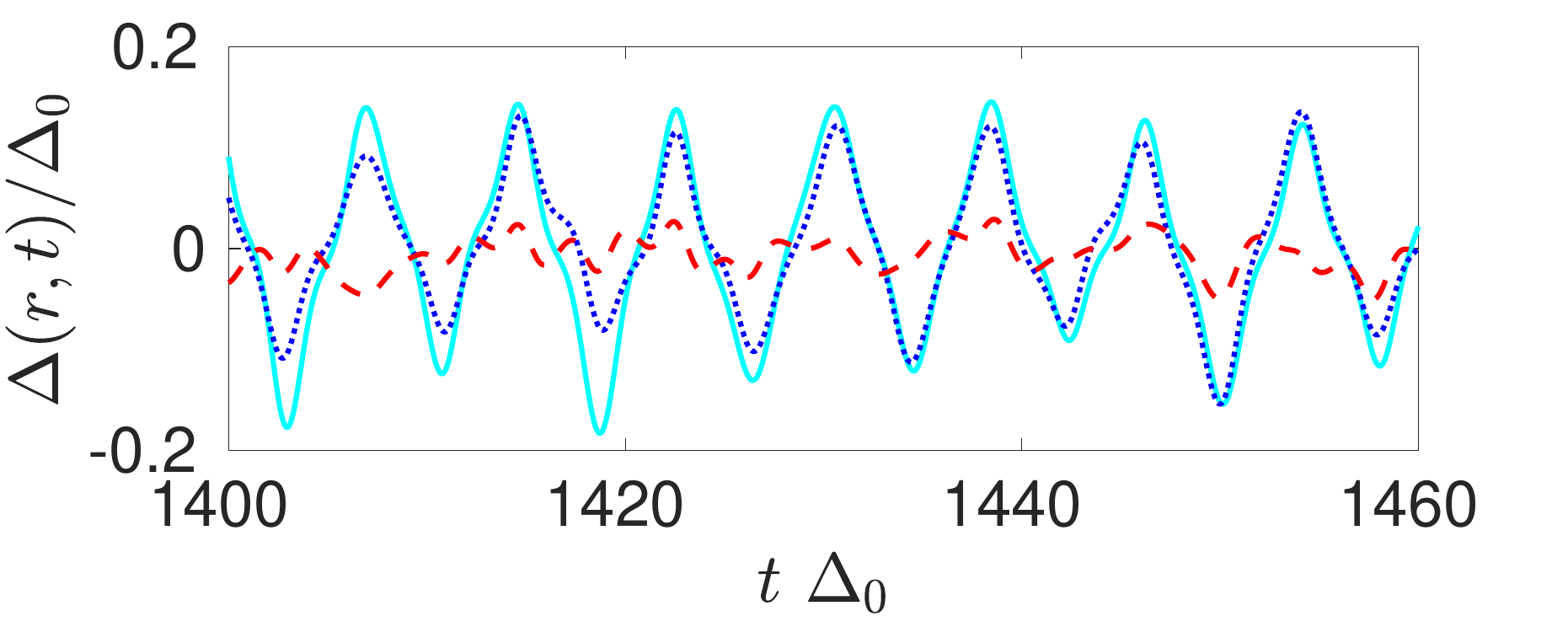} }
		\caption{ \subref{fig.BdG_V0p001_spatial}. The spatial distribution of the order parameter at different times (black dots in \subref{fig.BdG_V0p001_t1800}). \subref{fig.BdG_V0p001_t1800} $\sim$ \subref{fig.BdG_V0p001_t1460} Time dependence of the order parameter at positions marked in \subref{fig.BdG_V0p001_spatial}. The disorder strength is $V=0.001$ and the rest of parameters are the same as in Figure~\ref{Fig:L100_meangap_vs_Vweak}. 
		}\label{Fig:2D_BdG_V0p001}
	\end{center}
\end{figure}

In the presence of a very weak disorder $V=0.001$, the spatially averaged order parameter $\langle\Delta(r,t)\rangle$ experiences undamped oscillations, in agreement with the BCS result \cite{collado2021emergent,collado2023dynamical} up to around $t~\Delta_0 \sim 600$. For longer times, $\langle\Delta(r,t)\rangle$ suddenly experiences a rapid decay to a much smaller value with no change in the frequency of the oscillations, due to the spatial inhomogeneities induced by the driven dynamics, see video of the time dependence of the order parameter in the supplementary materials \cite{timecrystal_video} for further details. 
Figure~\ref{Fig:2D_BdG_V0p001} exhibits the spatial distribution of the order parameter at specific times. The most salient feature is the emergence at $t\Delta_0 \gtrsim 800$ of large islands characterized by an order parameter
rotated by $\pi$ with the respect to the largely homogeneous surroundings.

For a more quantitative description of the dynamics of these spatial inhomogeneities, we study the time evolution at three points: one in the spatially homogeneous region and two inside the islands characterized by a $\pi$ shift with respect to the homogeneous region. 
For $t~\Delta_0 \le 800$, the order parameter at those three sites oscillate synchronously. However, as mentioned earlier, when $t~\Delta_0 > 800$, the two sites inside the islands develop rather abruptly a phase shift $\pi$ with respect to the time crystal in the surrounding homogeneous region. We now have effectively three time crystals with the same frequency $\omega_d/2$ but different phases in the same sample. 
In this time scale, the decay of the spatial averaged order parameter $\langle \Delta(r,t) \rangle $ is much slower. Figure~\ref{Fig:fit_peak_var} and Appendix~\ref{app:peak_decay} reveal that the amplitude of the local maxima of the time crystal is approximately constant in this region. 

Finally, for $t~\Delta_0 \ge 1400$, $\langle\Delta(r,t)\rangle$ decays exponentially indicating that the time crystal gradually breaks down, likely due to finite size effects. 
The time crystal oscillation in the small island decays faster and finally breaks down. In the larger island, the time crystal oscillation is much more stable, although it finally synchronizes with the whole sample for sufficiently long times. Results in Figure~\ref{fig.BdG_V0p001_t1460} illustrate those features.

\begin{figure}[!htbp]
	\begin{center}
		\subfigure[]{ \label{fig.BdG_V0p6_spatial}
			\includegraphics[width=15cm]{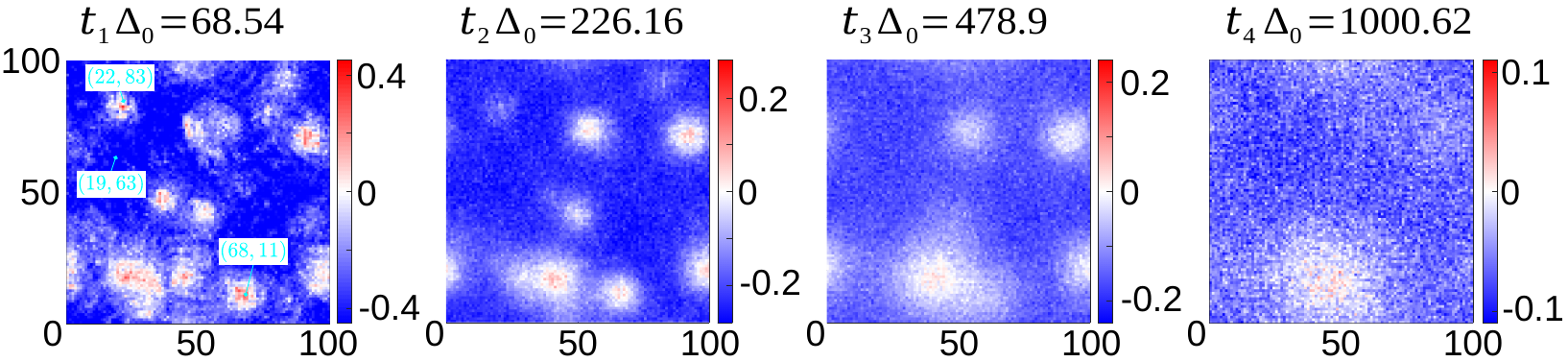}  }
		\subfigure[]{ \label{fig.BdG_V0p6_t1500}
			\includegraphics[width=7.5cm]{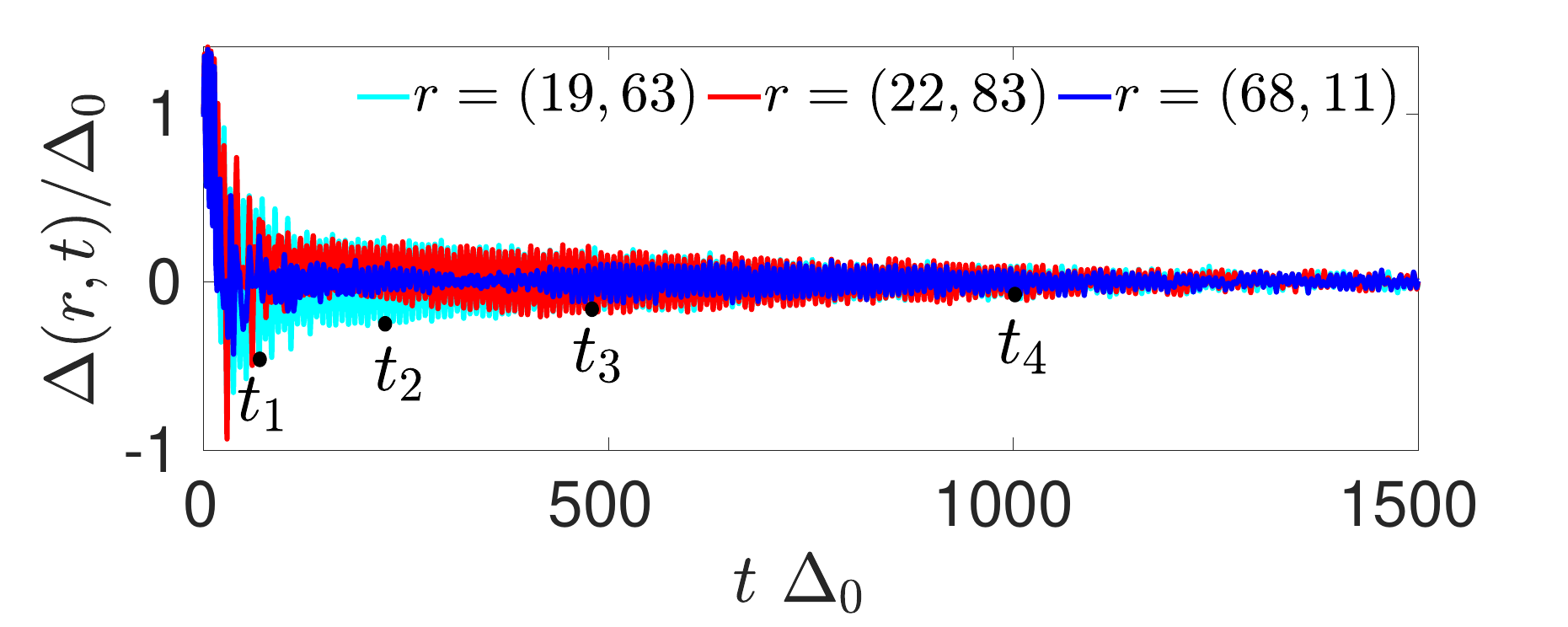}}
		\subfigure[]{ \label{fig.BdG_V0p6_t110}
			\includegraphics[width=7.5cm]{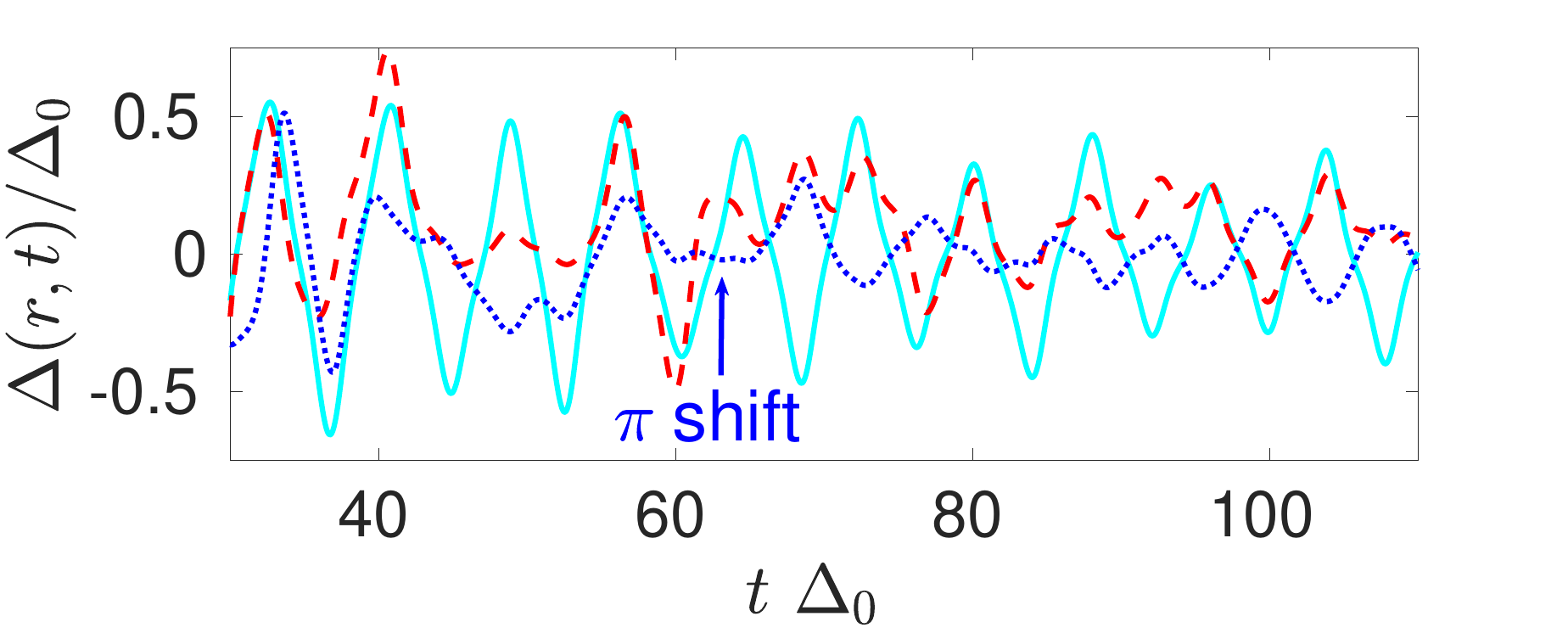}}
		\subfigure[]{ \label{fig.BdG_V0p6_t260}
			\includegraphics[width=7.5cm]{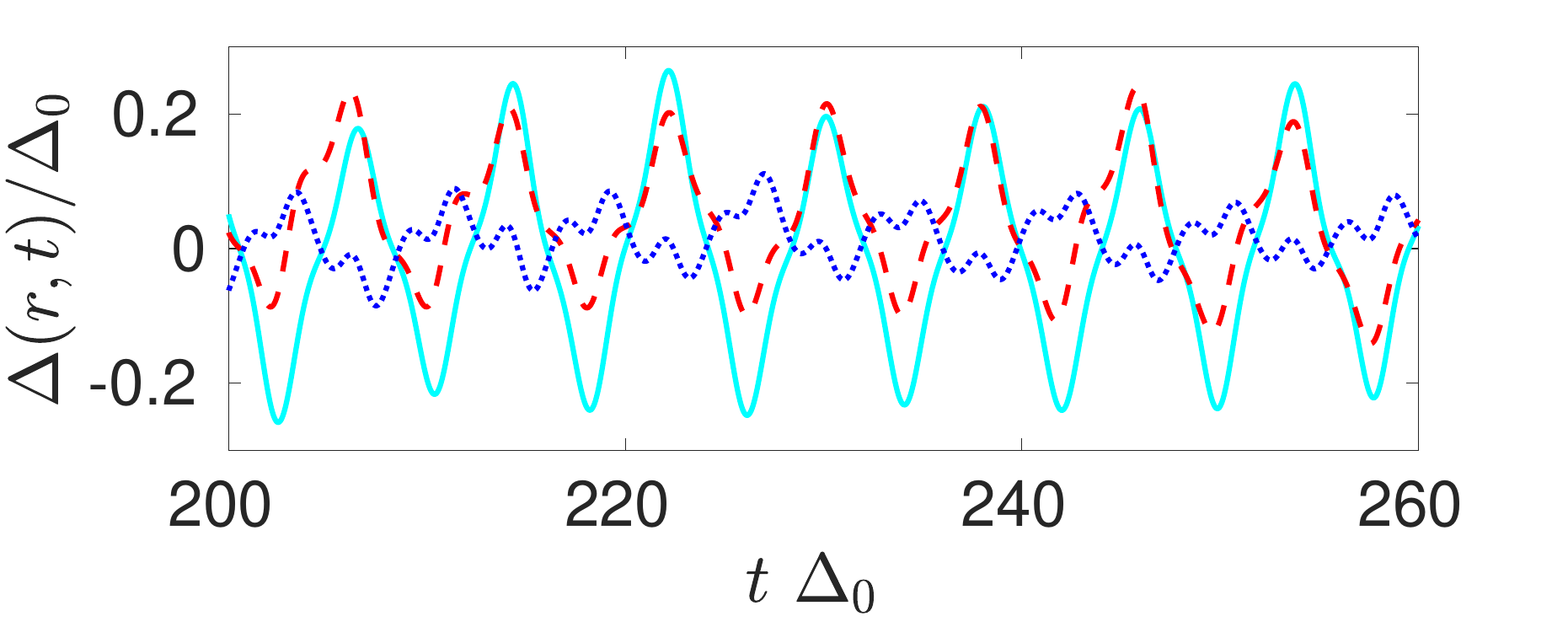}}
		\subfigure[]{ \label{fig.BdG_V0p6_t510}
			\includegraphics[width=7.5cm]{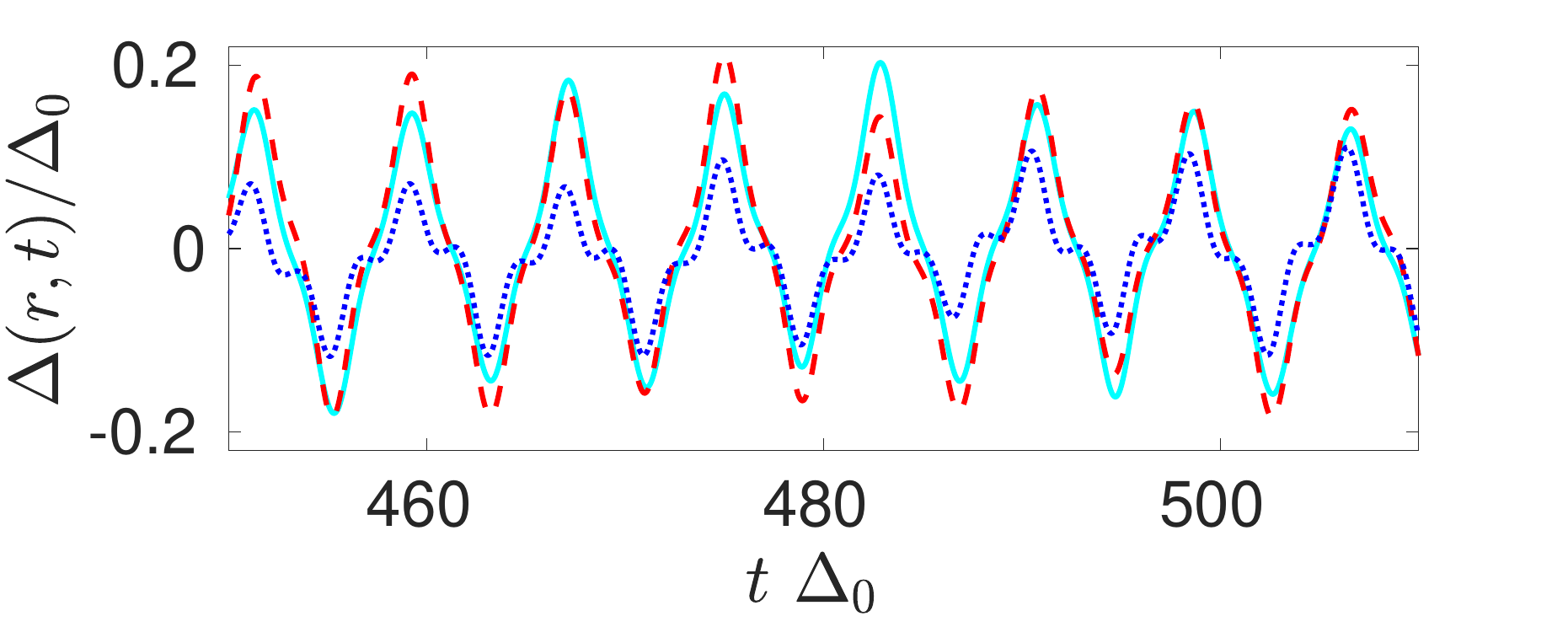} }
		\caption{ \subref{fig.BdG_V0p6_spatial}. The spatial dependence of the order parameter for $V = 0.6$ at different times, black points in \subref{fig.BdG_V0p6_t1500}. \subref{fig.BdG_V0p6_t1500} $\sim$ \subref{fig.BdG_V0p6_t510} Time dependence of the order parameter $\Delta(r,t)$ at positions marked in \subref{fig.BdG_V0p6_spatial}.  The other parameters are the same as in Figure~\ref{Fig:L100_meangap_vs_Vweak}. }\label{Fig:2D_BdG_V0p6}
	\end{center}
\end{figure}

For a stronger disorder $V = 0.6$, spatial inhomogeneities, characterized by the presence of many additional smaller islands, appear when the system enters into the time crystal phase, see Figure~\ref{Fig:2D_BdG_V0p6}. We also pick three sites in the sample, two in the islands and one in the normal region. The time crystal in the small island, position $r=(22,83)$, synchronizes with the whole sample very fast, at around $t~\Delta_0 \sim 200$. The corresponding video in the supplementary material  \cite{timecrystal_video} also shows that this island disappears when $t~\Delta_0 \ge 200$. Nonetheless, for position $r=(68,11)$, which is in a larger island, and near the islands with the same phase, the time crystal oscillation persists with a phase shift for a longer time, even with two small peaks at the shoulder, see Figure~\ref{fig.BdG_V0p6_t260}. The quantum coherence between those near islands protect the time crystal. However, it finally synchronizes with the whole sample rather abruptly at $t~\Delta_0 \ge 500$, as shown in Figure~\ref{fig.BdG_V0p6_t510}. Due to the synchronization, the time crystal oscillations continue for much longer times $t~\Delta_0 > 1000$ until finite size effects become important.

In all cases, the size of the islands gets reduced with respect to its initial value until the mentioned synchronization occurs.  

We show next that the growth in time of the spatial instability is exponential. The build up of this instability starts with extremely small spatial inhomogeneities at relatively early times which likely cannot be detected until much later times. 

\begin{figure}[!htbp]
	\begin{center}
		\subfigure[]{ \label{fig.fit_V0p001_tiny_var}
			\includegraphics[width=8.cm]{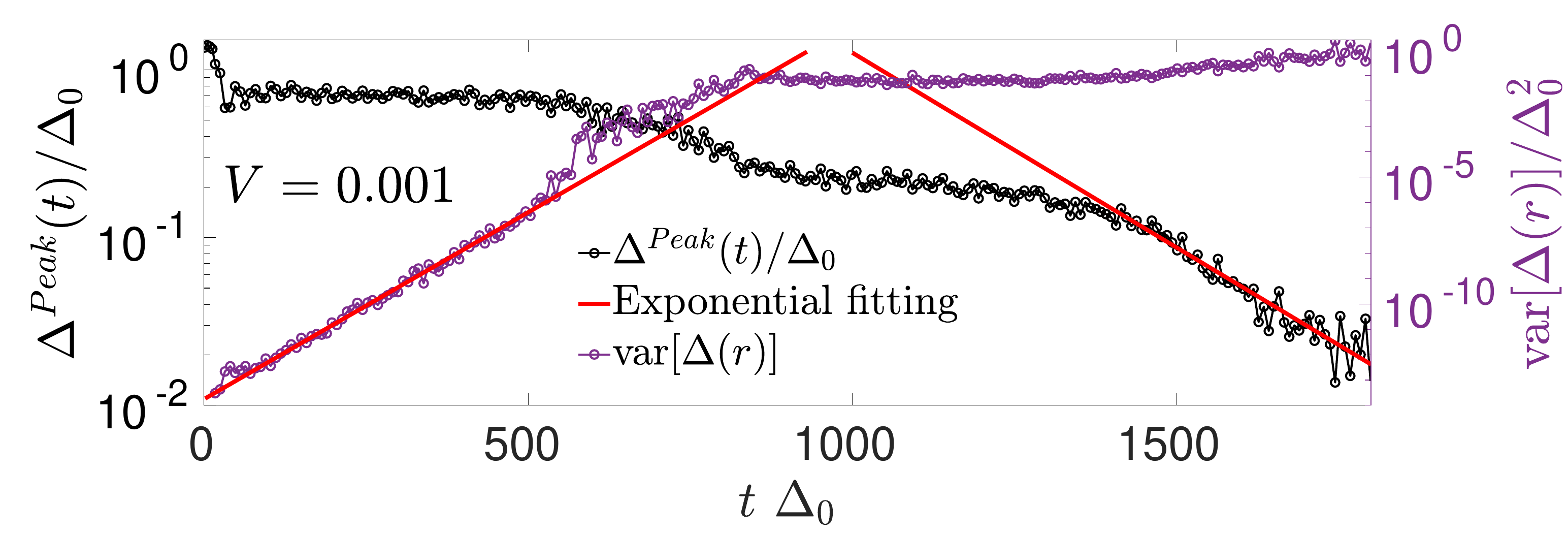}  }
		\subfigure[]{ \label{fig.fit_V0p01_tiny_var}
			\includegraphics[width=8.cm]{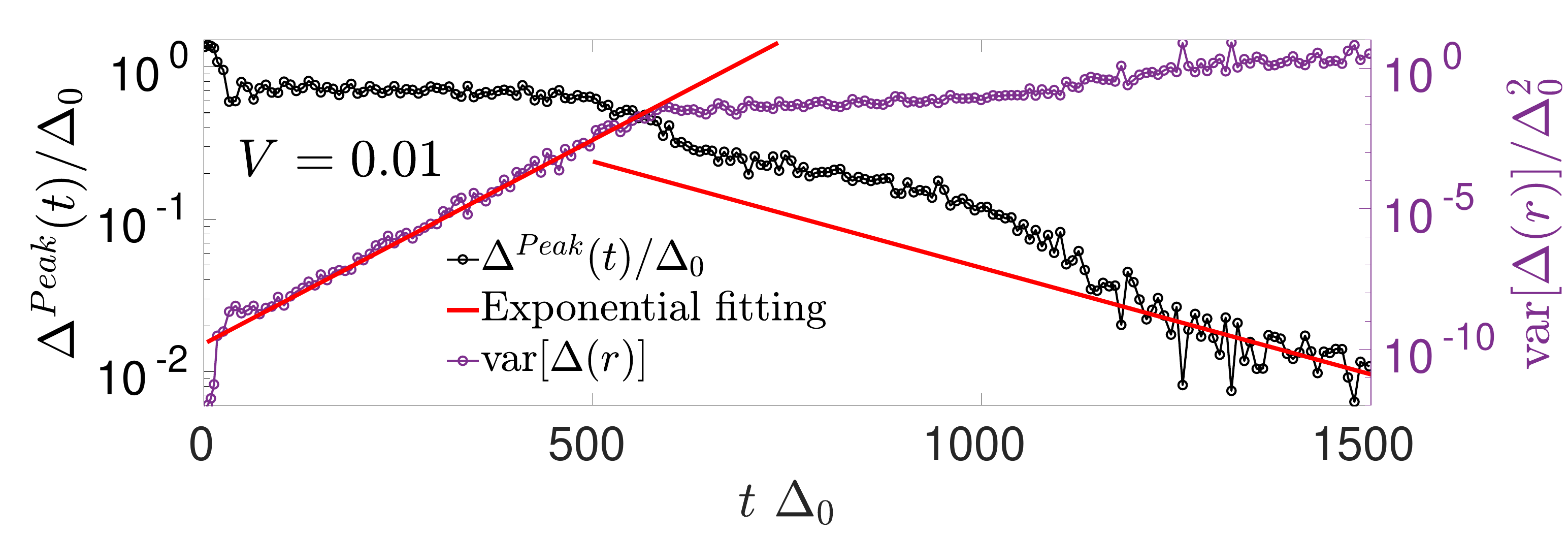}  }
		\subfigure[]{ \label{fig.fit_V0p1_weak_var}
			\includegraphics[width=8.cm]{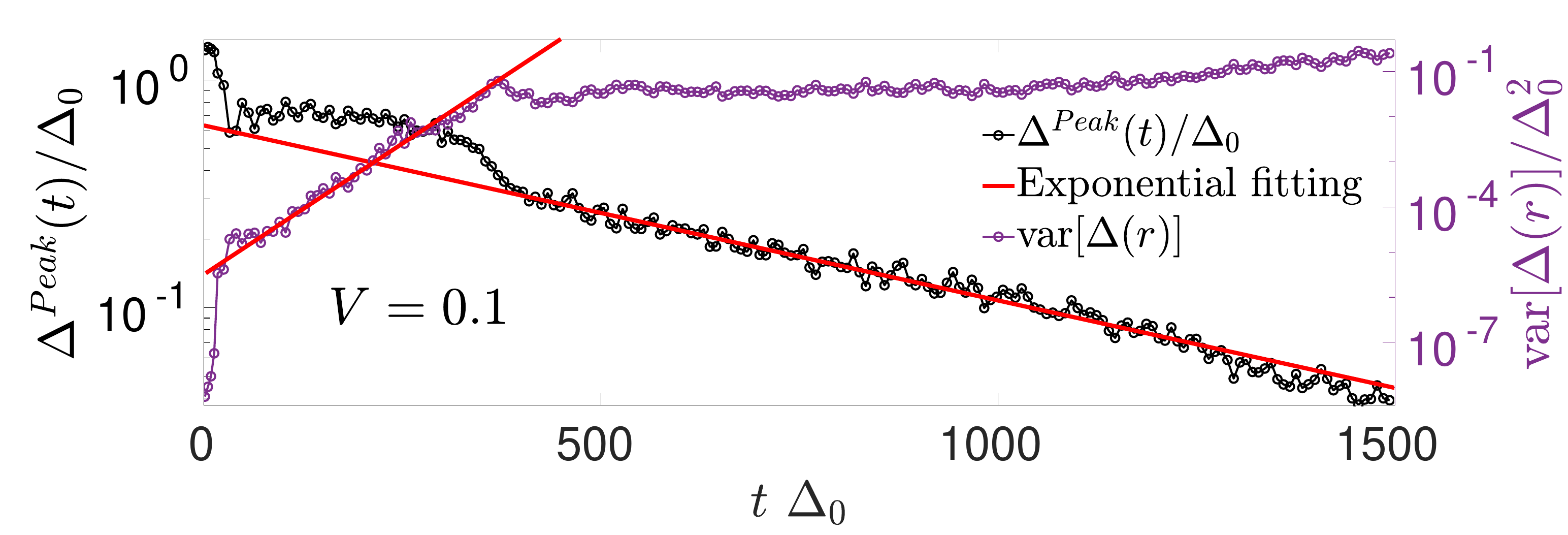}  }
		\subfigure[]{ \label{fig.fit_V0p6_weak_var}
			\includegraphics[width=8.cm]{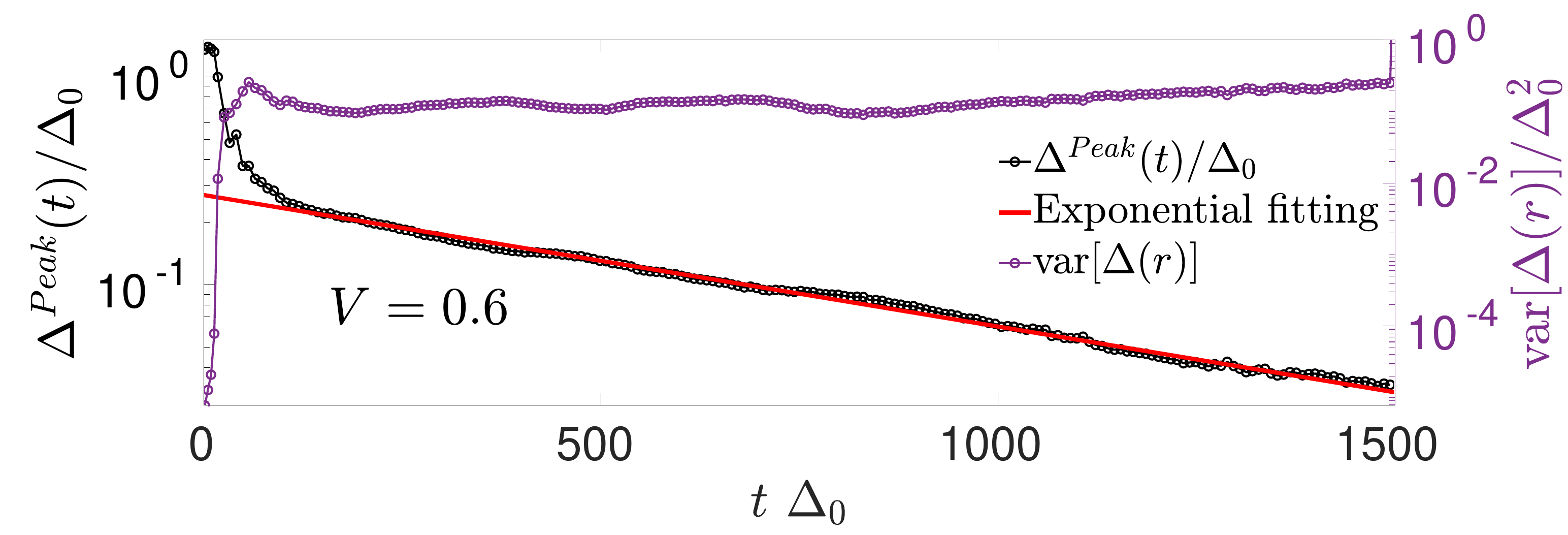}  }
		\caption{Left y axis: The peaks of the spatial average of the order parameter $\Delta^{Peak}(t)$, and the best exponential fittings (red line). Appendix~\ref{app:peak_decay} explains how to obtain the $\Delta^{Peak}(t)$ from the spatial averaged order parameter $\langle \Delta(r,t) \rangle$.
			Right y axis: the corresponding spatial variance of the order parameter Var$[\Delta(\bm{r})]$ as a function of time. 
		}\label{Fig:fit_peak_var}
	\end{center}
\end{figure}

\subsection{The exponential increase of the spatial instability}

In order to gain a better understanding of the relation between spatial instability and the decay of the time crystal oscillations, we compute the variance of the order parameter Var$|\Delta(\bm{r})| = \langle \Delta^2(\bm{r}) \rangle - \langle \Delta(\bm{r}) \rangle^2$ as a function of time, see Figure~\ref{Fig:fit_peak_var}. For a sufficiently weak disorder $V = 0.001$, there is a region in which the time crystal oscillations do not decay and spatial inhomogeneities are so small that are likely undetectable. However, Var$|\Delta(\bm{r})|$ has started to increase exponentially, indicating the build up of spatial instabilities \cite{dzero2009cooper} which is the seed for the eventual emergence of spatial patterns. The exponential growth of Var$|\Delta(\bm{r})|$ terminates when the spatial patterns are completely developed around $t \Delta_0 \ge 800$. Within the time range $800 \le t \Delta_0 \le 1400$, the peaks of the order parameter $\Delta^{Peak}(t)$, which characterize the amplitude of the time crystal oscillations, is approximately a constant, as shown in Figure~\ref{fig.fit_V0p001_tiny_var} and Appendix~\ref{app:peak_decay}. For even longer time $t~\Delta_0 \ge 1400$, $\Delta^{Peak}(t)$ decays exponentially indicating that the time crystal gradually breaks down, due to finite size effects.
For a stronger disorder, $V \ge 0.1$, Var$|\Delta(\bm{r})|$ increases exponentially until the spatial pattern is well formed. However, $\Delta^{Peak}(t)$ decays exponentially soon after the formation of the spatial patterns because finite size effects become important for earlier times due to the existence of smaller unstable islands.

\begin{figure}[!htbp]
	\begin{center}
		\subfigure[]{ \label{fig.BdG_V0p001_to_V0_2}
			\includegraphics[width=15cm]{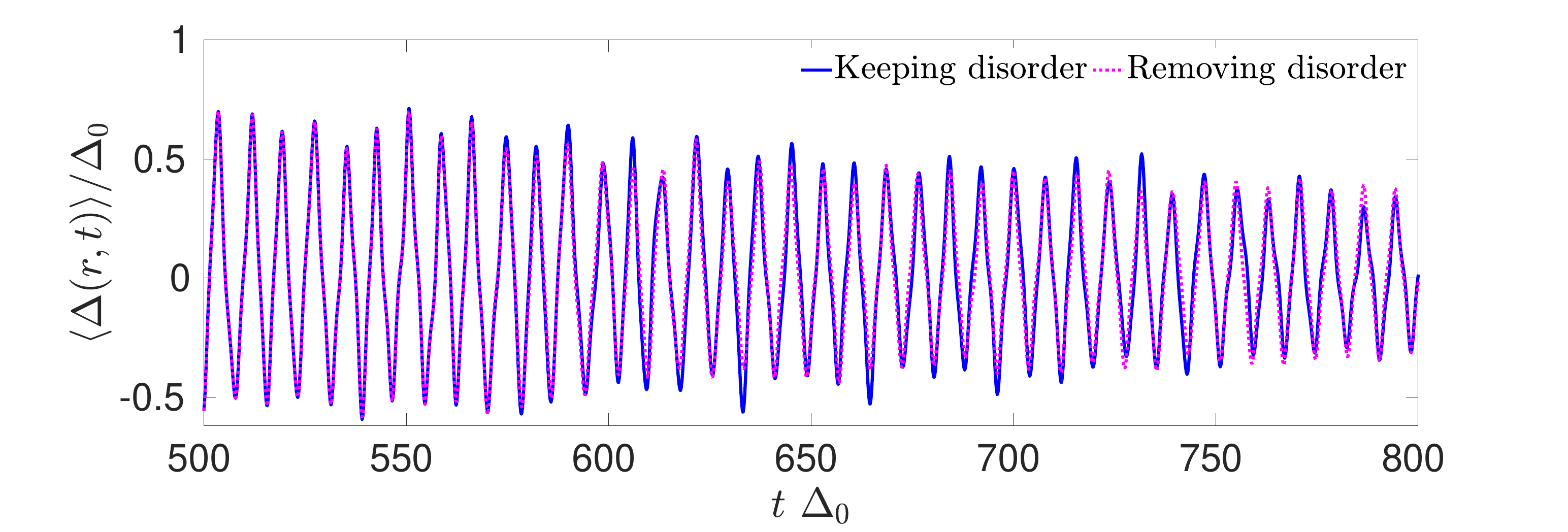}  }
		\subfigure[]{ \label{fig.BdG_V0p001_to_V0_3}
			\includegraphics[width=15cm]{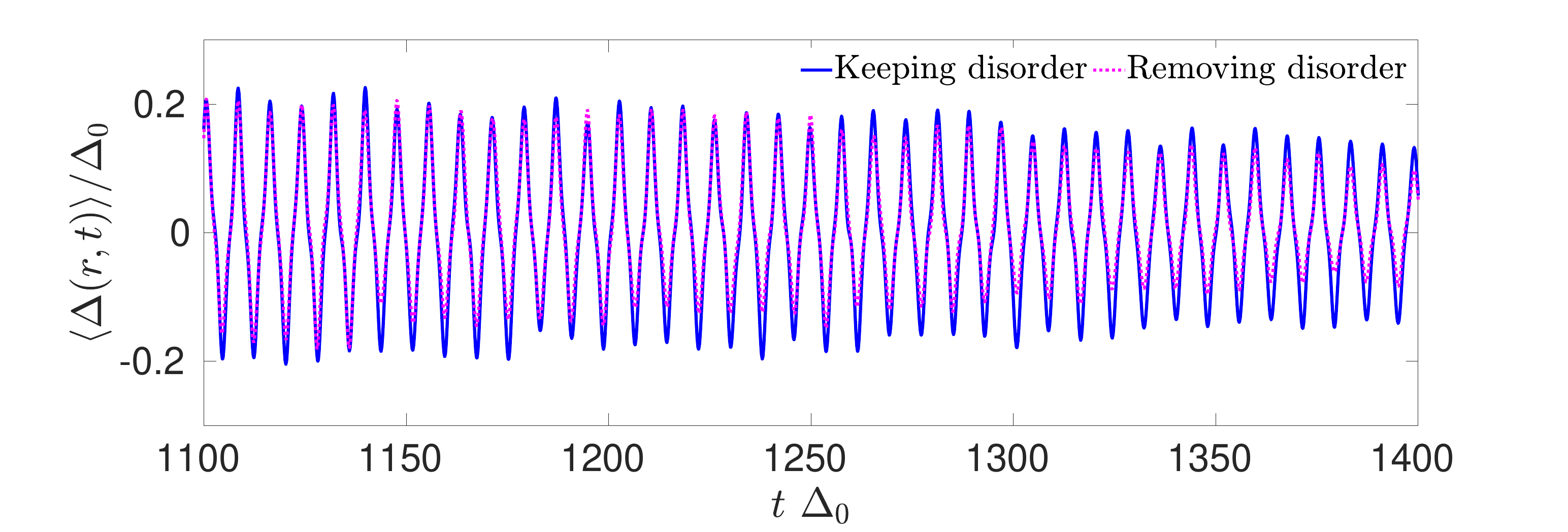}  }
		\caption{
			The spatial averaged order parameter as a function of time for a disorder strength $V=0.001$. The blue line corresponds to the time evolution keeping the disordered potential at all times. The pink dashed line describes the time evolution if this disordered potential ($V = 0.001$) is removed abruptly when the system enters the time crystal phase (pink dashed line). The system size is $N=100\times 100$ and the coupling constant $U_0 = -6$. The chemical potential is fixed at $\mu = 0$. The driving amplitude is $\alpha = 0.25$ and the driving frequency is $\omega_d = 0.8\times 2\langle\Delta(r)\rangle$.  }\label{Fig:2D_BdG_V0p001_to_V0_com}
	\end{center}
\end{figure}

\section{Time crystal phase after turning off disorder}\label{sec:turningoff}

In previous sections we have shown that though a very weak amount of disorder is needed to observe spatial inhomogeneities in the time crystal phase, the features of these emergent spatial inhomogeneities do not depend on disorder if the disorder strength is sufficiently weak. In other words, while a spatial perturbation is necessary for the generation of spatial patterns, those patterns do not depend much on the details of the perturbation. In this section, we provide further support to this fact by solving the time dependent BdG equations with a very weak disorder strength and then, when the homogeneous time crystal is formed, the disordered potential is turned off. The time evolution of the system after this point is therefore governed by the BdG formalism in the clean limit.     
In this way, the perturbations are introduced to the system self-consistently. The comparison of the spatial averaged order parameter in both cases are illustrated in Figure~\ref{Fig:2D_BdG_V0p001_to_V0_com}. In the early time evolution, the oscillations completely overlap with each other. While when $t~\Delta_0 \ge 600$, the amplitude of the oscillation in the absence of disorder is even smaller than the system with disorder all the time. This is because the effect of removing disorder is another quench in the dynamics which further weakens the amplitude of the oscillations.

\begin{figure}[!htbp]
	\begin{center}
		\includegraphics[width=16cm]{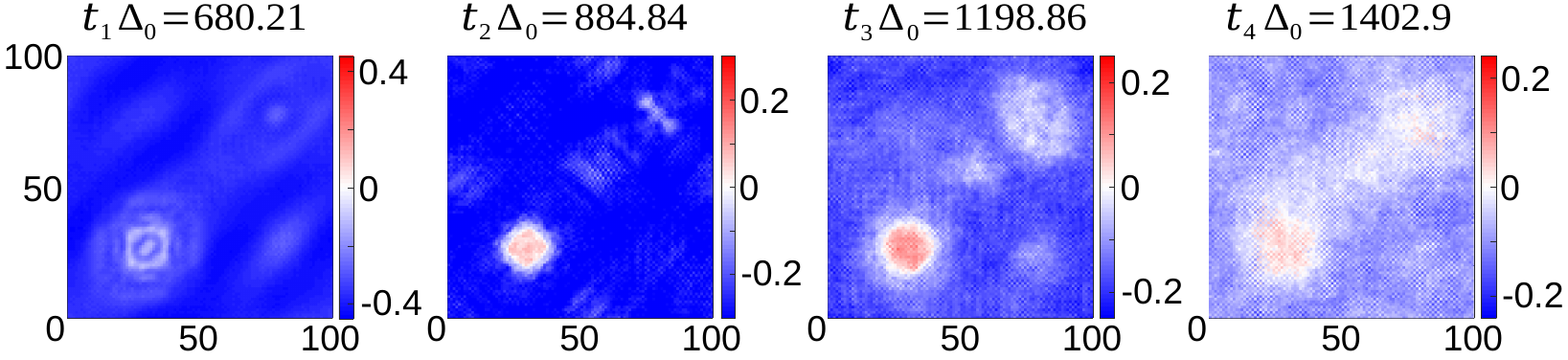} 
		\caption{The spatial distribution of the order parameter at the same times $t_1, t_2, t_3, t_4$ as in Figure~\ref{Fig:2D_BdG_V0p001}, so that we can compare both results directly. The initial state is the result of solving the static BdG equations with a random potential of  strength $V=0.001$. The disordered potential is suddenly removed when the system enters in the time crystal phase.
			The parameters are: system size $N=100\times 100$, coupling constant $U_0 = -6$, chemical potential $\mu = 0$, driving amplitude $\alpha = 0.25$ and driving frequency $\omega_d = 0.8\times 2\langle\Delta(r)\rangle$. }\label{Fig:2D_BdG_V0p001_to_V0}
	\end{center}
\end{figure}

\begin{figure}[!htbp]
	\begin{center}
		\includegraphics[width=16cm]{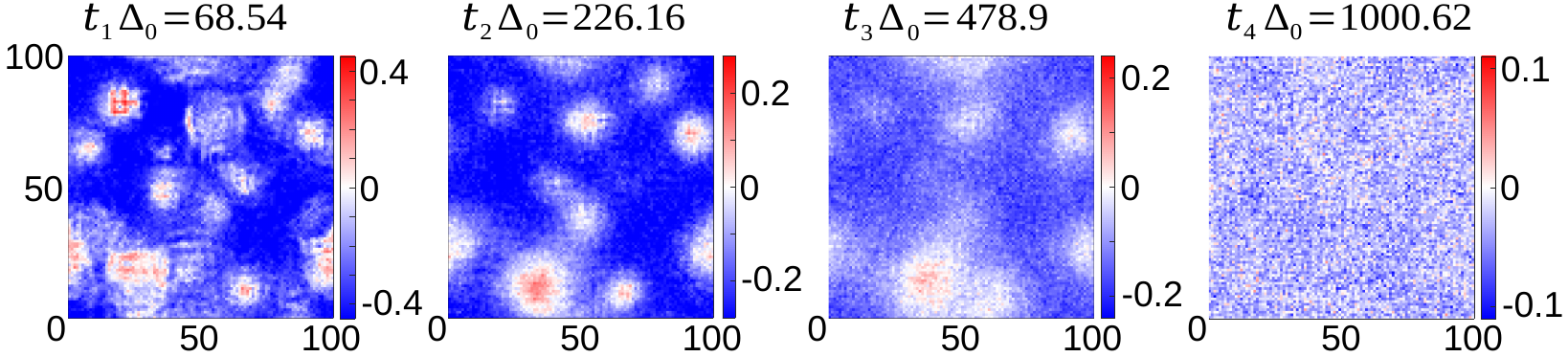} 
		\caption{The spatial distribution of the order parameter at four times $t_1,t_2, t_3, t_4$. The initial state is the solution of the BdG equations for $V=0.6$ as in Figure~\ref{Fig:2D_BdG_V0p6}. Disorder is removed when the system enters the time crystal phase. The other parameters are the same as in Figure~\ref{Fig:2D_BdG_V0p001_to_V0}. }\label{Fig:2D_BdG_V0p6_to_V0}
	\end{center}
\end{figure}

We now proceed with the comparison of the spatial distribution of the order parameter in both scenarios: with and without removal of disorder when the time crystal oscillations start at $t\Delta_0 \sim 19.6$ as is marked in Figure~\ref{fig.L100_meangap_vs_V_t100}. 
Results for $V = 0.001$ are depicted in Figure~ \ref{Fig:2D_BdG_V0p001_to_V0} (removing disorder $V = 0.001$  when time crystal oscillations start) and in Figure~\ref{Fig:2D_BdG_V0p001} (keeping the disorder all the time). For stronger disorder $V = 0.6$, results removing this disorder are presented in Figure~\ref{Fig:2D_BdG_V0p6_to_V0} and keeping it at all times in Figure~\ref{Fig:2D_BdG_V0p6}. 

Interestingly, the time evolution of the spatial distribution of the order parameter is qualitatively similar in both scenarios and for both disorder strengths. Although the details of the spatial structures are in general different, the islands with different time crystal phases appear at similar positions. Moreover, the different stages of the evolution: starting of the time crystal phase, formation of islands with out-of-phase time crystal oscillations and the eventual synchronization for long times occurs in all cases.  This suggests that the presence of disorder is not necessary to induce spatial patterns during the dynamics. 
Indeed, the specific details of the emergent spatial structure depend mostly on the chosen initial state. Disorder, provided it is not too strong that breaks the time crystal oscillations, only has a quantitative, not qualitative, effect on the dynamics.

\section{The breaking of the time crystal phase for stronger disorder}\label{sec:strongdis}

\begin{figure}[!htbp]
	\begin{center}
		\subfigure[]{ \label{fig.BdG_V0p7_t300}
			\includegraphics[width=7.5cm]{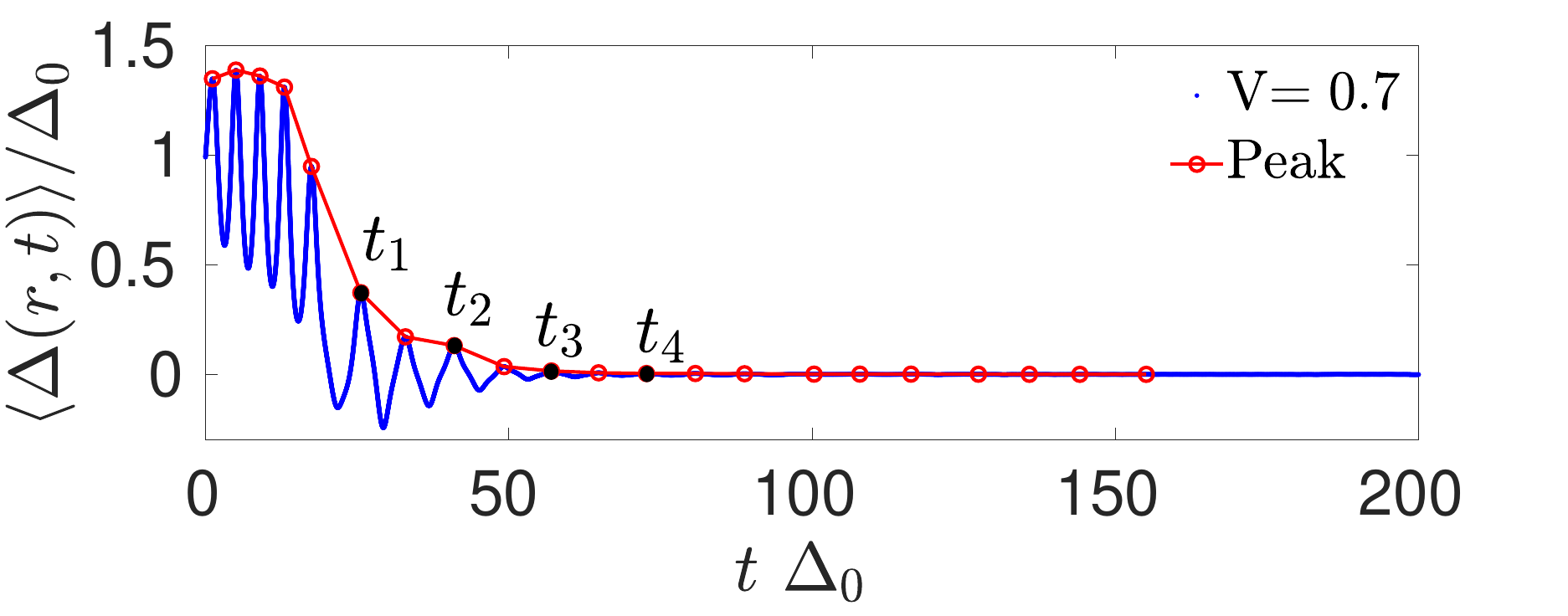}}
		\subfigure[]{ \label{fig.BdG_V0p7_t300_log}
			\includegraphics[width=7.5cm]{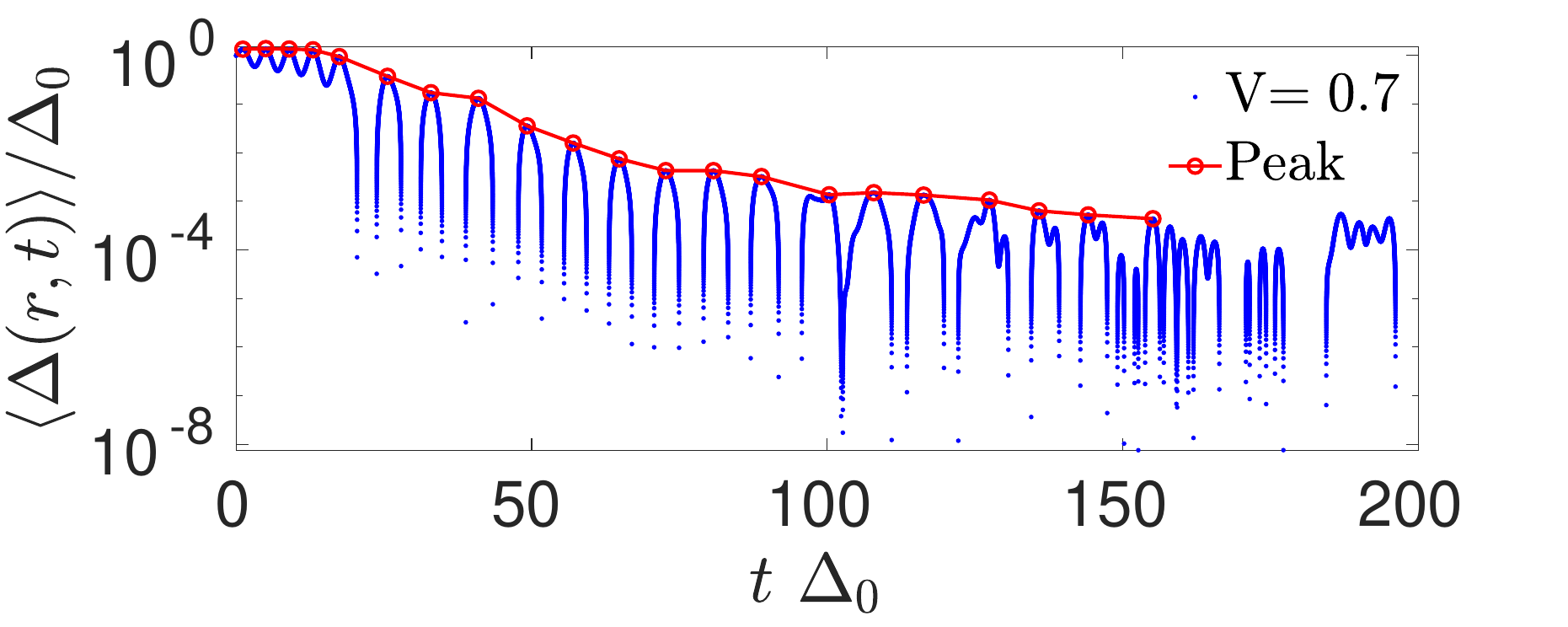}}
		\subfigure[]{ \label{fig.BdG_V0p7_spatial}
			\includegraphics[width=15cm]{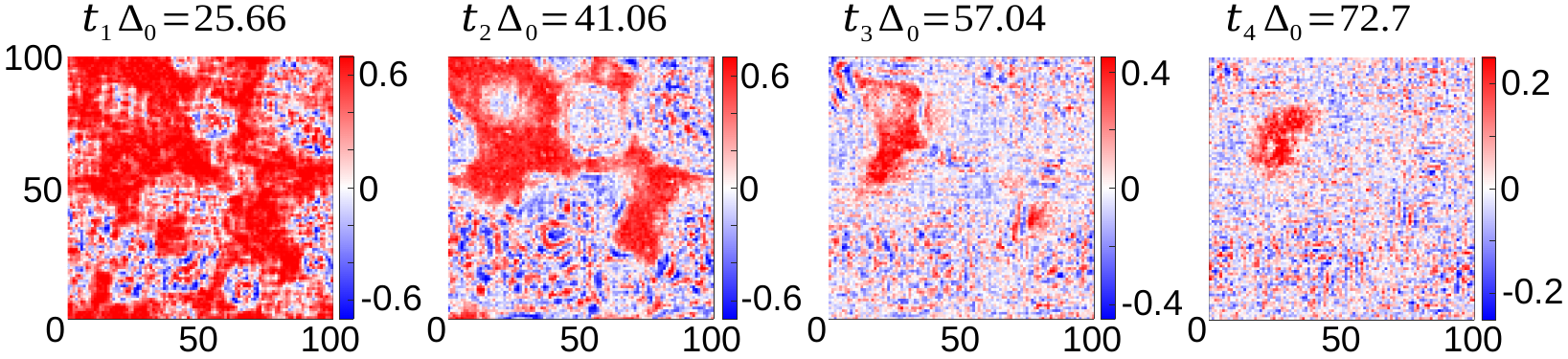}  }
		\caption{The time evolution of the spatial average order parameter $\langle \Delta(r,t) \rangle$ (blue dots), where the peaks (local maxima of the order parameter) are represented by red circles, at $V = 0.7$ which is very close to the critical disorder signaling the suppression of the time crystal phase. The rest of parameters are the same as in previous figures. \subref{fig.BdG_V0p7_t300} Linear scale. \subref{fig.BdG_V0p7_t300_log} Semi-logarithmic scale. Since few time crystal oscillations, namely, oscillations at half the driving frequency, are observed, the time crystal phase only exists for very short times. Even a small increase of disorder would suppress it completely. \subref{fig.BdG_V0p7_spatial} Spatial distribution of the order parameter at times $t_1, t_2, t_3, t_4$ in \subref{fig.BdG_V0p7_t300} that illustrates the gradually breaking of the time crystal phase for long times. 
		}\label{Fig:peak_V0p7}
	\end{center}
\end{figure}

For even stronger disorder ($V \gtrapprox 0.7$), but still deep in the metallic phase in the static limit, the time crystal breaks down. Results depicted in Figure \ref{Fig:peak_V0p7} indicate that at $V = 0.7$, around the critical disorder, the spatial average of the order parameter experiences for early times several oscillations still at the time crystal frequency. The time crystal gets frozen in some parts of the sample. These regions grow with time, and with disorder, eventually inducing the complete disappearance of the time crystal phase at any time. 
See the video in the supplemental material \cite{timecrystal_video} for further details.

\begin{figure}[!htbp]
	\begin{center}
		\subfigure[]{ \label{fig.meangap_V1p0}
			\includegraphics[width=5cm]{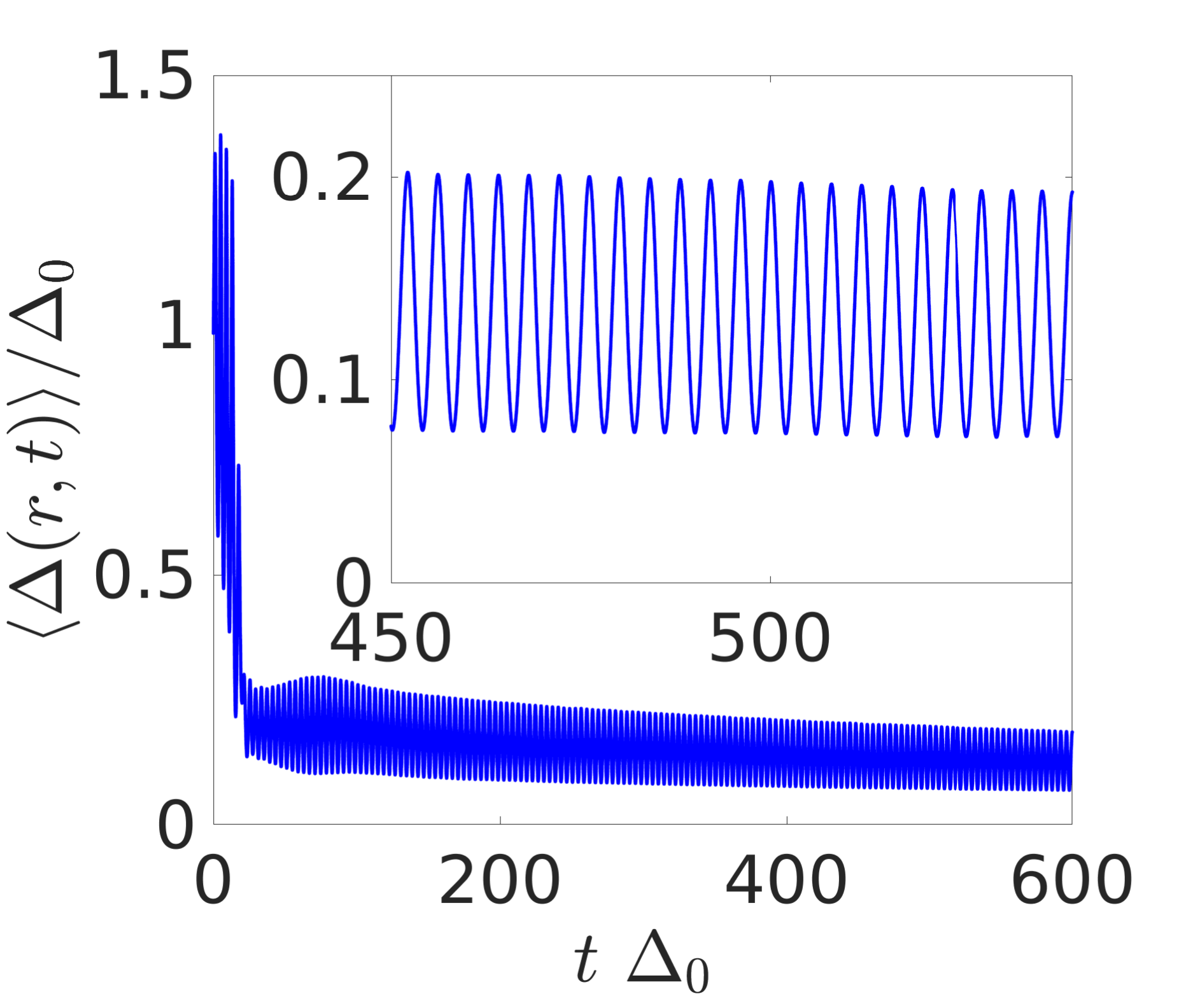}  }
		\subfigure[]{ \label{fig.spatialgap_V1p0}
			\includegraphics[width=5cm]{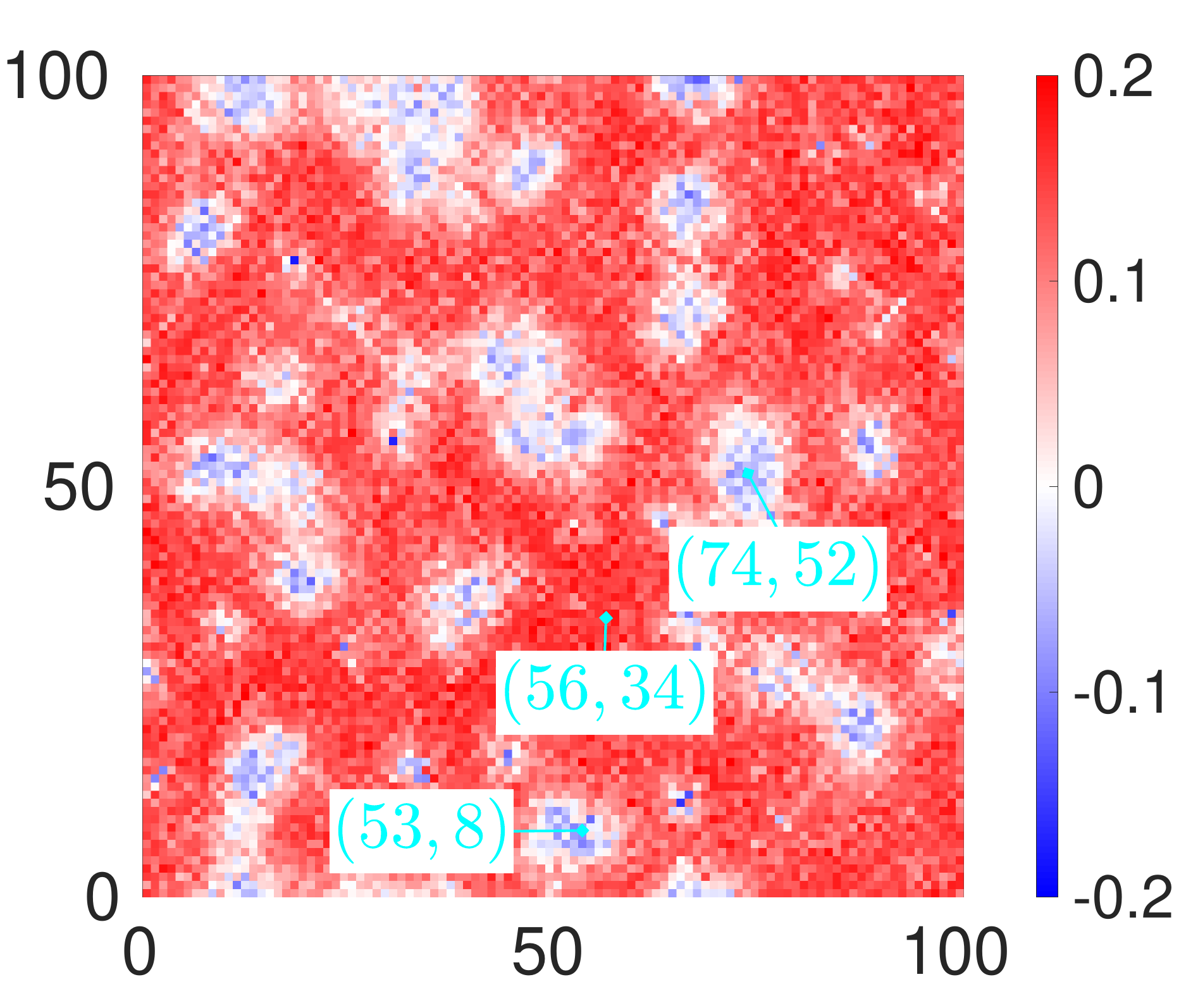}  }
		\subfigure[]{ \label{fig.positiongap_V1p0}
			\includegraphics[width=5cm]{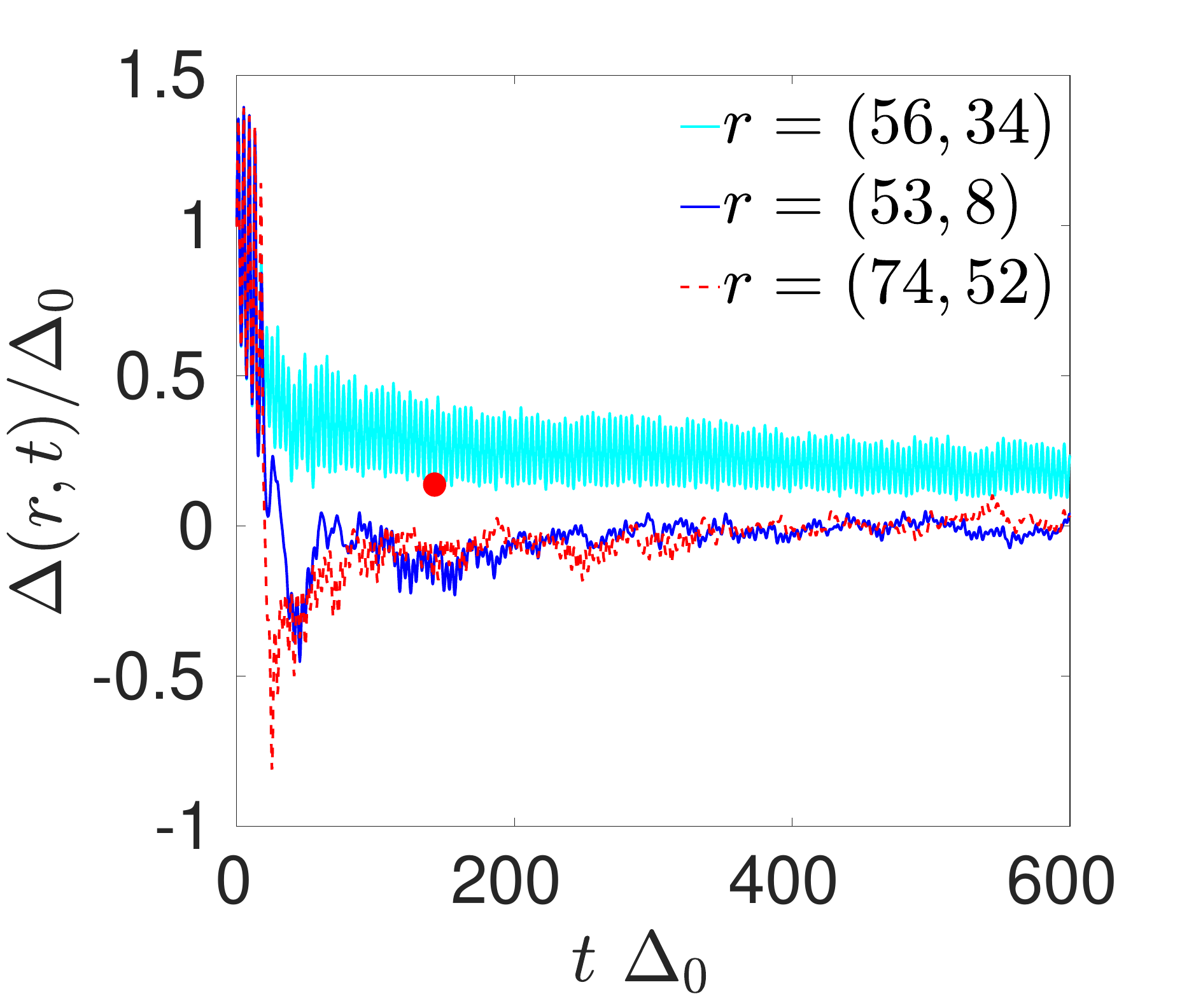}  }
		\caption{\subref{fig.meangap_V1p0}.   
			Time evolution of the spatial average order parameter $\langle \Delta(r,t) \rangle$ at $ V  = 1$ above the time crystal phase transition. \subref{fig.spatialgap_V1p0}. The spatial distribution of the order parameter at time $t \sim 142.88\Delta_0$ is marked by the red point in \subref{fig.positiongap_V1p0}. 
		We stress that at this disorder strength there is no time crystal phase and therefore $\langle \Delta(t) \rangle$ oscillates at the driving frequency. The other parameters are system size $N=100\times 100$, coupling constant $U_0 = -6$, chemical potential $\mu = 0$. The driving amplitude is $\alpha = 0.25$ and frequency $\omega_d = 0.8\times 2\langle\Delta(r)\rangle$. 
	}\label{Fig:2D_BdG_V1p0}
	\end{center}
\end{figure}

For stronger disorder strength, the time crystal phase is no longer observed. As mentioned earlier, this does not mean that the system undergoes an insulating
transition.  
In order to illustrate the nature of the superconducting state after the destruction of the time crystal phase, we depict in Figure~\ref{Fig:2D_BdG_V1p0} the time evolution of the order parameter at $V = 1$, still on the metallic side. 
A video of the full time evolution at each site can be found in the supplementary material \cite{timecrystal_video}. We still observe oscillations in time of the spatial average of the order parameter but at the driving frequency confirming thus the termination of the time crystal phase. As expected, due to the stronger disorder,  space inhomogeneous becomes more prevalent. Many islands emerges, whose order parameter has a negative sign and its size is much larger than the superconducting coherence length in the equilibrium state which is about one lattice spacing in the strong coupling limit investigated. Time oscillations are suppressed in these regions and the amplitude of the order parameter is heavily suppressed which likely indicates that long-order is lost and Anderson localization effects \cite{anderson1958absence}
induced by disorder start to become important. We speculate that eventually these islands will become larger and insulating with superconductivity completely lost.  By contrast, in the largely homogeneous regions surrounding the islands the order parameter oscillates at the driving frequency.

\section{Conclusions}\label{sec:con}

 We have studied the space-time dynamics of a BdG superconductor driven by an oscillating coupling constant. For no disorder, we reproduce the full phase diagram obtained from a simpler BCS approach \cite{collado2021emergent,collado2023dynamical}
 as a function of the amplitude and frequency of the driving. Some relatively small discrepancies with the BCS result are traced back to the simplified spectrum used in those studies. In the rest of the paper, we have focused on the range of parameters leading to a time crystal phase characterized by oscillations in time of the order parameter at a frequency half of the driving frequency. We have shown that the time crystal phase is robust in the presence of a sufficiently weak disordered potential. 
 
 For sufficiently long times, that for weak disorder do not depend on the disorder strength, we have reported the spontaneous generation of spatial patterns modeled by large islands embedded in the otherwise homogeneous order parameter. Each of these islands, whose typical length is much larger than the superconducting coherence length at equilibrium, is itself a time-crystal with the same frequency as the homogeneous one but with a initial time shift, namely a phase difference, of $\pi$. After its emergence, these islands-like structures of different sizes becomes gradually smaller, though with the same phase shift, and eventually they become synchronized with the homogeneous regions, namely, for sufficiently long times the island disappear so that the only spatial dependence is that induced directly by disorder. The dynamics for stronger disorder $V \lesssim 0.7$ is qualitatively similar, the time crystal phase persists but both the shape of the spatial inhomogeneities, and the time scale at which they are observed, becomes controlled by disorder. The breaking of the time crystal phase, that occurs at a disorder strength $V \approx 0.7$, is characterized by a rather abrupt diminution of the number of oscillations in time at half the driving frequency so the time crystal is only observed for short times. Moreover, the amplitude of the oscillations is sharply reduced as the critical disorder strength is approached. Interestingly, close to this critical disorder, larger parts of the sample become frozen in the sense that oscillations in time are restricted to specific spatial regions.  
 For even stronger disorder, still in the metallic phase, we observe oscillations at the driving frequency (not a time crystal), inside largely homogeneous regions of the sample combined with an increasing number of small islands with a reduced value of the amplitude of the order parameter, and almost no oscillations in time. The latter feature suggests that incipient localization effects may be already at play.
 
 Natural extensions of this study include the study of the time crystal phase in p-wave and d-wave BdG superconductors and the experimental confirmation of these results either by using the QED cavity techniques \cite{young2024observing} mentioned in the introduction or by tuning a cold atom setting close to a Feschbach resonance, in order to reach the strong coupling limit, where disorder is modeled by a quasi-periodic optical lattice.   
  
 \vspace{1.0cm}
\centerline{\bf Acknowledgments} 
A. M. G. G. and B. F. acknowledge support from the
National Natural Science Foundation of China (NSFC): 
Individual Grant No. 12374138, Research Fund for International Senior Scientists No. 12350710180, 
National Key R$\&$D Program of China (Project ID:2019YFA0308603), and 
China Postdoctoral Science Foundation (Grant numbers: 2023M732256, 2023T160409). 
A. M. G. G. acknowledges support from a Shanghai talent program. 
Z. C. acknowledges support from the National Key Research and Development Program of China (Grant No. 2020YFA0309000), NSFC of  China (Grant No.12174251), Natural Science Foundation of Shanghai (Grant No.22ZR142830),  Shanghai Municipal Science and Technology Major Project (Grant No.2019SHZDZX01).

\bibliographystyle{unsrt}
\bibliography{time_crystal.bib} 
\newpage

\appendix

\section{Dynamical phase diagram in the clean limit} \label{app:BCS_result}

\begin{figure}[!htbp]
	\begin{center}
		\subfigure[$N = 100\times 100, U_0 = -6$]{
			\includegraphics[width=8cm]{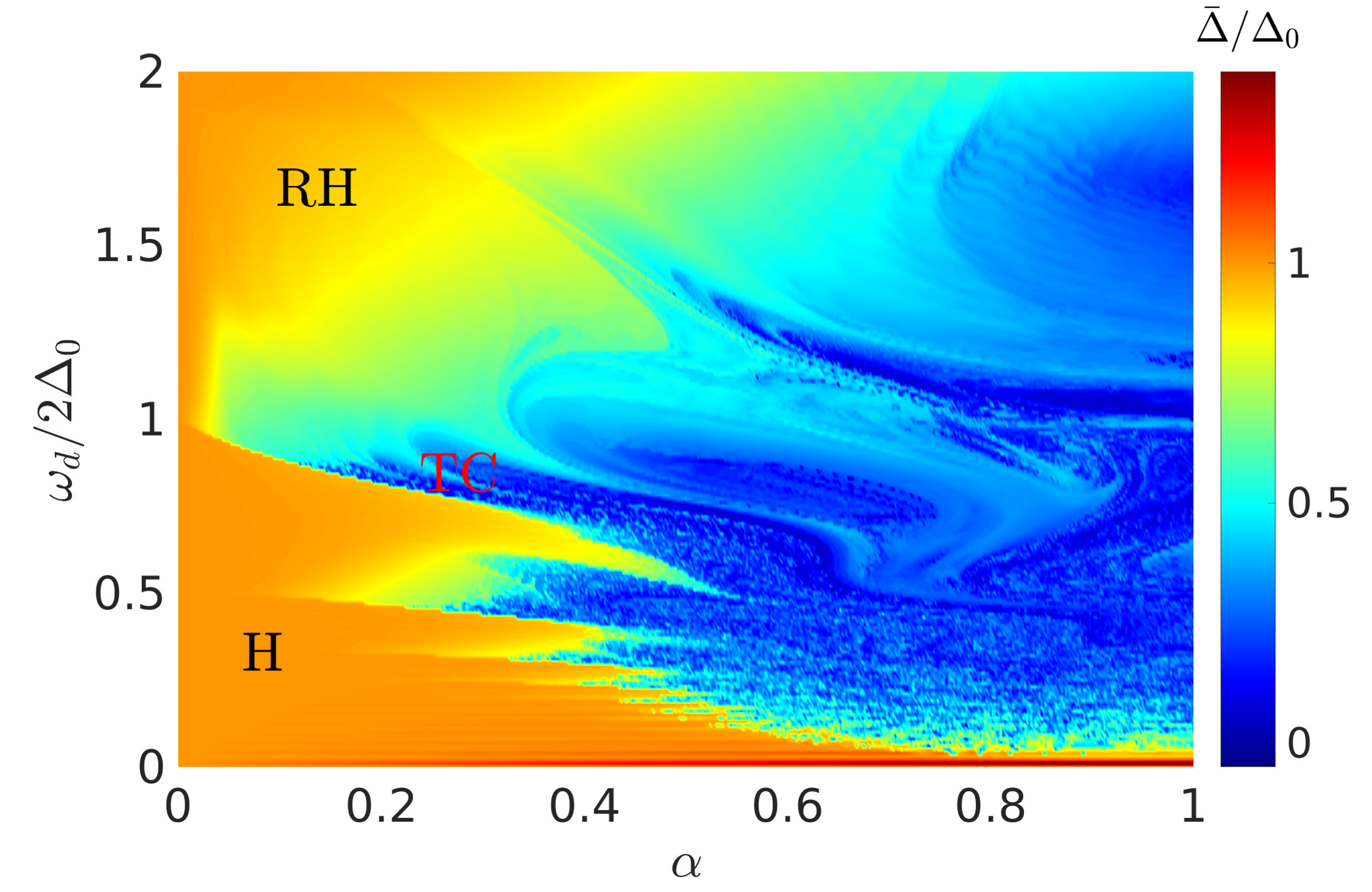}}
		\subfigure[$N = 400\times 400, U_0=-1.5$ ]{
			\includegraphics[width=8cm]{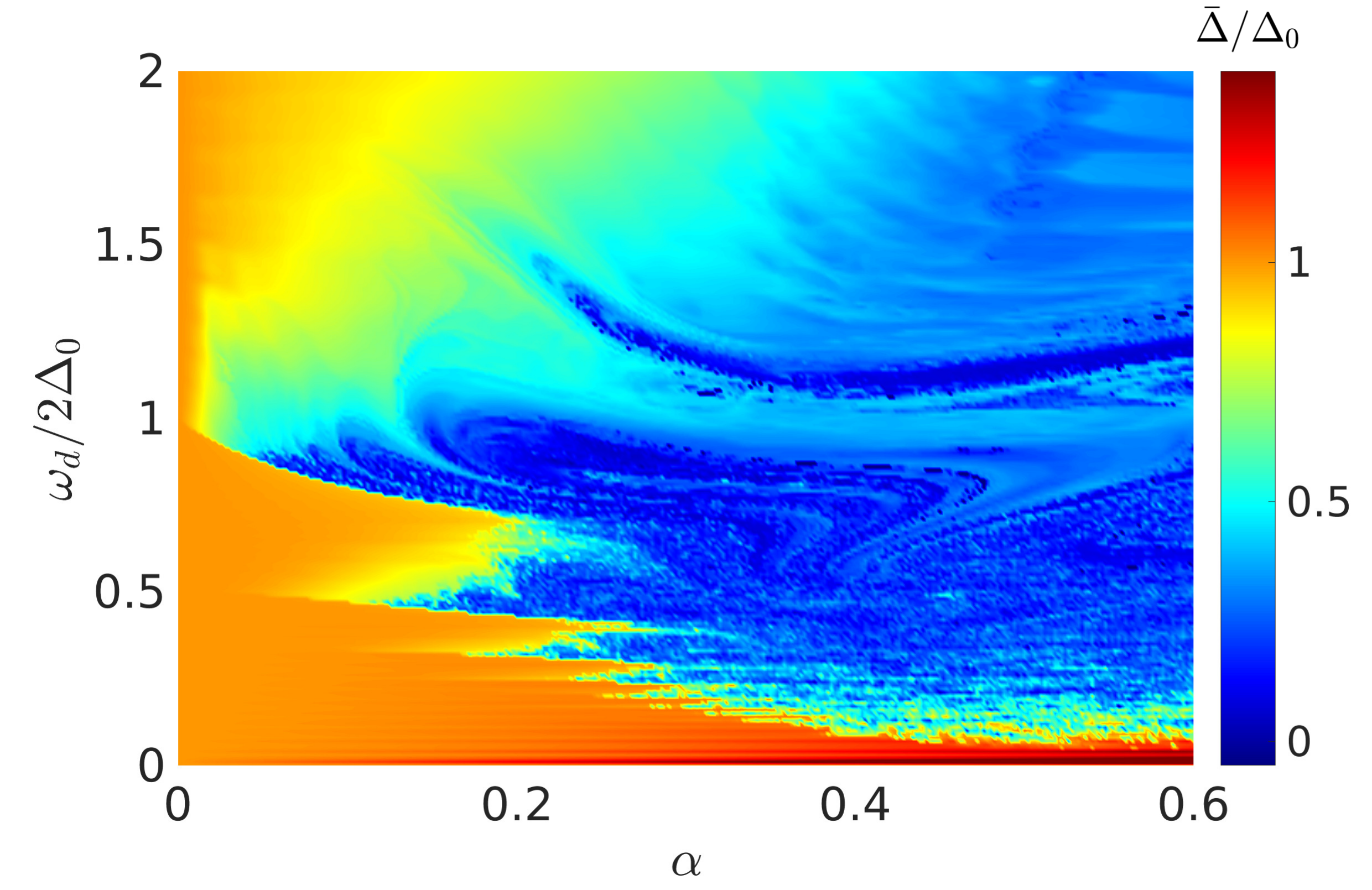}}
		\caption{Temporal average of the superconducting order parameter $\bar{\Delta}$ as a function of the driving amplitude $\alpha$ and the frequency $\omega_d$. The label ` H ' stands for Synchronized Higgs phase, ` TC ' stands for the time crystal phase and ` RH ' stands for the Rabi-Higgs phase.  The time window considered for the time evolution is $t\Delta_0 \in [0,200]$. The chemical potential is fixed at $\mu=0$. } \label{Fig:2D_n1_phase}
	\end{center}
\end{figure}

In this appendix, we present the temporal average of the order parameter in Figure~\ref{Fig:2D_n1_phase}, which indicates the phase diagram of the superconducting system under driven. There are three different dynamical phases: Synchronized Higgs phase, time-crystal phase, Rabi-Higgs phase. This phase diagram is similar to Ref.~\cite{collado2023dynamical}, except we do not observe the real gapless phase. 
In a two dimensional square lattice, we perfectly reproduced those three phases reported in Ref.~\cite{collado2023dynamical}, in which they were just simply modeled with equally spaced eigenenergies. 
The time crystal phase is a narrow region near the driving frequency $\omega_d \sim 0.8\times2\Delta_0$. In this study, we focused on the time crystal phase region, and fixed the driving amplitude to $\alpha = 0.25$ and the driving frequency to $\omega_d = 0.8\times 2\langle \Delta(r)\rangle$, where $\langle\Delta(r)\rangle$ is the spatially averaged order parameter before driving. 

In addition, we provide a diagram of the system under a weaker coupling strength of $U = -1.5$ and a much larger system size $N = 400\times 400$ in Figure~\ref{Fig:2D_n1_phase}. When comparing it to the strong coupling limit of $U = -6$, we observe that a smaller driving amplitude and a wider range of driving frequencies will induce the time crystal phase. This indicates that the realization of a time crystal in experiments is more feasible, because it is relatively easier to fabricate a large sample, and a weaker coupling strength is more realistic.

\begin{figure}[!htbp]
	\begin{center}
		\subfigure[Synchronized Higgs phase {$\alpha=0.1,\omega_d = 0.4\times2\Delta_0$}]{\label{fig.2D_BdG_time_fourier_V0_a0p1_wd0p4}
			\includegraphics[width=5cm]{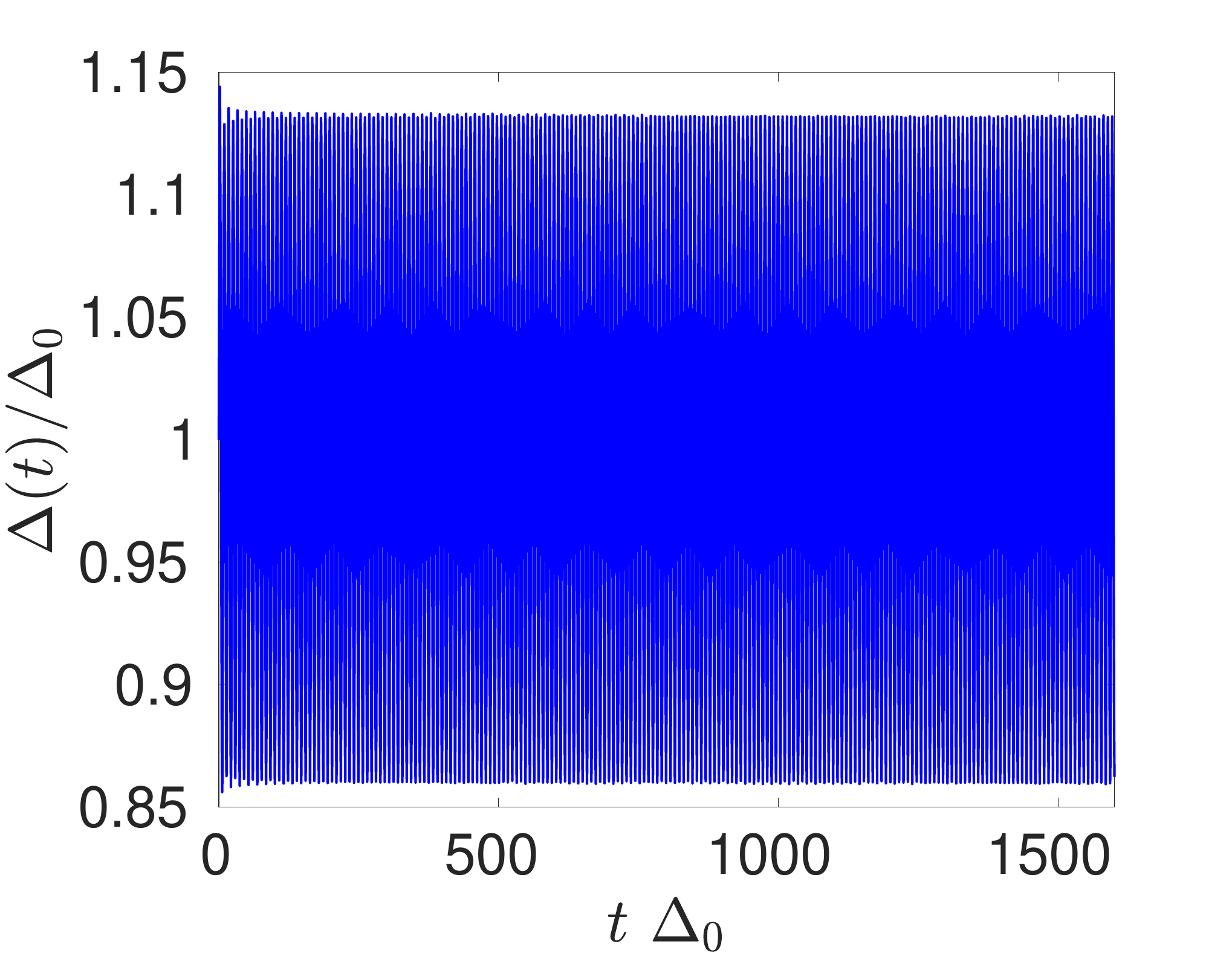}
			\includegraphics[width=5cm]{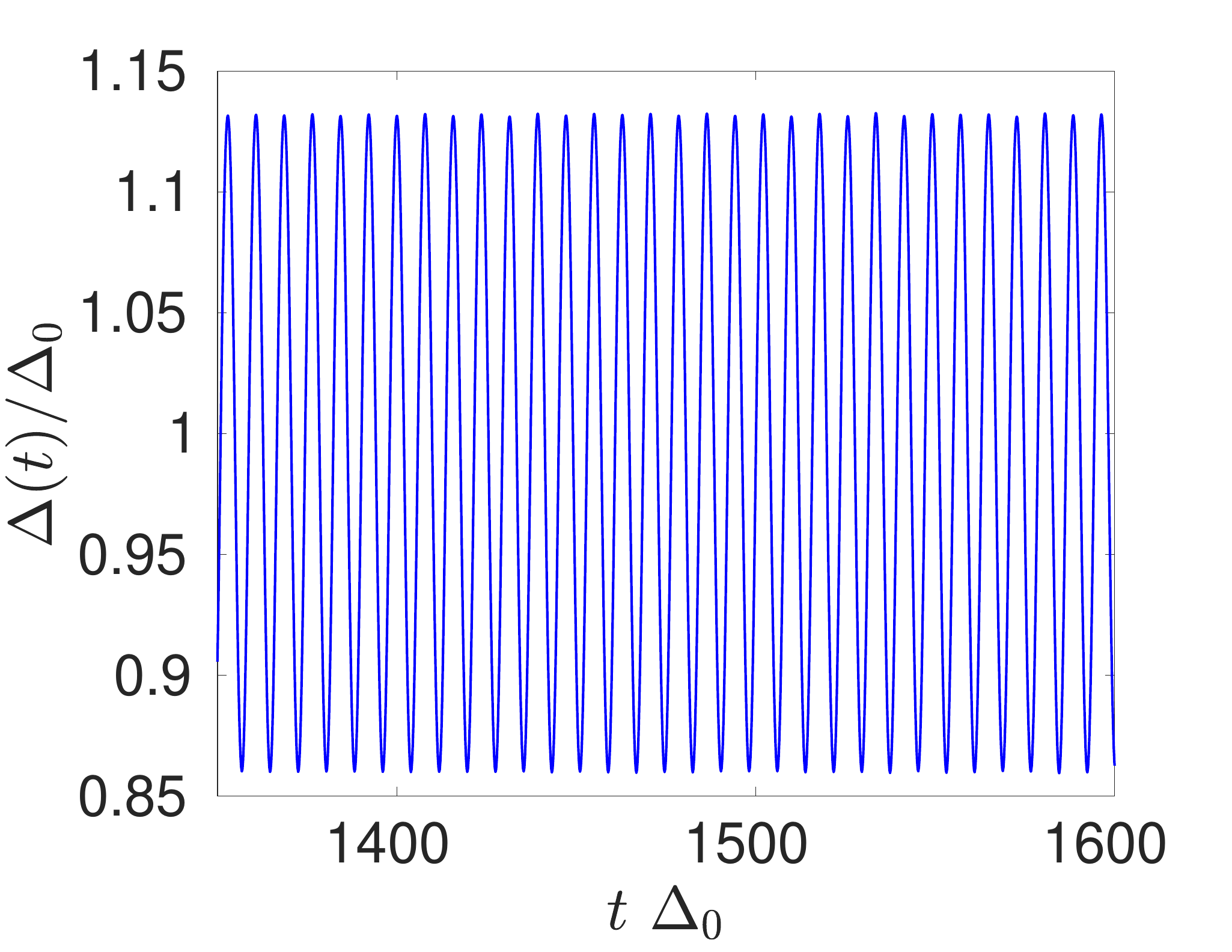}
			\includegraphics[width=5cm]{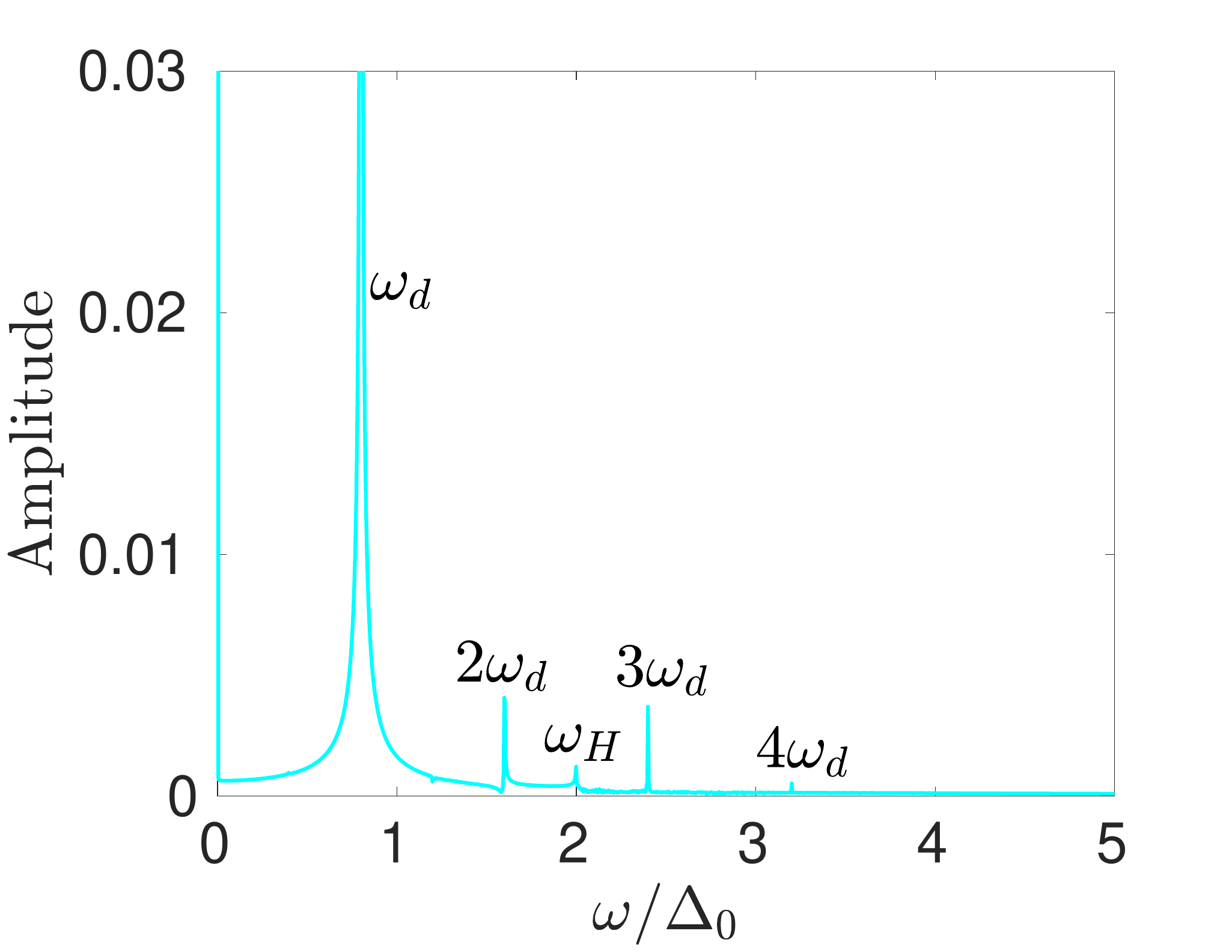}
		} 
		\subfigure[Rabi-Higgs phase {$\alpha=0.1,\omega_d = 1.8\times2\Delta_0$}]{\label{fig.2D_BdG_time_fourier_V0_a0p1_wd1p8}
			\includegraphics[width=5cm]{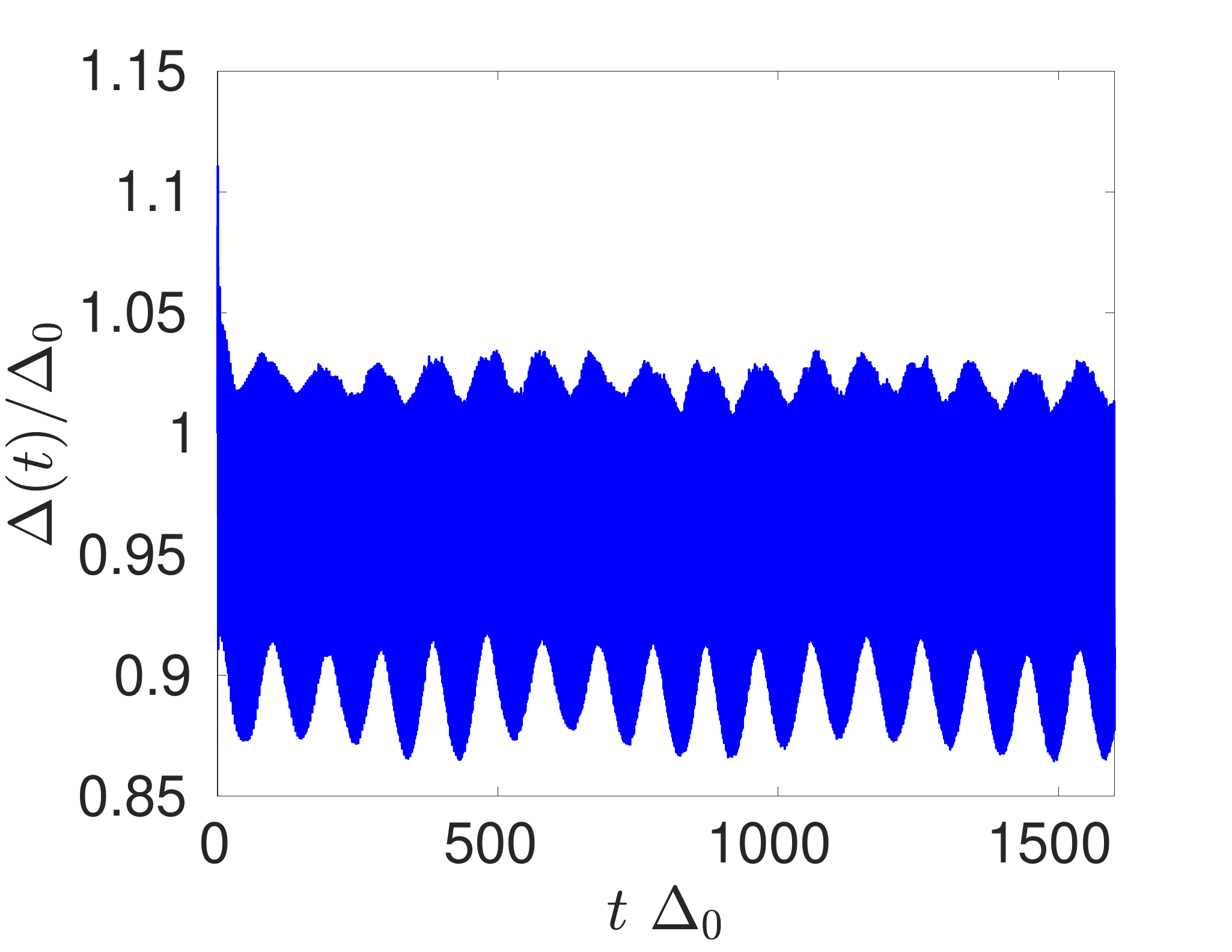}
			\includegraphics[width=5cm]{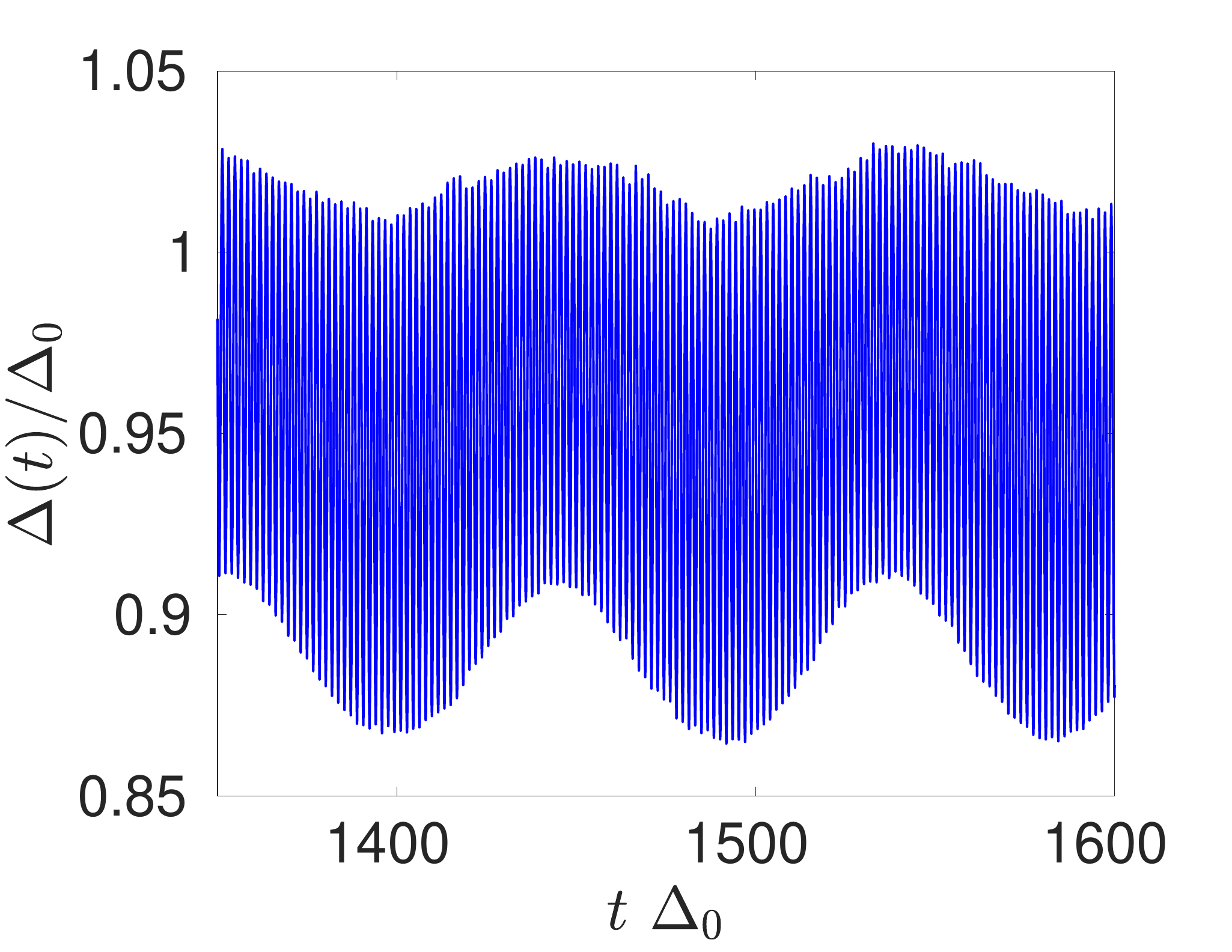}
			\includegraphics[width=5cm]{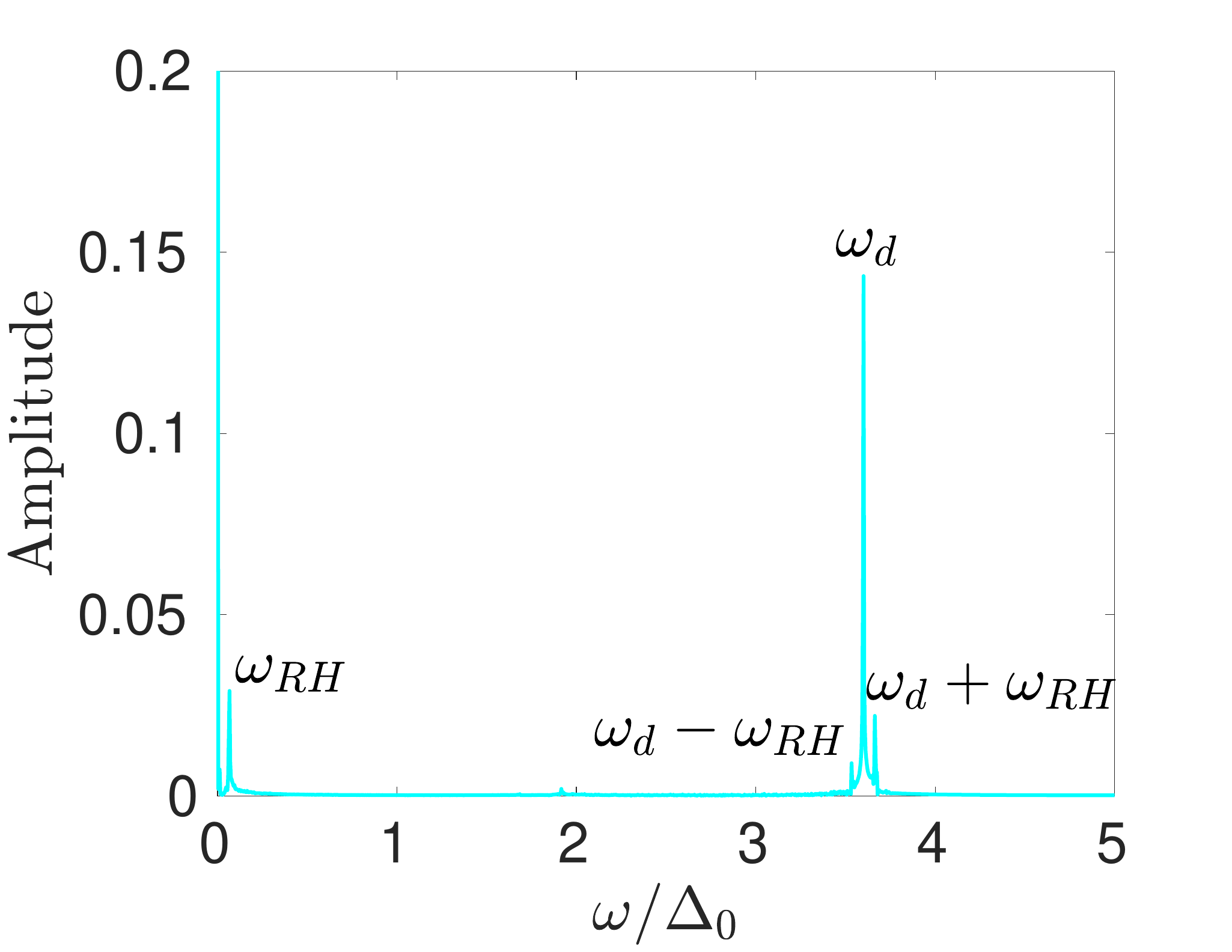}
		}
		\subfigure[Rabi-Higgs phase ? {$\alpha=0.95,\omega_d = 1.8\times2\Delta_0$}]{\label{fig.2D_BdG_time_fourier_V0_a0p95_wd1p8}
			\includegraphics[width=5cm]{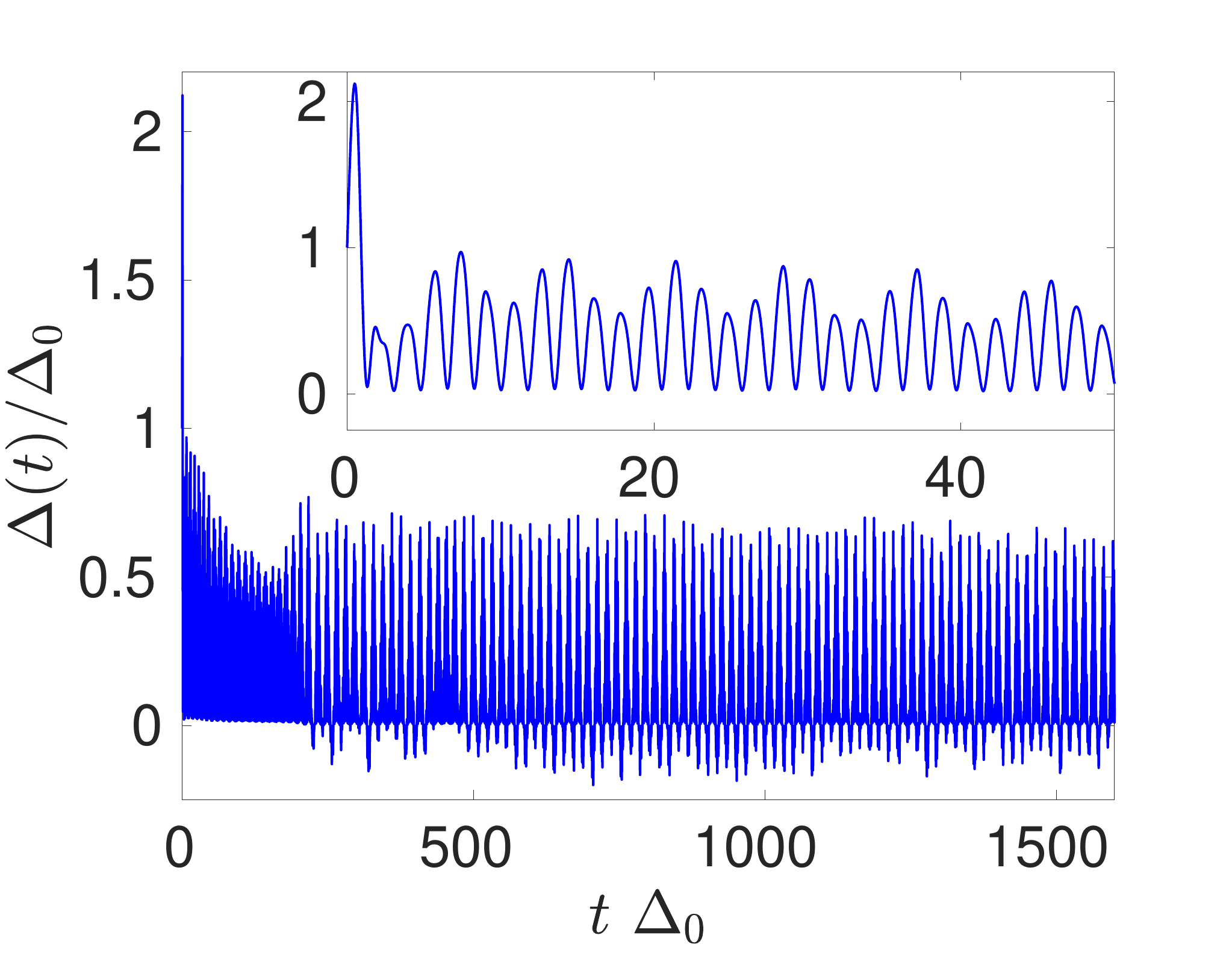}
			\includegraphics[width=5cm]{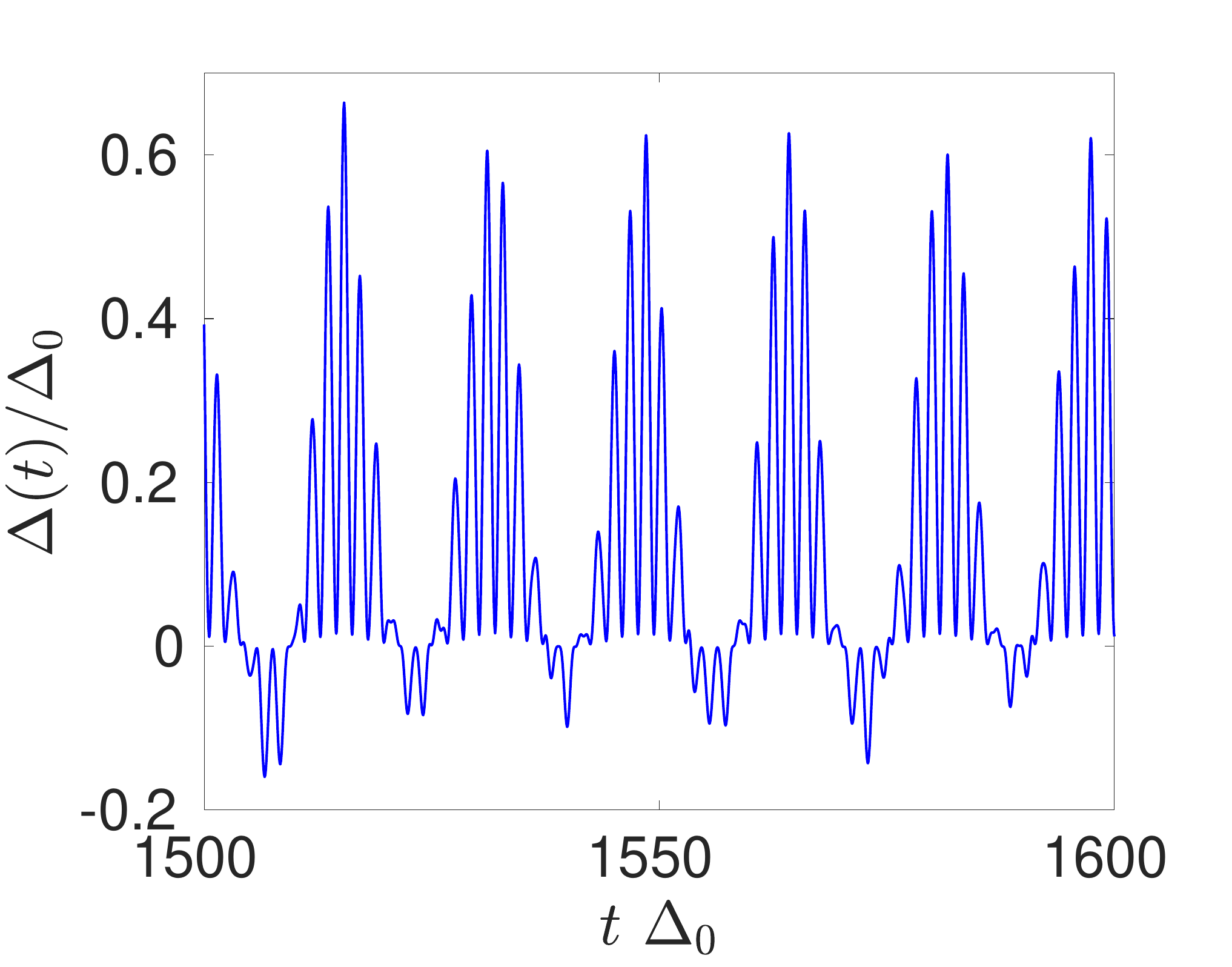}
			\includegraphics[width=5cm]{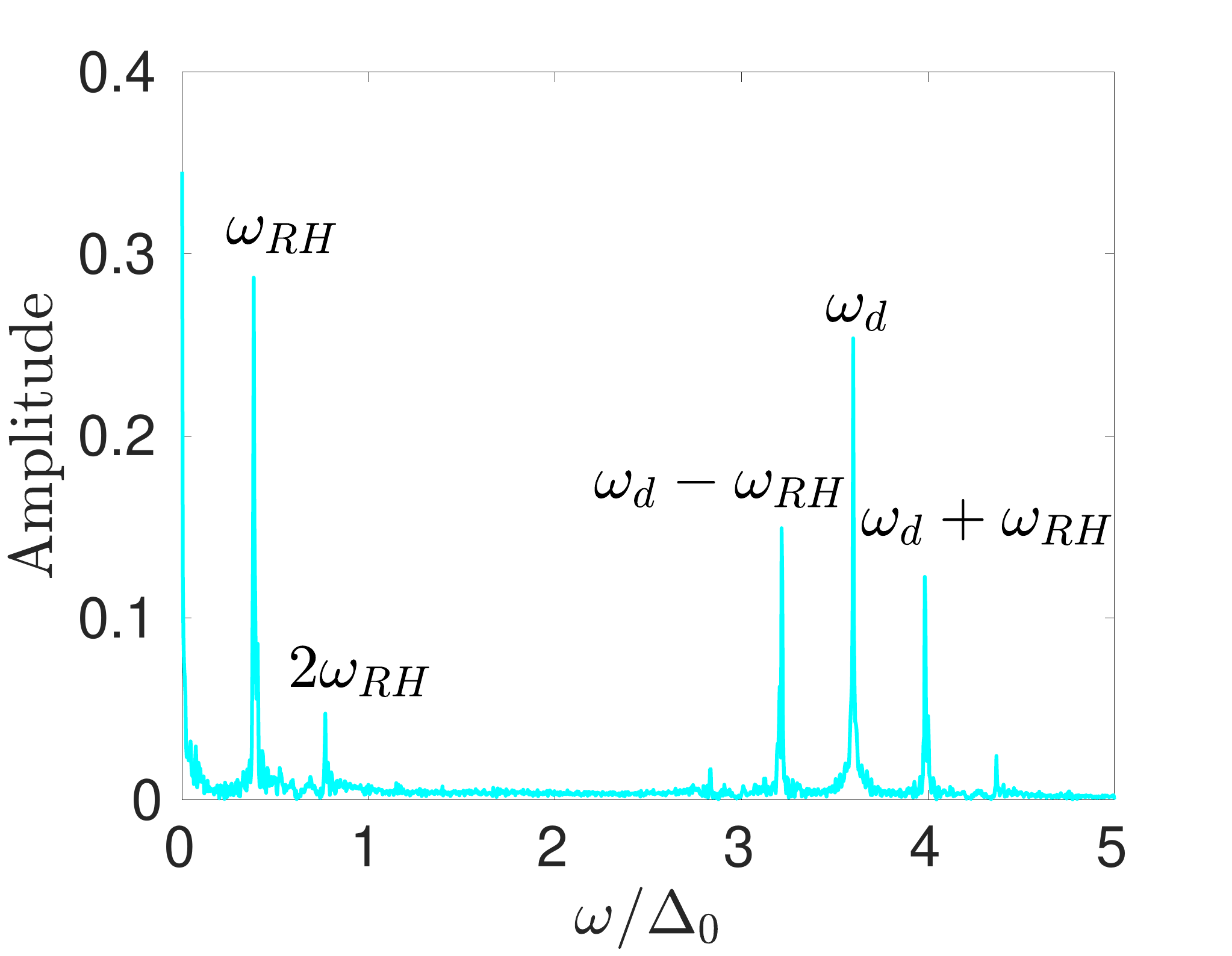}
		}
		\caption{The time evolution of the order parameter and the corresponding FT for different phases: Synchronized Higgs phase and Rabi-Higgs phase. 
			The system size is $N=200\times200$, the equilibrium coupling constant $U_0 = -6$ and the chemical potential is fixed at $\mu =0$.  Our results have shown that the accumulated numerical error could be negligible even when $t~\Delta_0>2000$ for these two phases. 
		}\label{Fig:2D_BdG_time_fourier}
	\end{center}
\end{figure}

However, we note that, although the phase diagram shows similarity with Ref.\cite{collado2023dynamical}, in Figure~\ref{fig.2D_BdG_time_fourier_V0_a0p95_wd1p8}, we do not obtain a gapless phase without oscillation, which has been reported in their study. In this region, the order parameter decays to zero, but it oscillates between zero and maximum peak around $0.6\Delta_0$.
 Although the dynamic order parameter shows significant difference when comparing with Figure~\ref{fig.2D_BdG_time_fourier_V0_a0p1_wd1p8}, the Fourier transform shows that they share similar properties. It is unclear whether it is still the Rabi-Higgs phase. The reason of not observing a gapless phase without oscillations, might be that Ref.~\cite{collado2023dynamical} use equally spaced eigenenergies, and a constant eigenenergy degeneracy which does not take into account any dimensionality or degeneracy due to different sets of quantum numbers $\{n_x,n_y\}$. It is too simplified to describe real two dimensional system. These settings are also used in previous studies \cite{barankov2006synchronization,zhoubcs,yuzbashyan2006dynamical}, and the order parameter decays rapidly into zero without oscillation, if the coupling constant of the system changes significantly and satisfies $\Delta_i \ge \exp(\pi/2)\Delta_f$, where $\Delta_i$ is the amplitude order parameter before the quench, and $\Delta_f$ is te order parameter of the final equilibrium state after the quench.

 \begin{figure}[!htbp]
 	\begin{center}
 		\includegraphics[width=12cm]{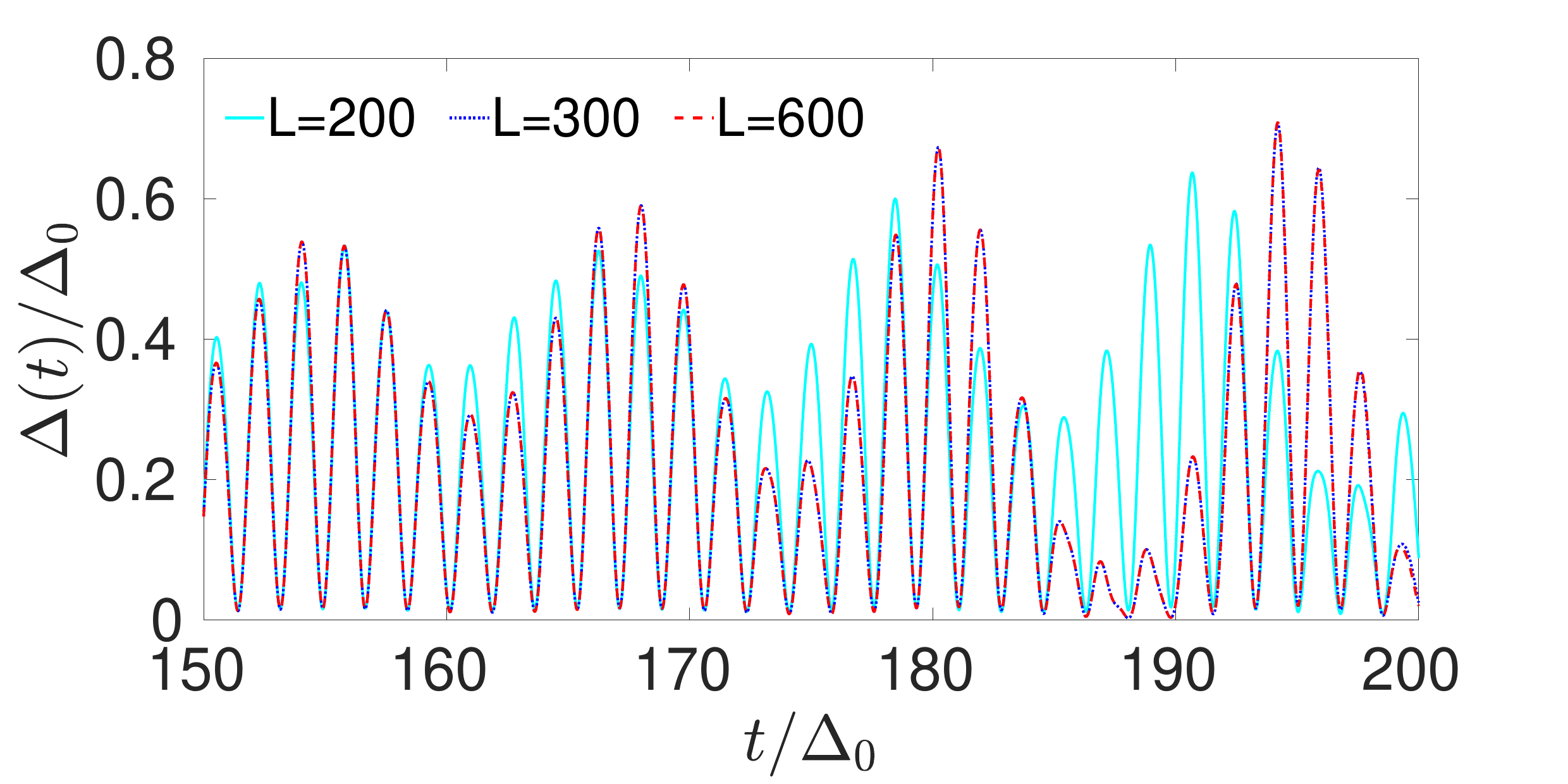}
 		\caption{The size dependence of the order parameter time evolution in the region reported to be the gapless phase \cite{collado2023dynamical}. The driving parameters are the same with Figure~\ref{fig.2D_BdG_time_fourier_V0_a0p95_wd1p8}. It shows that when $N = 200\times200$, finite size effect is still significant in this region. But when the system size $N\ge 300\times300$, the dashed line and the dotted line completely overlap with each other, which means there is no size effect, and the oscillations persist in the infinite system.  }\label{Fig:a0p95_wd1p8_vs_L}
 	\end{center}
 \end{figure}

 In Ref.~\cite{hannibal2018dynamical} it was shown that in a two dimensional system, finite size effects could also result in oscillations in the gapless phase. However, we tried different system size from $N=50\times50$ to $N=600\times600$ with the same parameters in the range of $t~\Delta_0 \in [0,200]$, and no evident signal of oscillations was found so we believe that in this region there will be no oscillations in the infinite size limit. Instead, the oscillation remains unchanged when $N\ge 300\times300$, as is shown in Figure~\ref{Fig:a0p95_wd1p8_vs_L}. Therefore, we would expect that this oscillation is due to the degeneracy of different sets of quantum numbers $\{n_x,n_y\}$.

\section{The quantum dynamics for $\mu \neq 0$} \label{app:difmu}

\begin{figure}[!htbp]
	\begin{center}
		\subfigure[]{
			\includegraphics[width=8.cm]{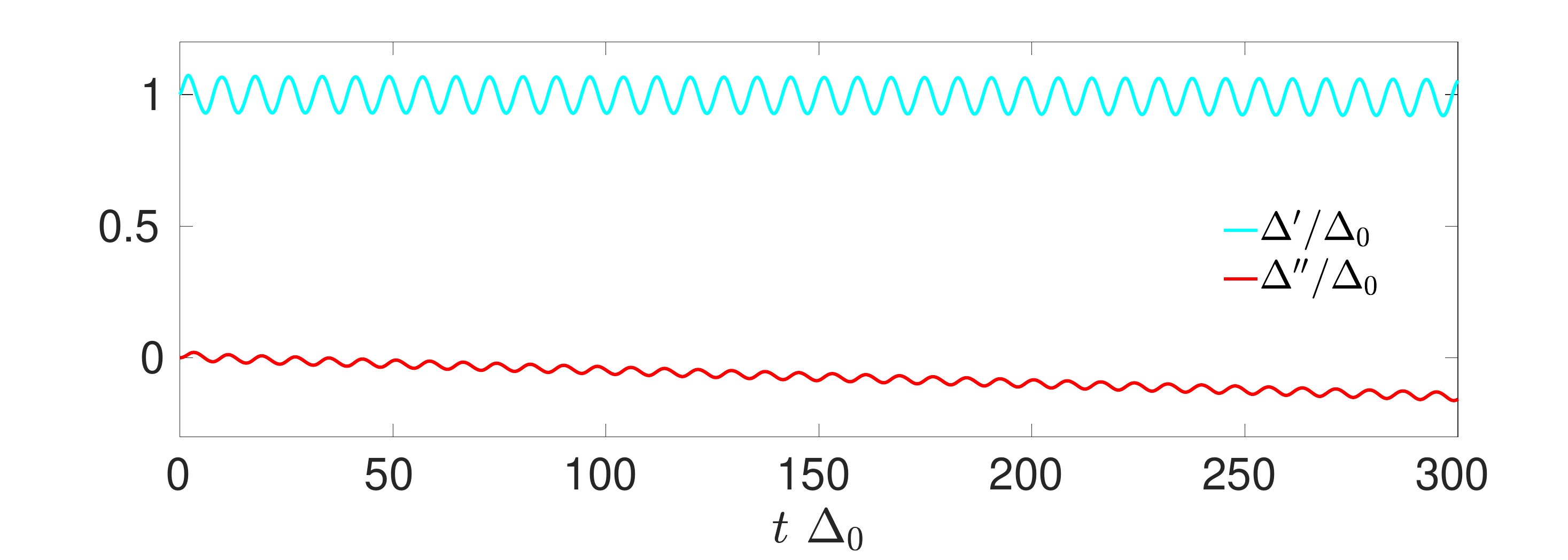} }
		\subfigure[]{
			\includegraphics[width=8.cm]{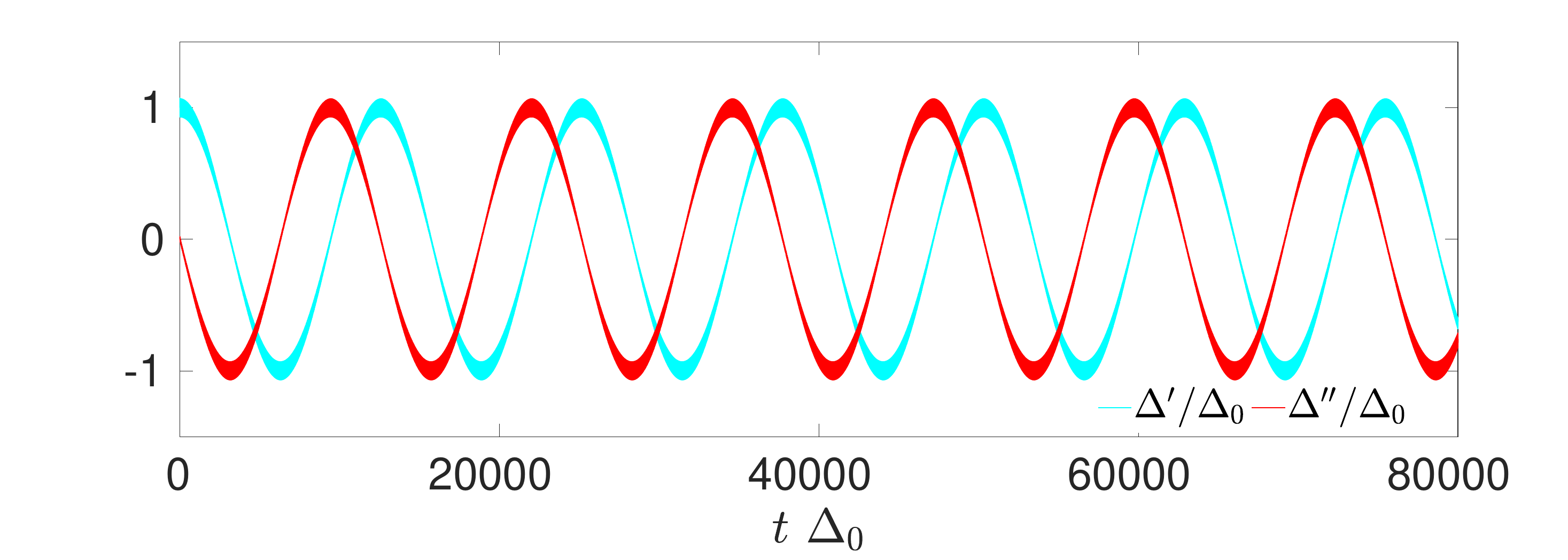} }
		\subfigure[]{
			\includegraphics[width=8.cm]{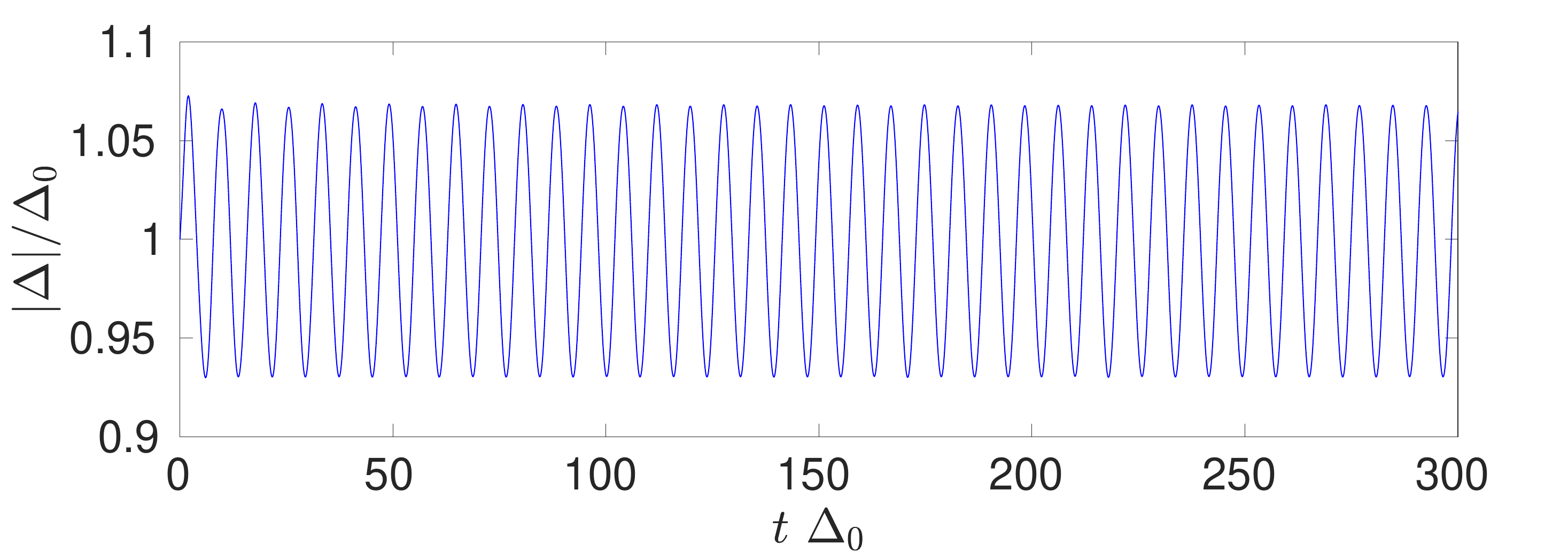} }
		\subfigure[]{
			\includegraphics[width=8.cm]{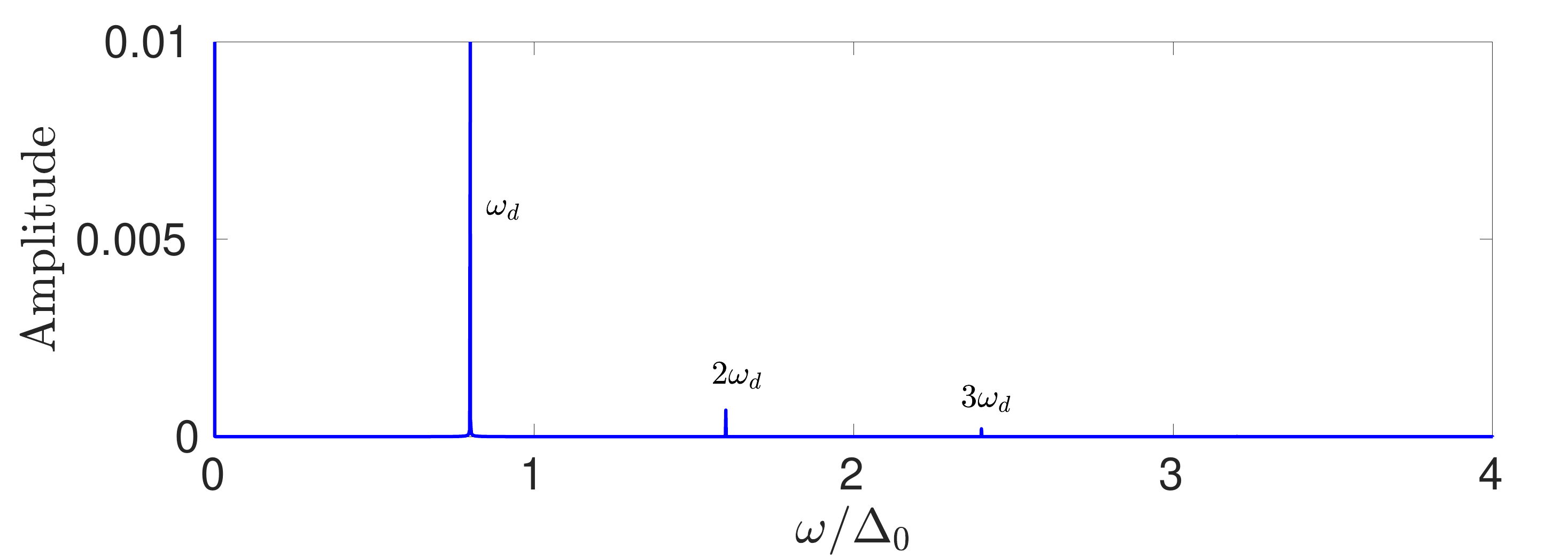} }
		\caption{The upper panel: the real time evolution of the real $\Delta^\prime$ and the imaginary $\Delta^{\prime \prime}$ part of the complex order parameter at different timescale. Below are the modulus $|\Delta|$ of the order parameter and the corresponding Fourier transform. The driving amplitude $\alpha = 0.05$ and the driving frequency $\omega_d = 0.4 \times 2\Delta_0$, which corresponds to the Synchronized Higgs phase in Figure~\ref{Fig:2D_BdG_time_fourier}.} \label{Fig:2D_n0p875_H}
	\end{center}
\end{figure}

In the main text, we only studied the system with chemical potential $\mu = 0$, so that the system is particle-hole symmetric. Therefore, the order parameter is always real. In this appendix, we consider the system with chemical potential $\mu = -0.4989$, which corresponds to charge density $\langle n \rangle = 0.875$ in the static system. In this case, the imaginary part of the order parameter becomes non-zero in the dynamic system.  The complex order parameter can be written as $\Delta = \Delta^\prime + i\Delta^{\prime\prime}$ where $\Delta^\prime$ is the real part of the order parameter and $\Delta^{\prime\prime}$ is the imaginary part. We present the results of three different parameters, in Figure~\ref{Fig:2D_n0p875_H}$\sim$\ref{Fig:2D_n0p875_TC}, which correspond to the three different phases: Synchronized Higgs phase, Rabi-Higgs phase and time crystal phase.

\begin{figure}[!htbp]
	\begin{center}
		\subfigure[]{\label{fig.BdG_L100_n0p875_RH_1}
			\includegraphics[width=8cm]{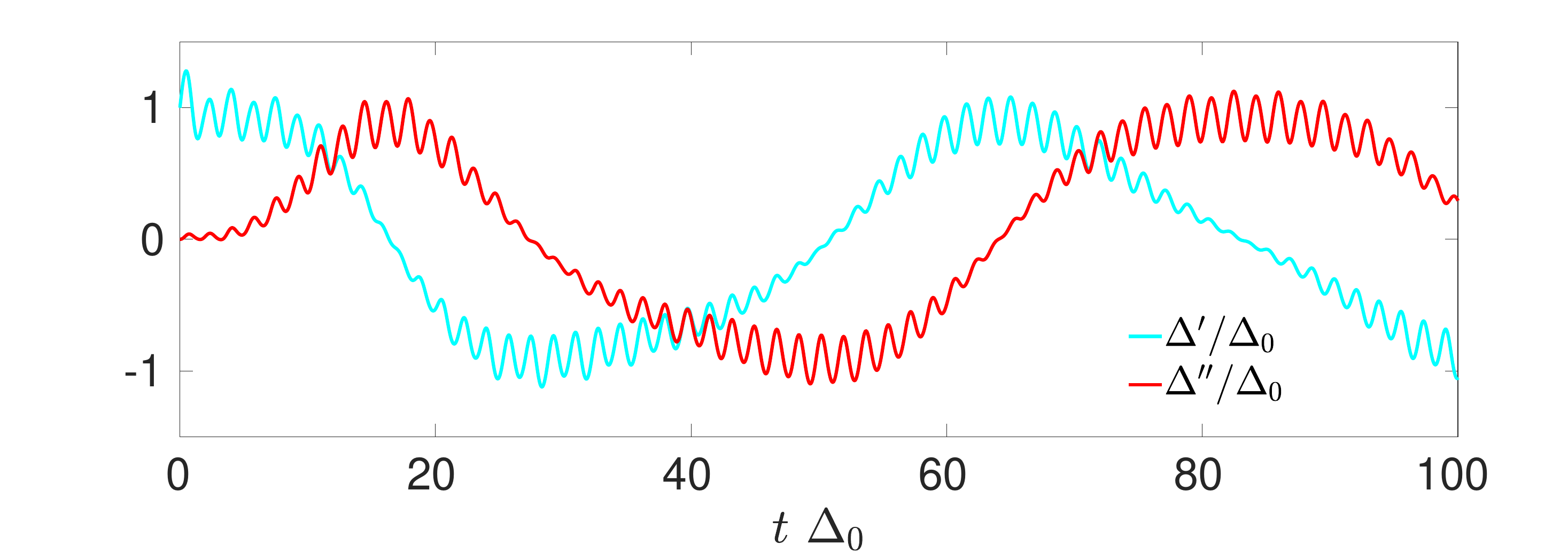} }
		\subfigure[]{\label{fig.BdG_L100_n0p875_RH_2}
			\includegraphics[width=8cm]{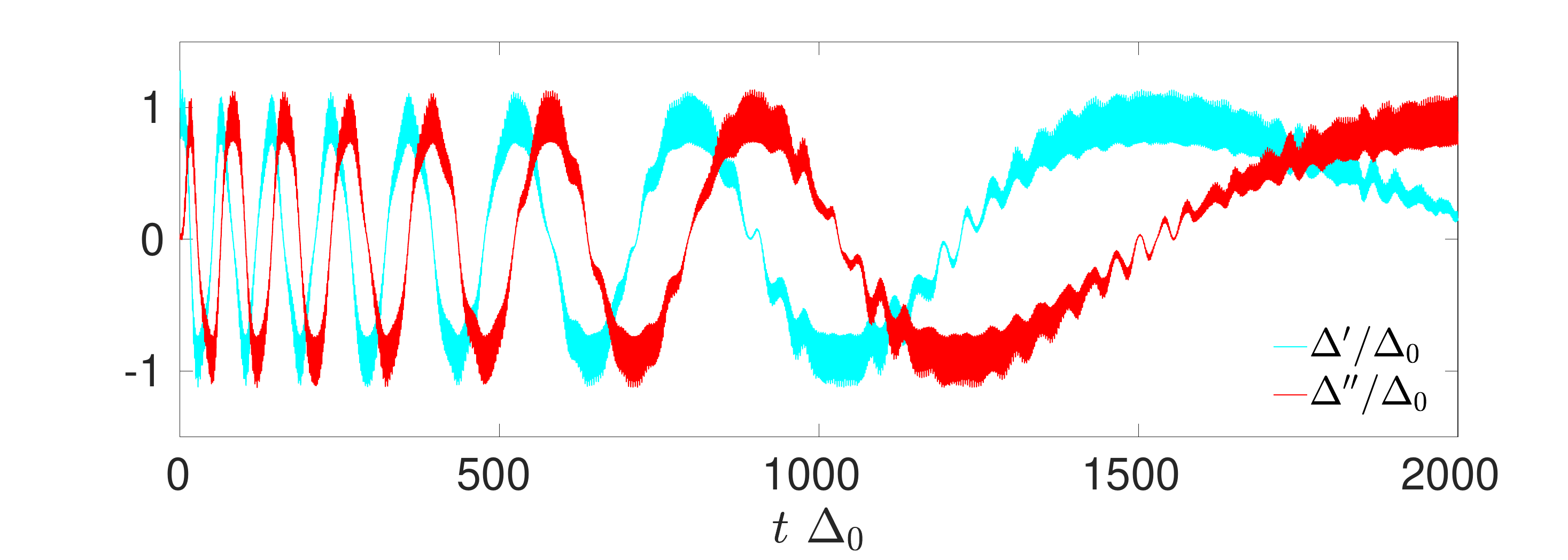} }
		\subfigure[]{\label{fig.BdG_L100_n0p875_RH_3}
			\includegraphics[width=8cm]{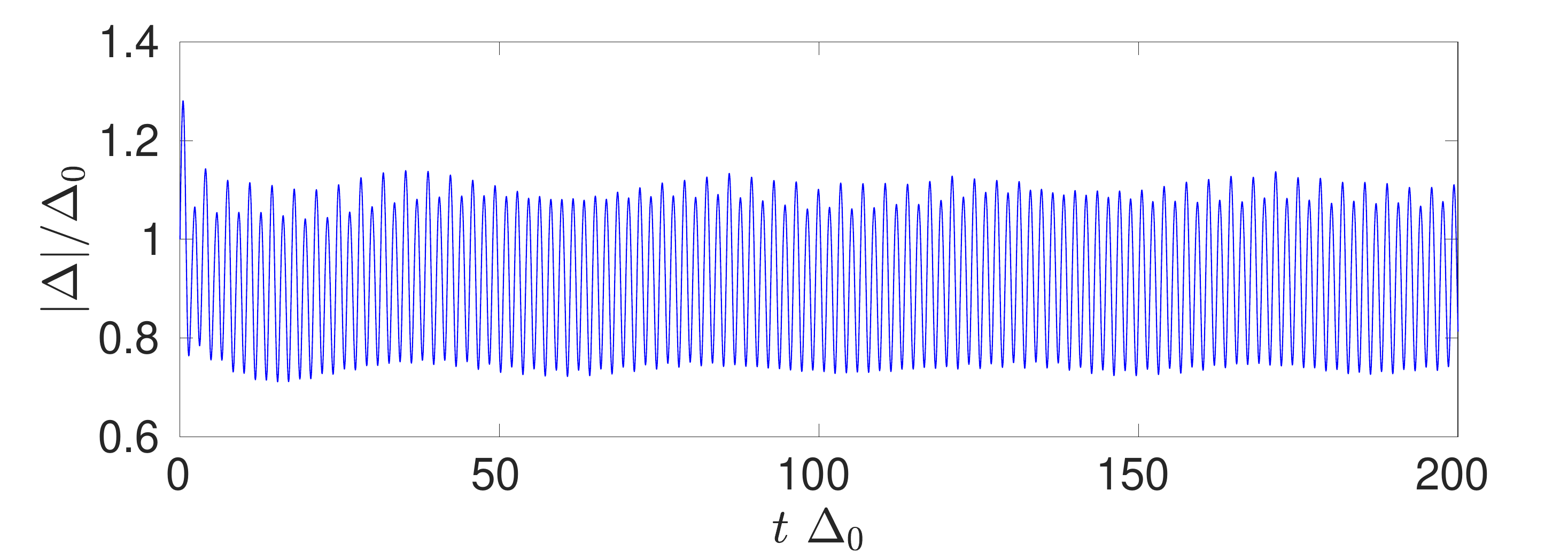}}
		\subfigure[]{\label{fig.BdG_L100_n0p875_RH_4}
			\includegraphics[width=8cm]{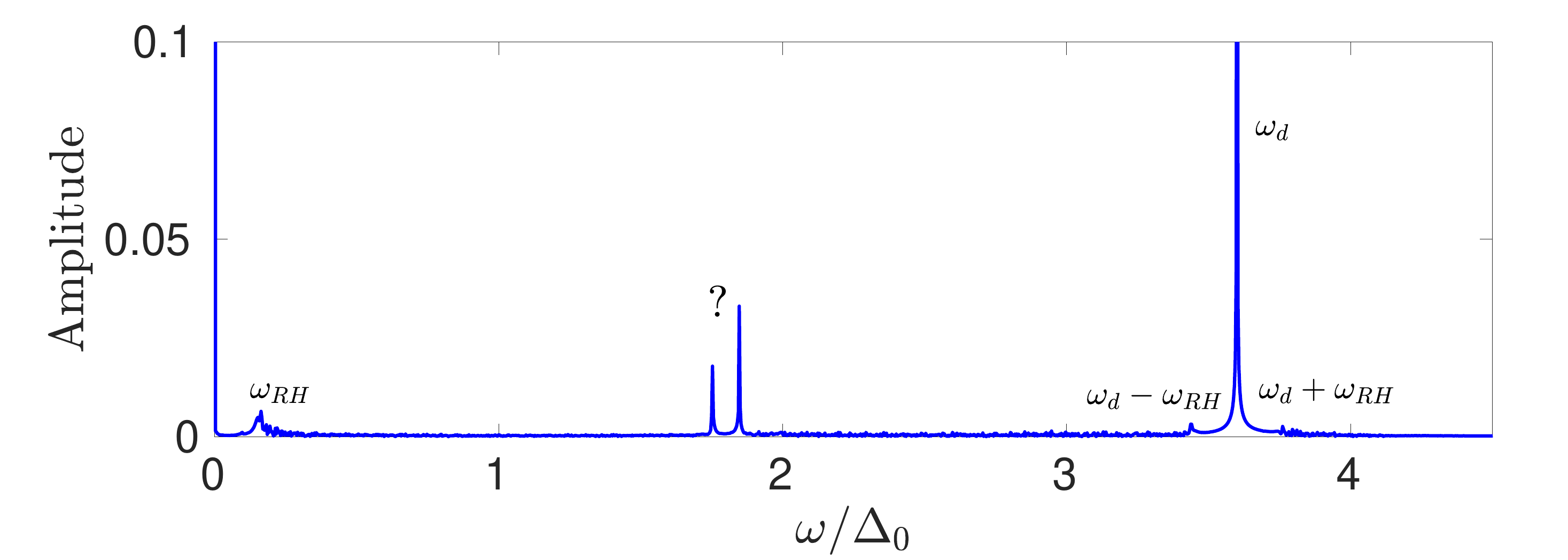} }
		\caption{The real time evolution of \subref{fig.BdG_L100_n0p875_RH_1} the real part $\Delta^\prime$, \subref{fig.BdG_L100_n0p875_RH_2} imaginary part $\Delta^{\prime \prime}$ and \subref{fig.BdG_L100_n0p875_RH_3} modulus $|\Delta|$ of the order parameter and \subref{fig.BdG_L100_n0p875_RH_4} the corresponding Fourier transform in the Rabi-Higgs phase. The driving amplitude $\alpha = 0.25$ and the driving frequency $\omega_d = 1.8 \times 2\Delta_0$. The question mark means the physical meaning of the two frequency peaks is unknown.} \label{Fig:2D_n0p875_RH}
	\end{center}
\end{figure}

When the driving amplitude $\alpha = 0.05$ and the driving frequency $\omega_d = 0.4 \times 2\Delta_0$, although $\Delta^\prime$ and $\Delta^{\prime\prime}$ oscillate with hybrid frequency, the modulus of the complex order parameter $|\Delta| = \sqrt{{\Delta^\prime}^2 + {\Delta^{\prime\prime}}^2}$ oscillate with the driving frequency $\omega_d = 0.4 \times 2\Delta_0$, which is the same as the Synchronized Higgs phase in Figure~\ref{Fig:2D_BdG_time_fourier}. For the parameters in the Rabi-Higgs phase, $|\Delta|$ show similar behavior with the $\mu =0$ system, except in the $\mu \ne 0$ system there are two extra unknown oscillation frequencies marked by the question mark in \ref{fig.BdG_L100_n0p875_RH_4}. For the parameters in the time crystal phase, $\Delta^\prime$ and $\Delta^{\prime\prime}$ both oscillate with a frequency different from the driven frequency. However, we note that in the $\mu \ne 0$ system, the oscillation frequency is also different from $\omega_d/2$, which is the emergent time crystal frequency in the main text. Since the peak in Fourier space are quite broad, it is unclear whether this is really a true time crystal phase.

\begin{figure}[!htbp]
	\begin{center}
		\subfigure[]{\label{fig.BdG_L100_n0p875_TC_1}
			\includegraphics[width=8cm]{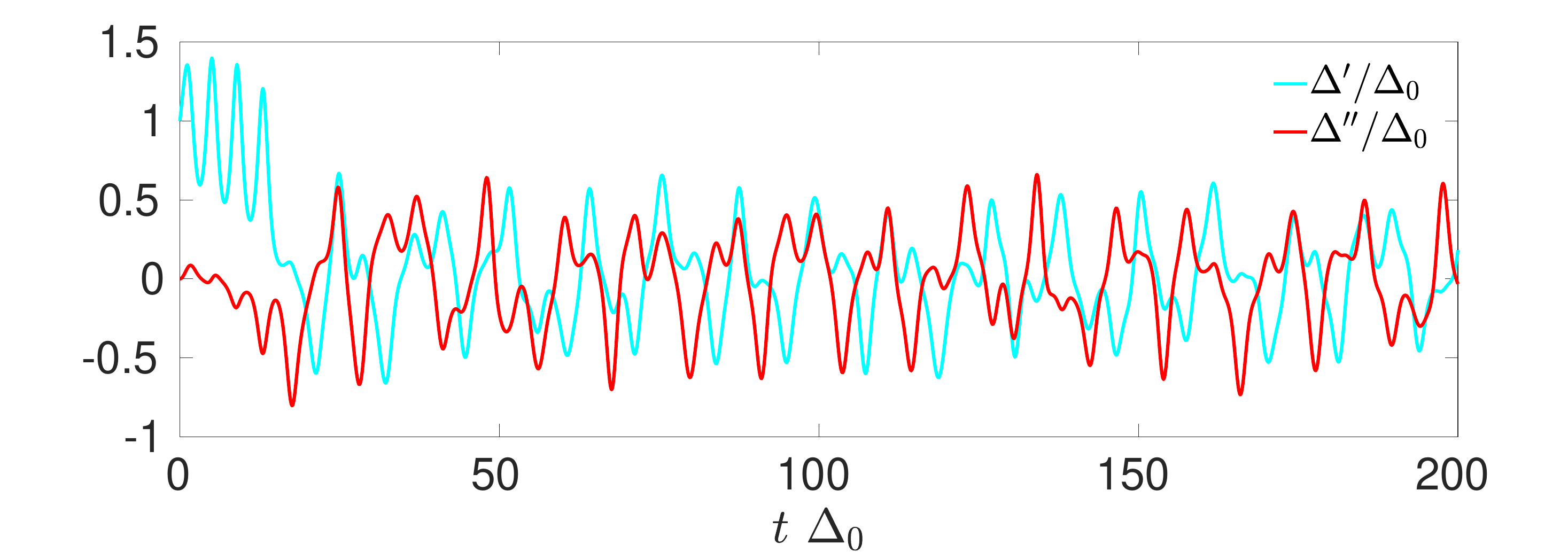} }
		\subfigure[]{\label{fig.BdG_L100_n0p875_TC_2}
			\includegraphics[width=8cm]{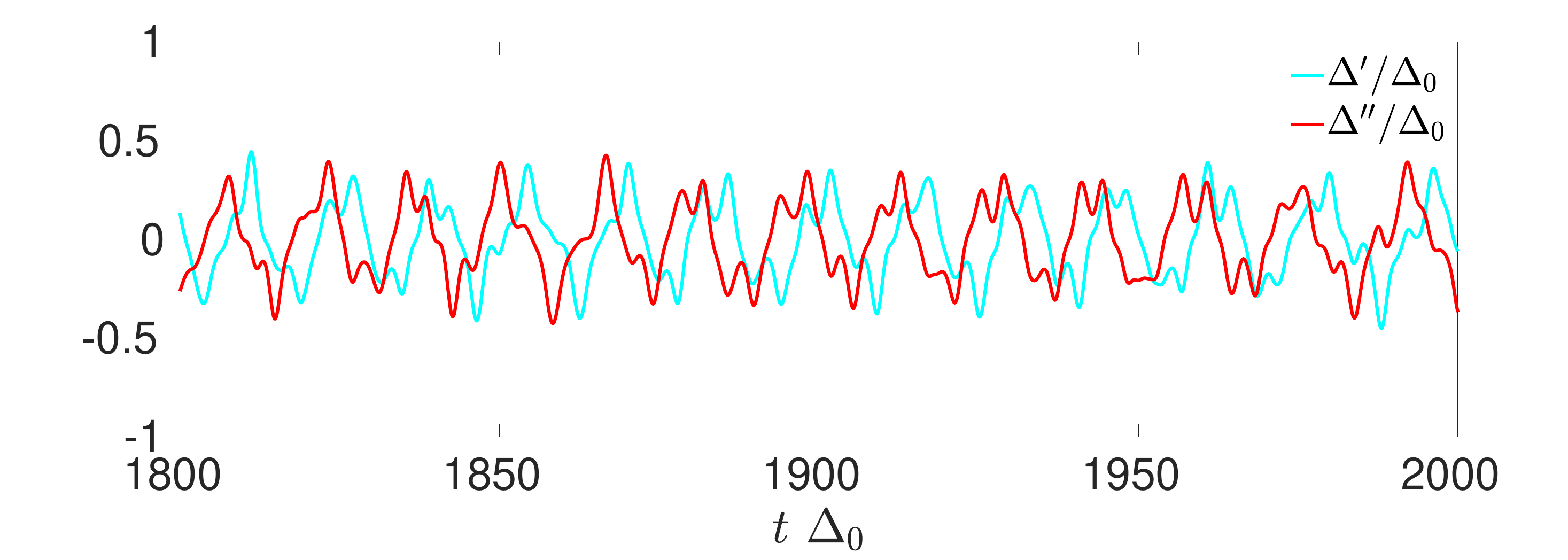}}
		\subfigure[]{\label{fig.BdG_L100_n0p875_TC_3}
			\includegraphics[width=8cm]{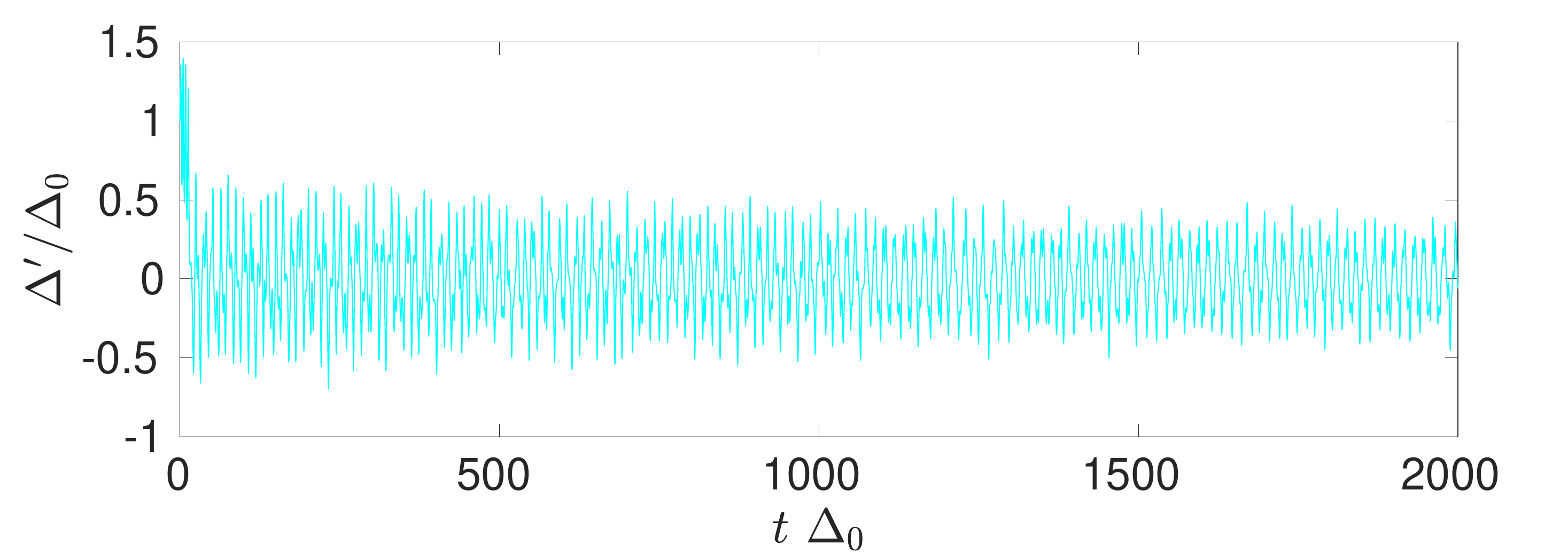} }
		\subfigure[]{\label{fig.BdG_L100_n0p875_TC_4}
			\includegraphics[width=8cm]{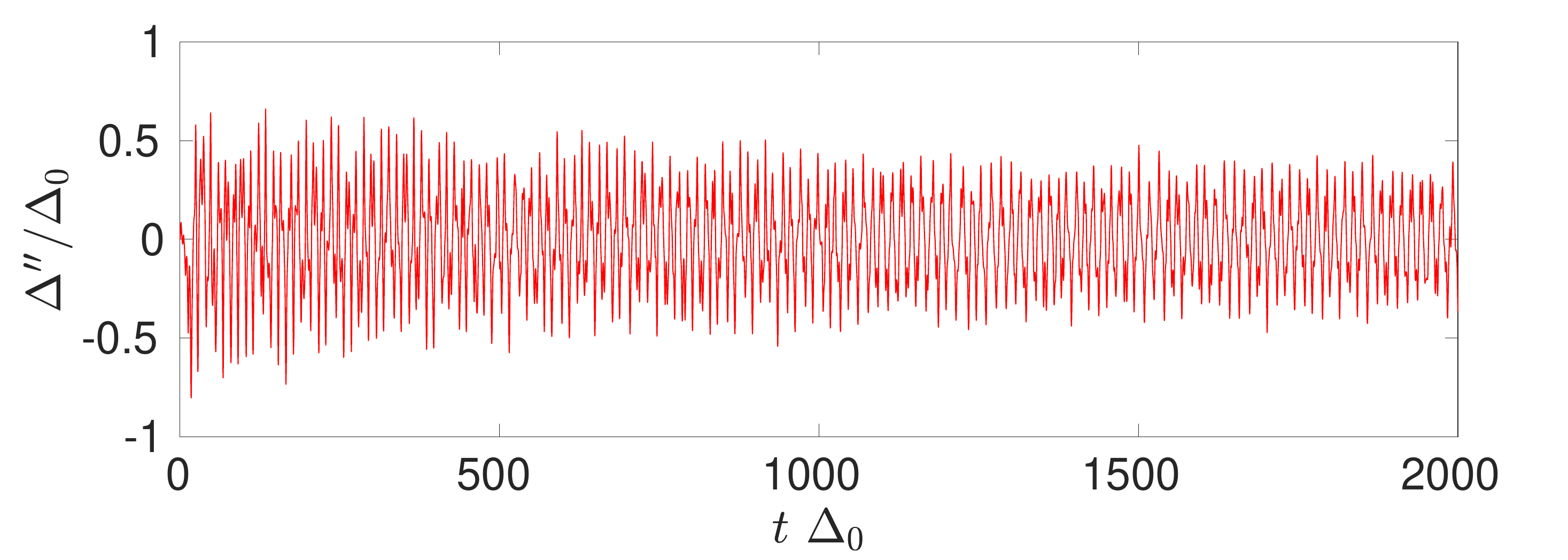}  }
		\subfigure[]{\label{fig.BdG_L100_n0p875_TC_5}
			\includegraphics[width=8cm]{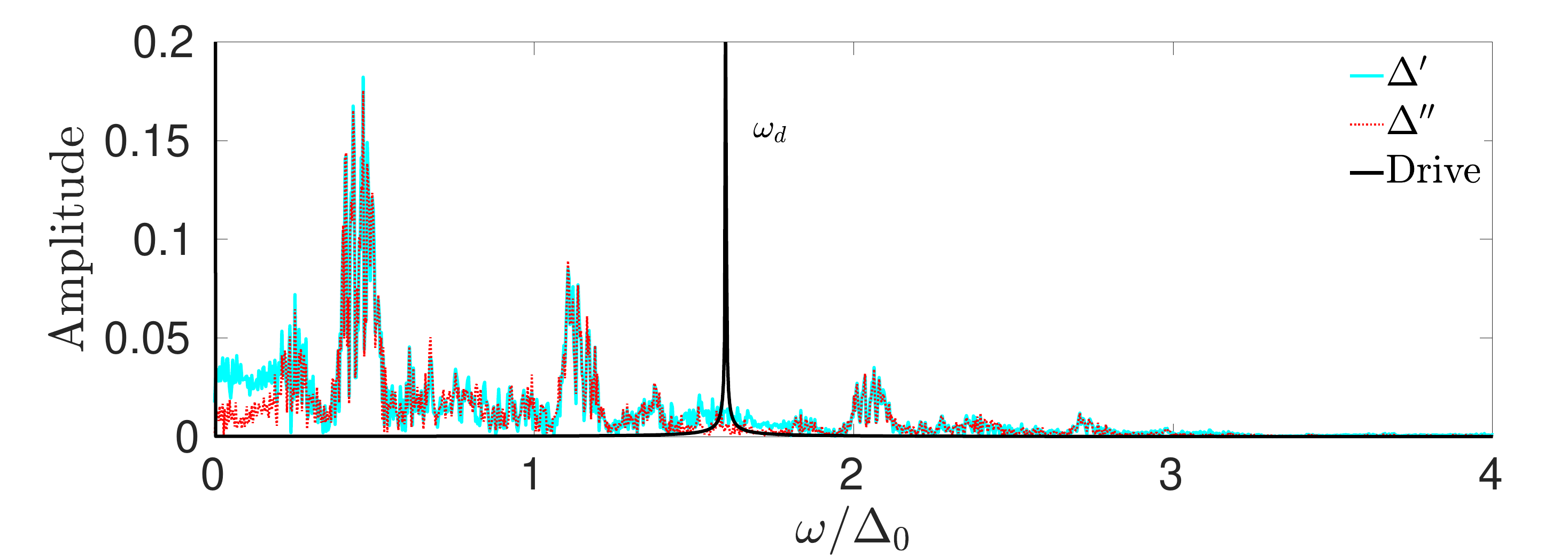} }
		\caption{\subref{fig.BdG_L100_n0p875_TC_1} $\sim$ \subref{fig.BdG_L100_n0p875_TC_4} The real time evolution of  the real $\Delta^\prime$, imaginary $\Delta^{\prime \prime}$  part of the order parameter and \subref{fig.BdG_L100_n0p875_TC_5} is the corresponding Fourier transform.
			The driving amplitude $\alpha = 0.25$ and the driving frequency $\omega_d = 0.8 \times 2\Delta_0$, which corresponds to the time crystal phase in Figure~\ref{Fig:identimecrystal}.} \label{Fig:2D_n0p875_TC}
	\end{center}
\end{figure}

\section{Identify the peaks of the time crystal oscillations and the fittings of its decay}\label{app:peak_decay}

\begin{figure}[!htbp]
	\begin{center}
		\subfigure[]{\label{fig.peak_V0p001}
			\includegraphics[width=17cm]{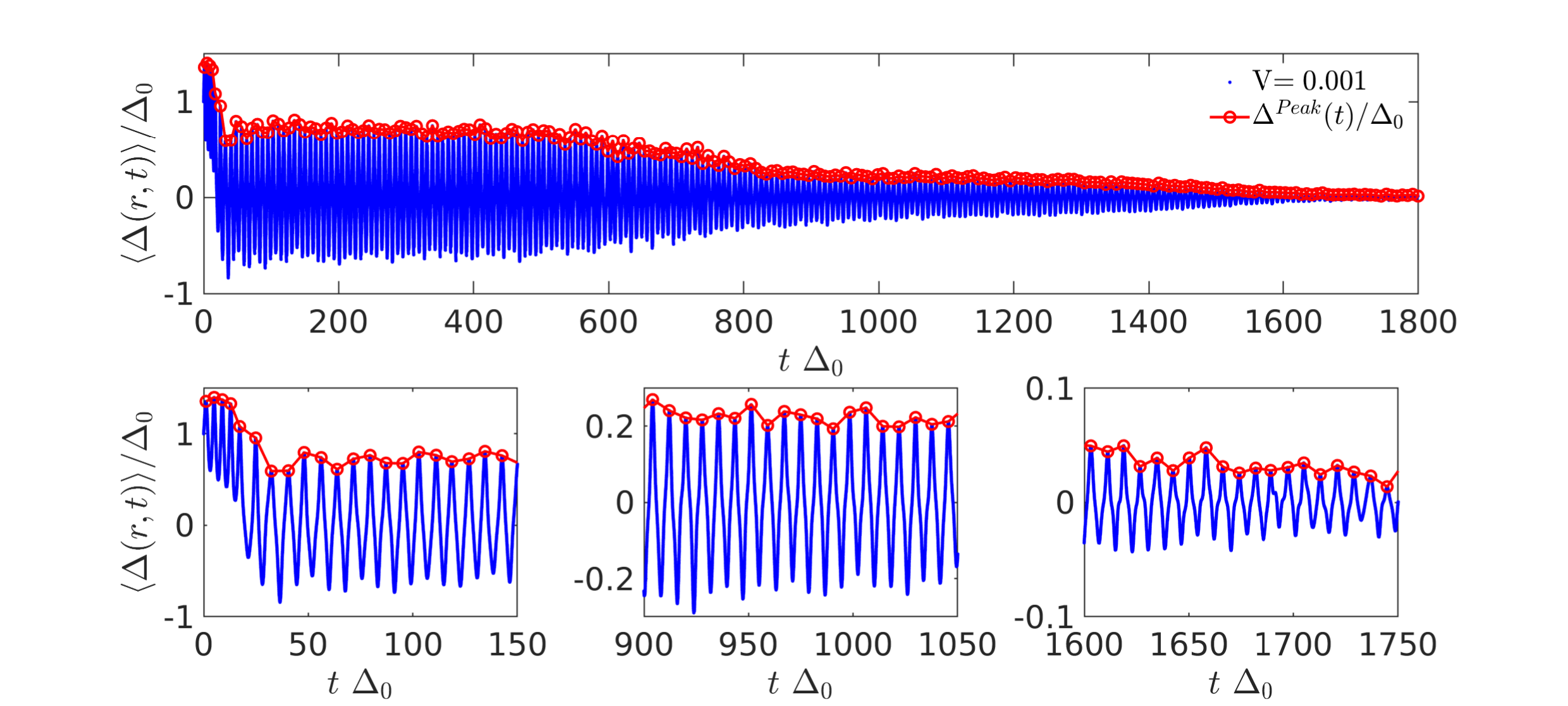}}
		\subfigure[]{\label{fig.peak_V0p6}
			\includegraphics[width=17cm]{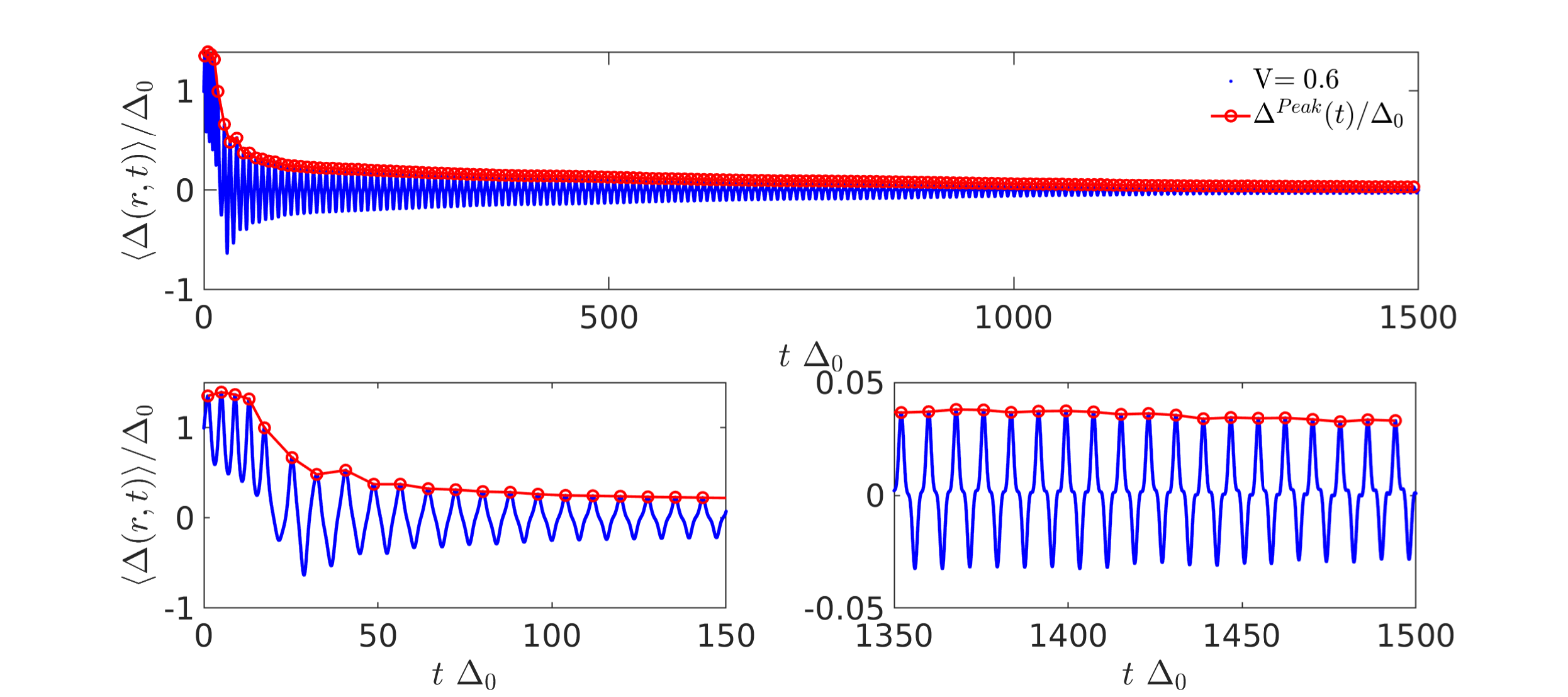} }
		\caption{The time evolution of the spatial average order parameter $\langle \Delta(r,t) \rangle$ (blue dots) and the corresponding peaks $\Delta^{Peak}(t)$ (red circle). The disorder is \subref{fig.peak_V0p001} $V=0.001$ and \subref{fig.peak_V0p6} $V=0.6$. The other parameters are $N=100\times 100$, $U_0 = -6$ and $\mu = 0$. The driving amplitude $\alpha = 0.25$ and frequency $\omega_d = 0.8\times 2\langle\Delta(r)\rangle$.  }\label{Fig:extract_peak}
	\end{center}
\end{figure}

\begin{figure}[!htbp]
	\begin{center}
		\subfigure[]{ \label{fig.fit_V0p001_tiny}
			\includegraphics[width=8.cm]{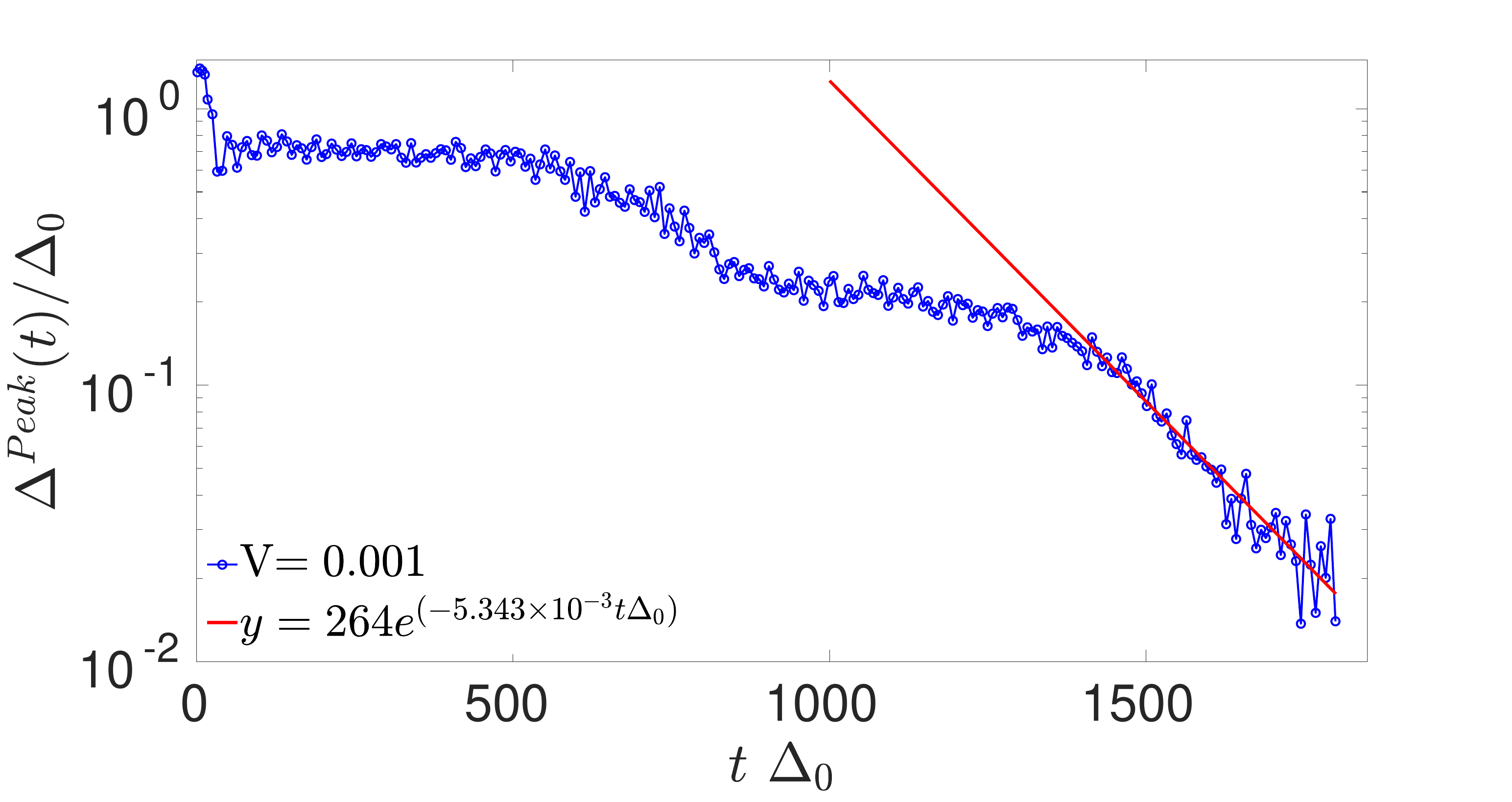}  }
		\subfigure[]{ \label{fig.fit_V0p01_tiny}
			\includegraphics[width=8.cm]{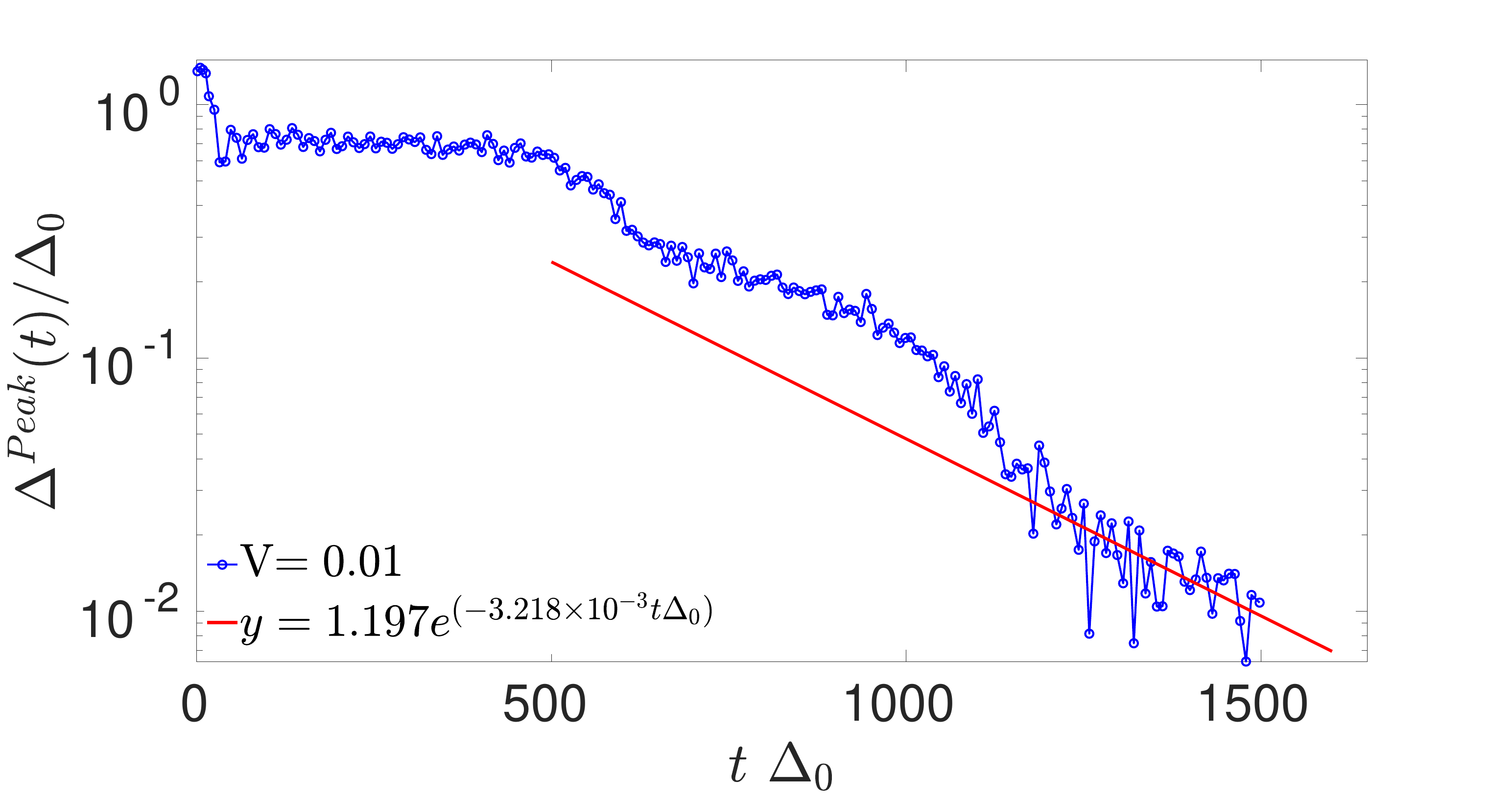}  }
		\subfigure[]{ \label{fig.fit_V0p1}
			\includegraphics[width=8.cm]{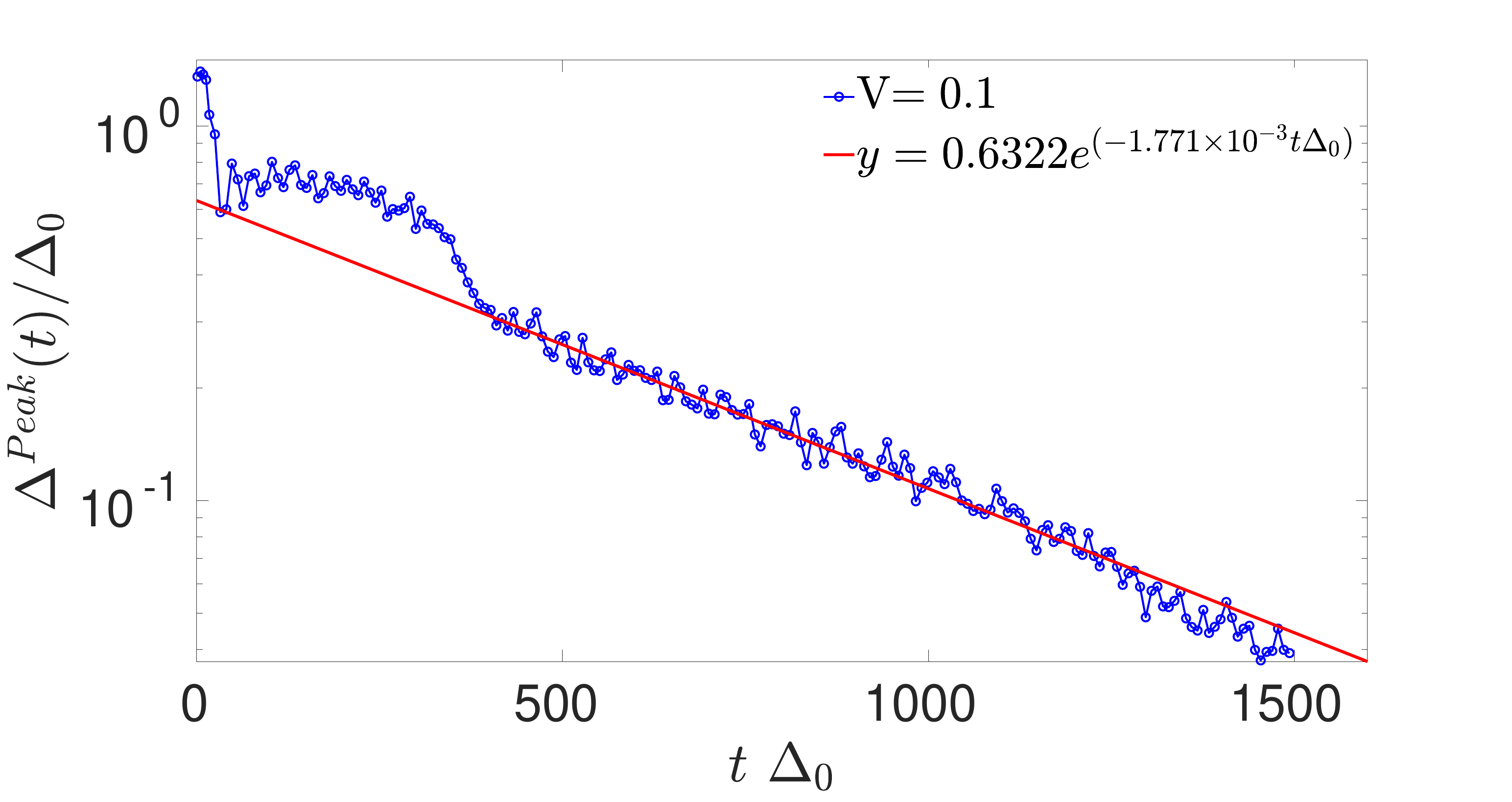}  }
		\subfigure[]{ \label{fig.fit_V0p6}
			\includegraphics[width=8.cm]{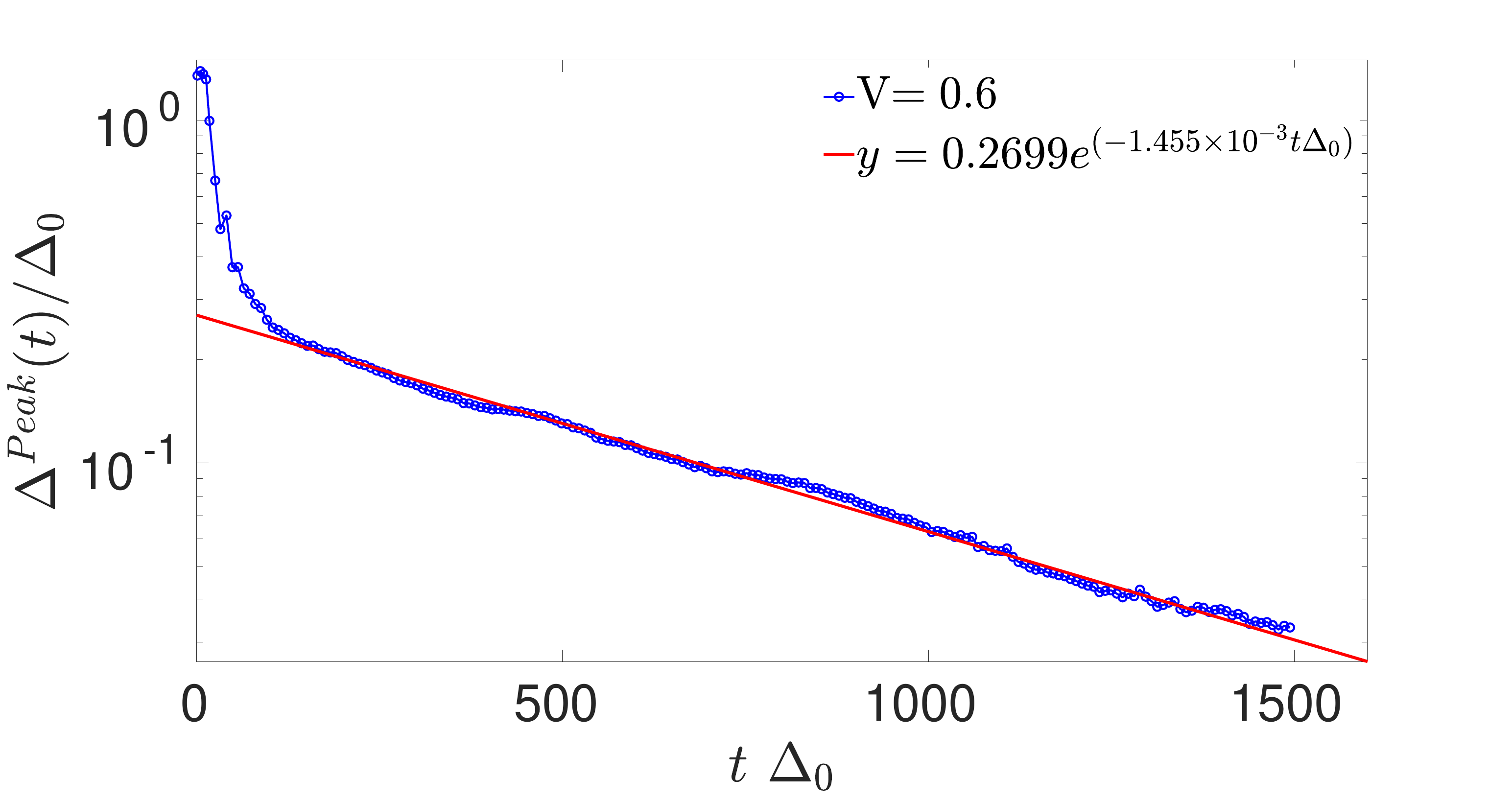}  }
		\caption{ 
			The peaks (local maxima) of the spatial averaged order parameter describing the decay in time of oscillations amplitude. The red line is the best exponential fit for long times. The disorder strength is \subref{fig.fit_V0p001_tiny}. $V=0.001$ and \subref{fig.fit_V0p01_tiny}. $V=0.01$. The rest of parameters are the same with Figure~\ref{Fig:extract_peak}  
		}\label{Fig:fit_peak}
	\end{center}
\end{figure}

In the main text, we have shown that in the presence of disorder, the amplitude of time crystal oscillation are damped. When disorder increases, the amplitude of the oscillation is suppressed at earlier time. However, the frequency of the oscillation do not change.
In this Appendix, we present the results of those peaks  $\Delta^{Peak}(t)$ and the corresponding oscillatory behavior of $\langle \Delta(r,t) \rangle$ in Figure~\ref{Fig:extract_peak}. In Figure~\ref{Fig:fit_peak}, we show that the exponential fitting function $\langle\Delta(t)\rangle = A \exp(-D t \Delta_0)$, where $A$ is related to its amplitude and $D$ is the decay rate, describes well the decay in time of the order parameter local maxima $\Delta^{Peak}(t)$.

In the weak disorder limit $V \le 0.01$, there seems to have three different stages. The order parameter is spatially homogeneous in the first stage when evolution time $t~\Delta_0 \le 500$. Later, it will induce inhomogeneity in space due to this very weak disorder. However, it will enter into the second stage, within the time range $800 \le t \Delta_0 \le 1400$, which is a metastable state that coexist with spatial patterns. In this stage, there are some islands have a phase shift of $\pi$, but still time crystal. The amplitude of the time crystal oscillation also decays slowly, which is approximately a constant, see Figure~\ref{fig.fit_V0p001_tiny}. In the third stage, those time crystal gradually breaks due to the finite size effects, and the amplitude of oscillation decays exponentially. 
In a stronger disordered system $V \ge 0.1$, $\Delta^{Peak}(t)$ decays exponentially when the pattern is formed.

\begin{figure}[!htbp]
	\begin{center}
		\subfigure[]{ \label{fig.BdG_L100_peak_xy}
			\includegraphics[width=8.cm]{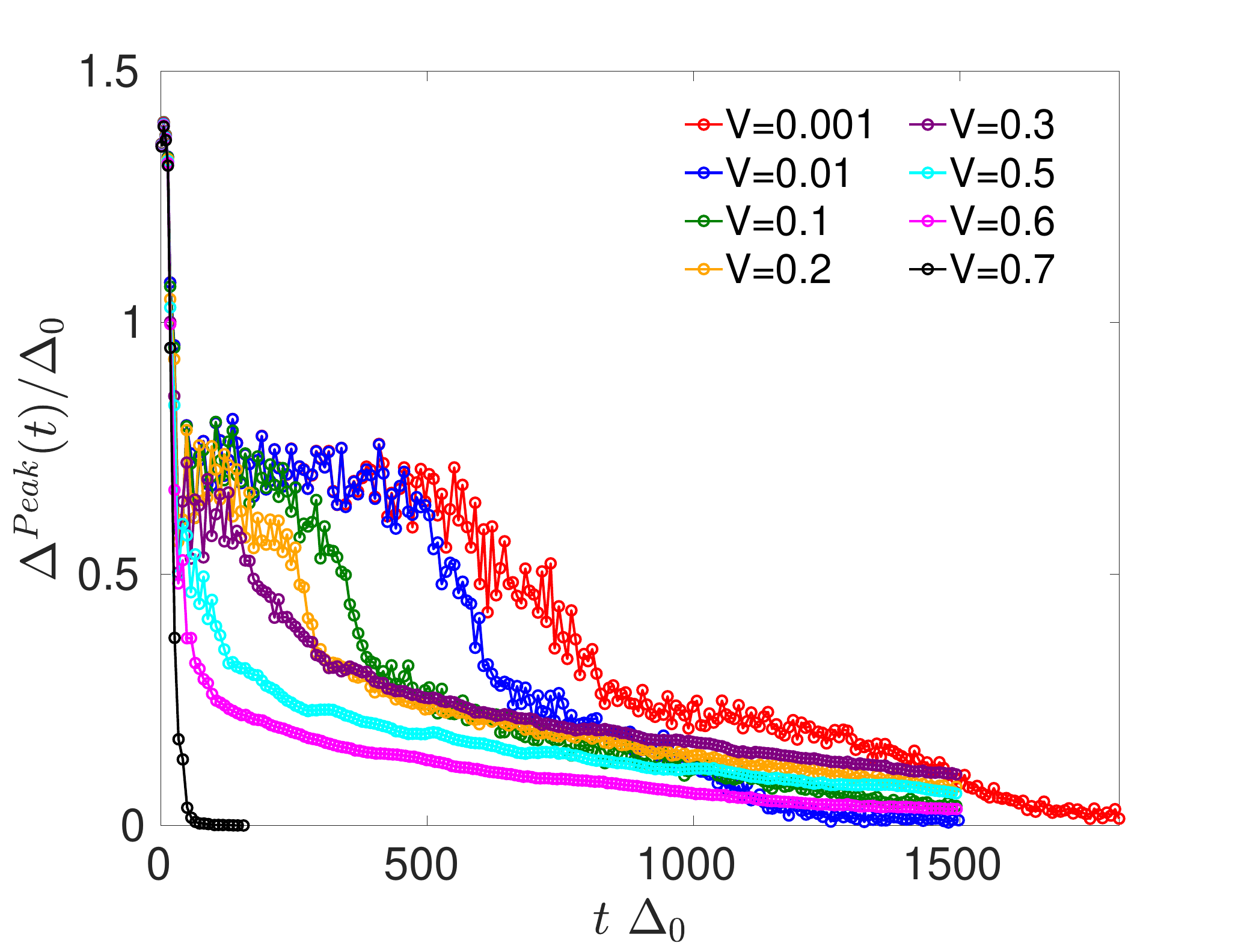}  }
		\subfigure[]{ \label{fig.BdG_L100_peak_xylg}
			\includegraphics[width=8.cm]{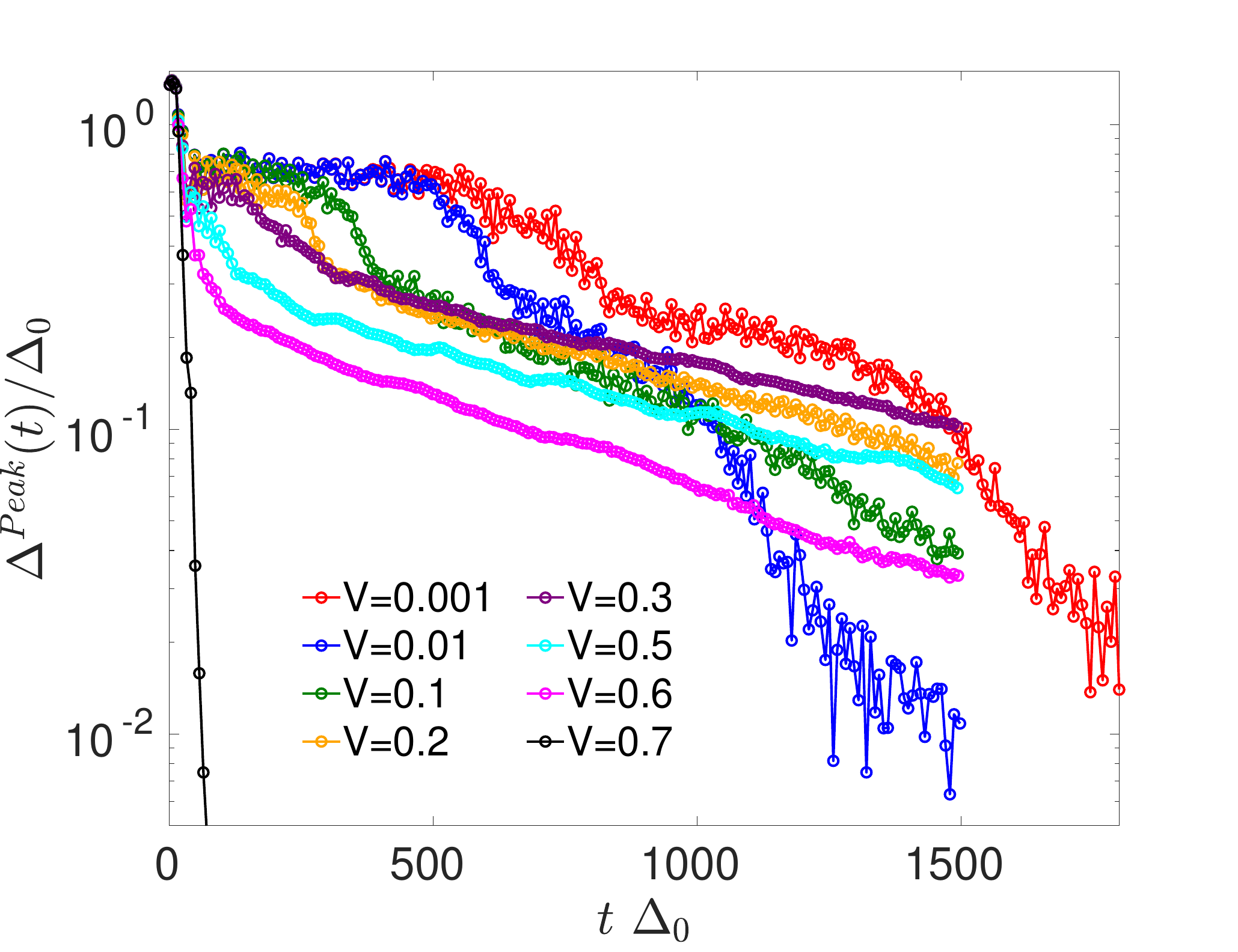}  }
		\caption{Local maxima of the oscillations in time $t$ of the order parameter $\Delta^{Peak}(t)$ for different disorder strengths in both linear (left) and log (right) scale. }\label{Fig:BdG_L100_peak}
	\end{center}
\end{figure}

We extract the peaks of time crystal oscillations, and depict it in Figure~\ref{Fig:BdG_L100_peak}. It shows that by increasing disorder, the $\Delta^{Peak}(t)$ starts the damped oscillation in a shorter time. But its decay exponent $D$ becomes smaller, and therefore, the amplitude even become larger in the long time evolution. Of course, keep increasing disorder when $V\ge 0.5$, it will decay fast with a larger exponent $D$. Finally, when $V>0.7$, the time crystal breaks down and there is no time crystal oscillations.

\begin{figure}[!htbp]
	\begin{center}
		\includegraphics[width=8cm]{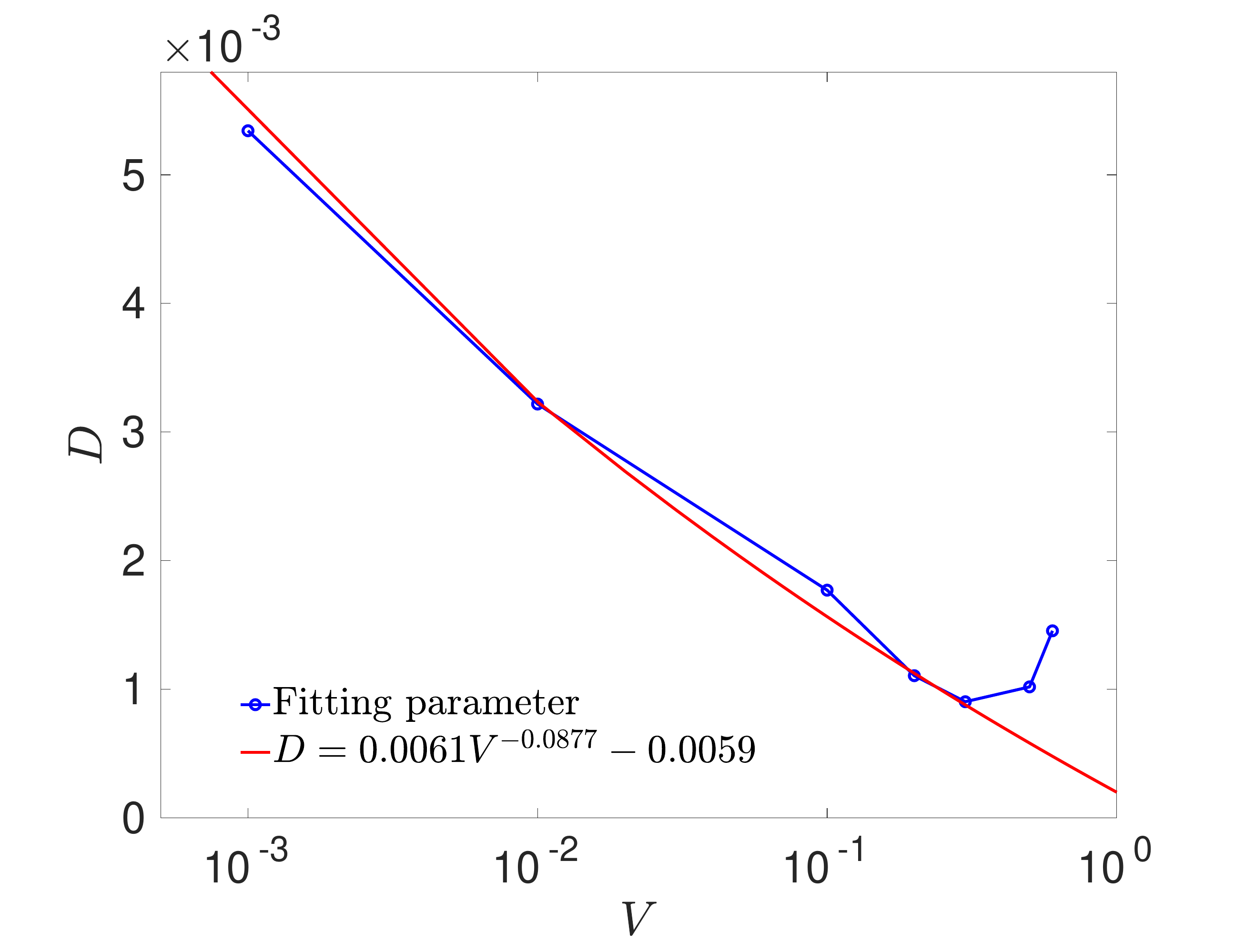} 
		\caption{The decay rate $D$ (blue), describing the exponential suppression of $\Delta^{Peak}(t)$ as a function of the disorder strength $V$. The best fit (red) corresponds a weak power-law dependence on $V$.}\label{Fig:BdG_L100_peak_decay}
	\end{center}
\end{figure}

To study the effects of disorder to the decay rate of the damped time crystal oscillations, we plot the exponent $D$ as a function of disorder strength $V$ in Figure~\ref{Fig:BdG_L100_peak_decay}. It shows that $D$ decreases with increasing disorder $V$ when $V \le 0.3$, and then increases.
More interestingly, $D$ have an approximately power-law relation with $V$ when $V \le 0.3$. The results suggest that a proper disorder strength make the time crystal more robust against the finite size effects, so that it survives longer time.

\section{Finite size effects} \label{app:size}

\begin{figure}[!htbp]
	\begin{center}
		\includegraphics[height=4cm]{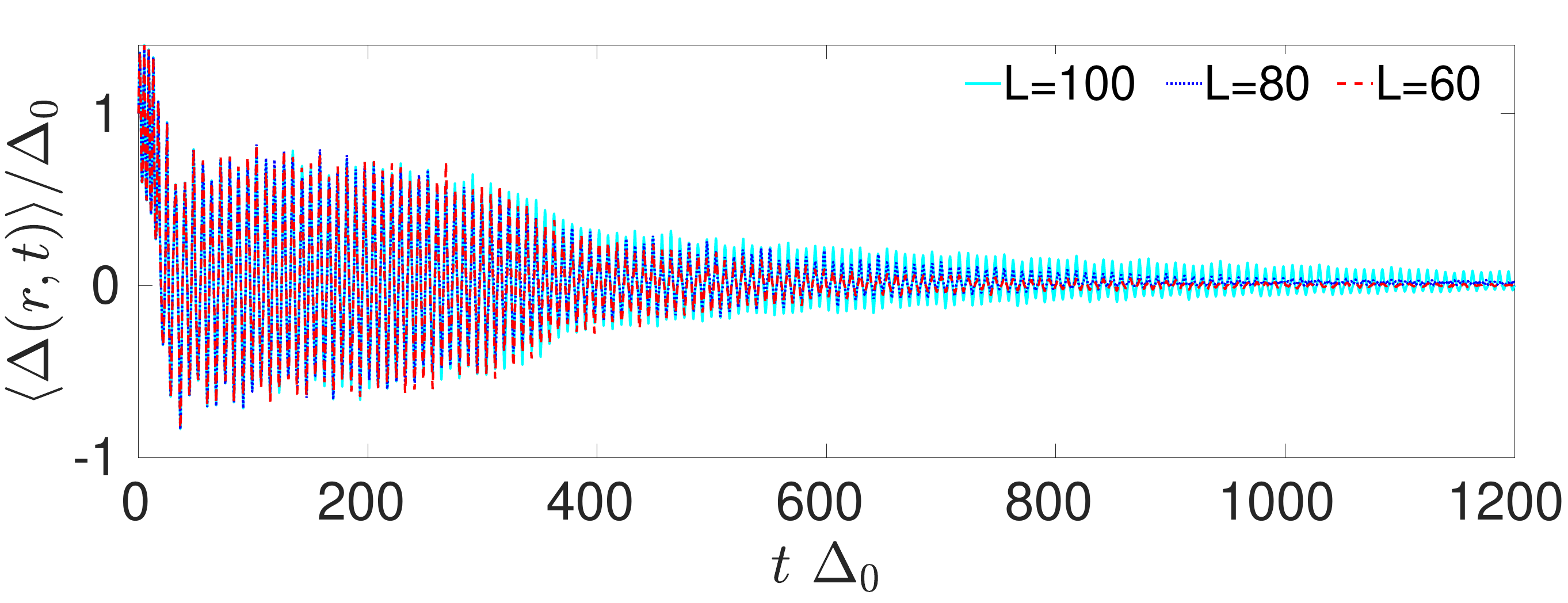} 
		\caption{The time evolution of the spatial averaged order parameter $\langle\Delta(r,t)\rangle$ for different system sizes. For a larger system size, the time crystal phase remains longer time. The corresponding parameters are system size $N=L\times L$, coupling constant $U_0 = -6$, disorder strength $V=0.1$ and chemical potential $\mu = 0$. $\alpha = 0.25$ and $\omega_d = 0.8 \times 2\langle\Delta(r)\rangle$ are the driving amplitude and frequency. }\label{Fig:V0p1_meangap_vs_L}
	\end{center}
\end{figure}

\begin{figure}[!htbp]
	\begin{center}
		\subfigure[]{ \label{fig.fit_L36}
			\includegraphics[width=8cm]{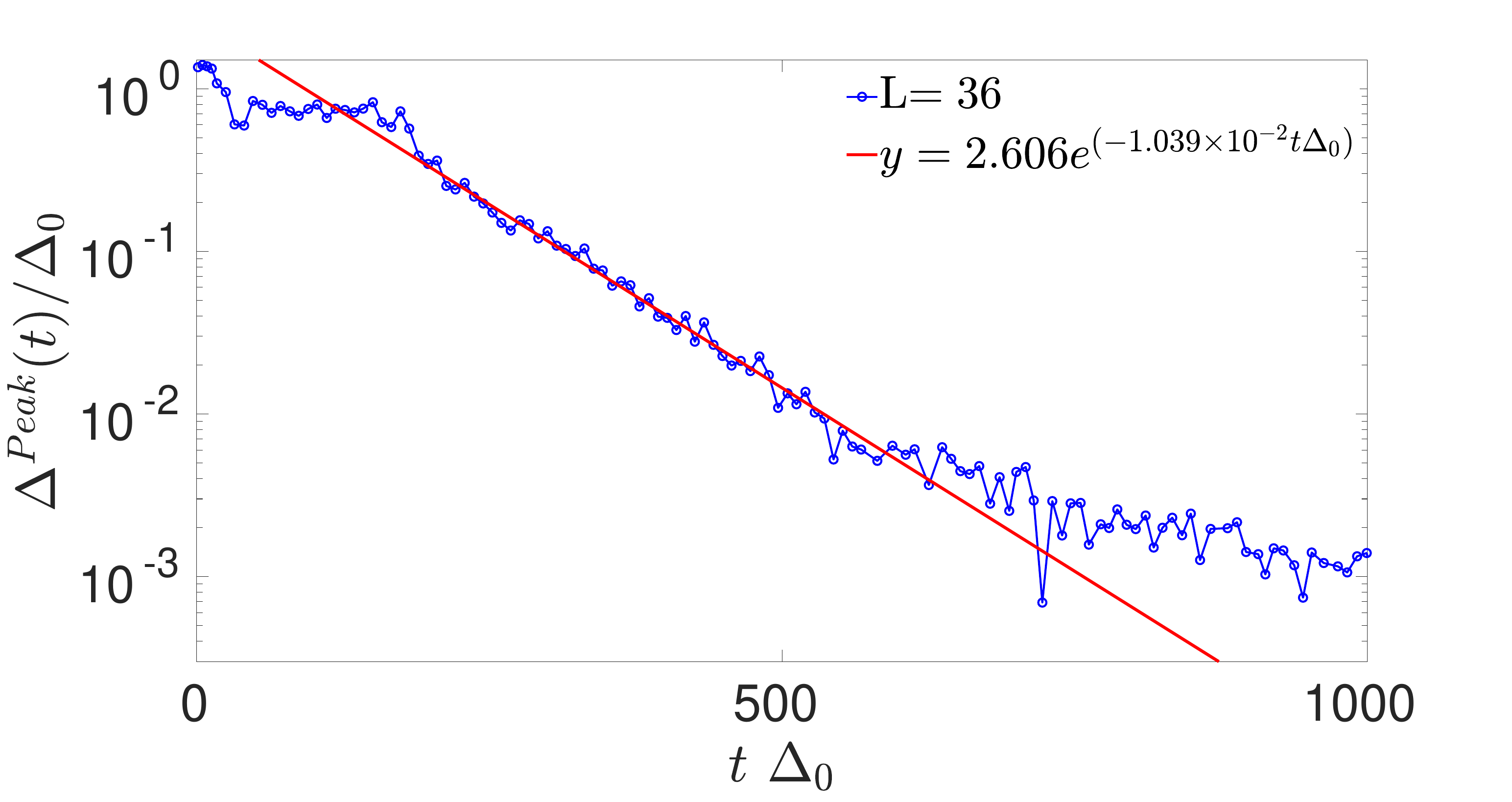}  }
		\subfigure[]{ \label{fig.fit_L120}
			\includegraphics[width=8cm]{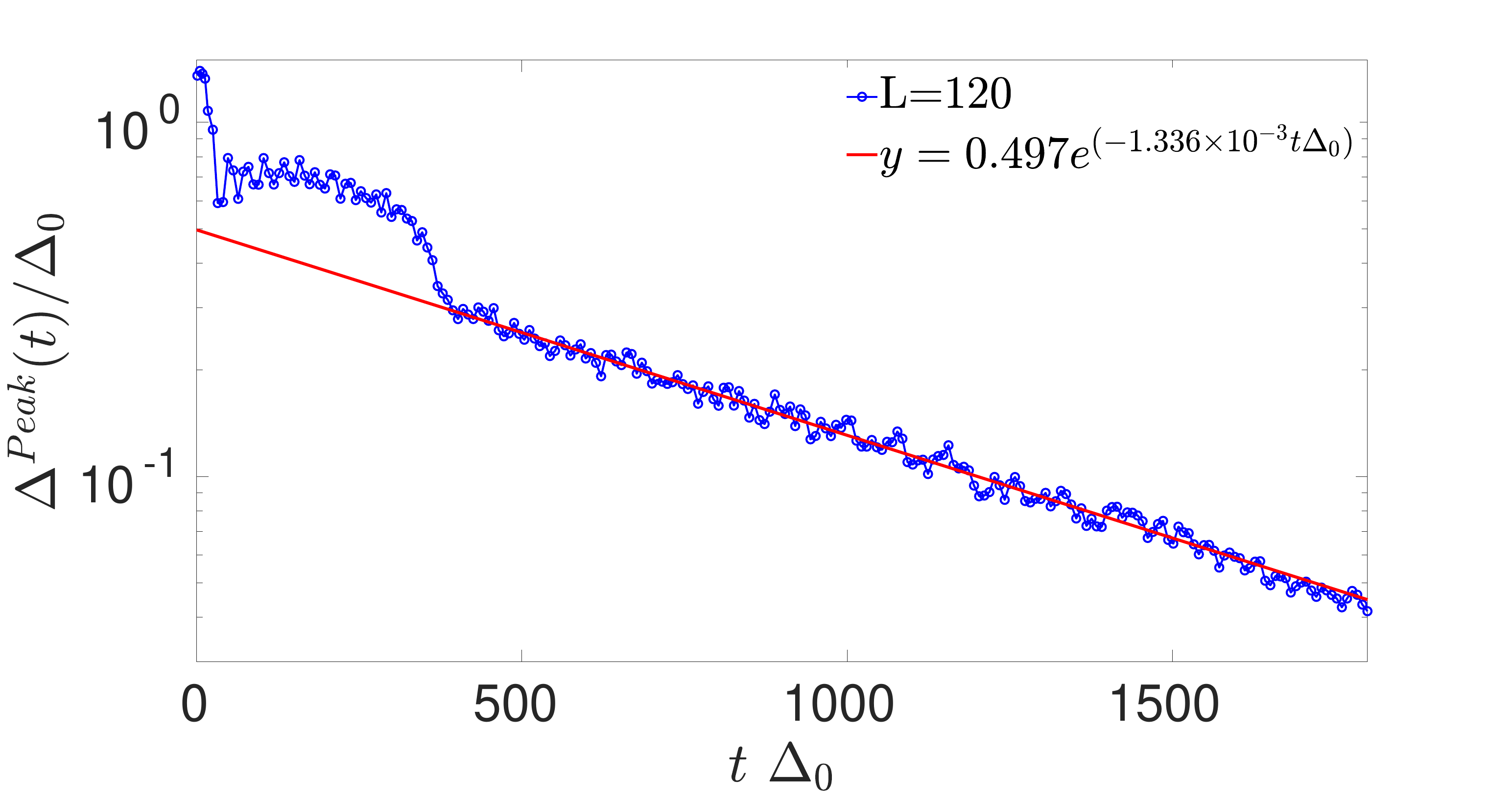}  }
		\caption{The peaks of spatial average order parameter $\Delta^{Peak}(t)$ and the corresponding fittings for different system size $N=L\times L$. The other parameters remain the same to Fig.~\ref{Fig:V0p1_meangap_vs_L}. }\label{Fig:fit_peak_vs_L}
	\end{center}
\end{figure}

In this Appendix, we conducted an analysis of the finite size effects of the time crystal phase, specifically focusing on a fixed weak disorder strength of $V = 0.1$. In the presence of a weak disorder, the order parameter evolves spatially inhomogeneous, and the oscillation decays at earlier time, which is suitable to study finite size effects. 
Figure~\ref{Fig:V0p1_meangap_vs_L} compares the time evolution of the spatial averaged order parameter $\langle\Delta(r,t)\rangle$ for different system sizes.
For a fixed weak disorder $V = 0.1$, when the order parameter evolves spatially inhomogeneous, the oscillation decays slower for larger system size, which means the time crystal is more robust. The fitting results in Figure~\ref{Fig:fit_peak_vs_L} demonstrate that the oscillation peaks decay in an exponential way. 

\begin{figure}[!htbp]
	\begin{center}
		\subfigure[]{ \label{fig.BdG_V0p1_peak_xy}
			\includegraphics[width=7cm]{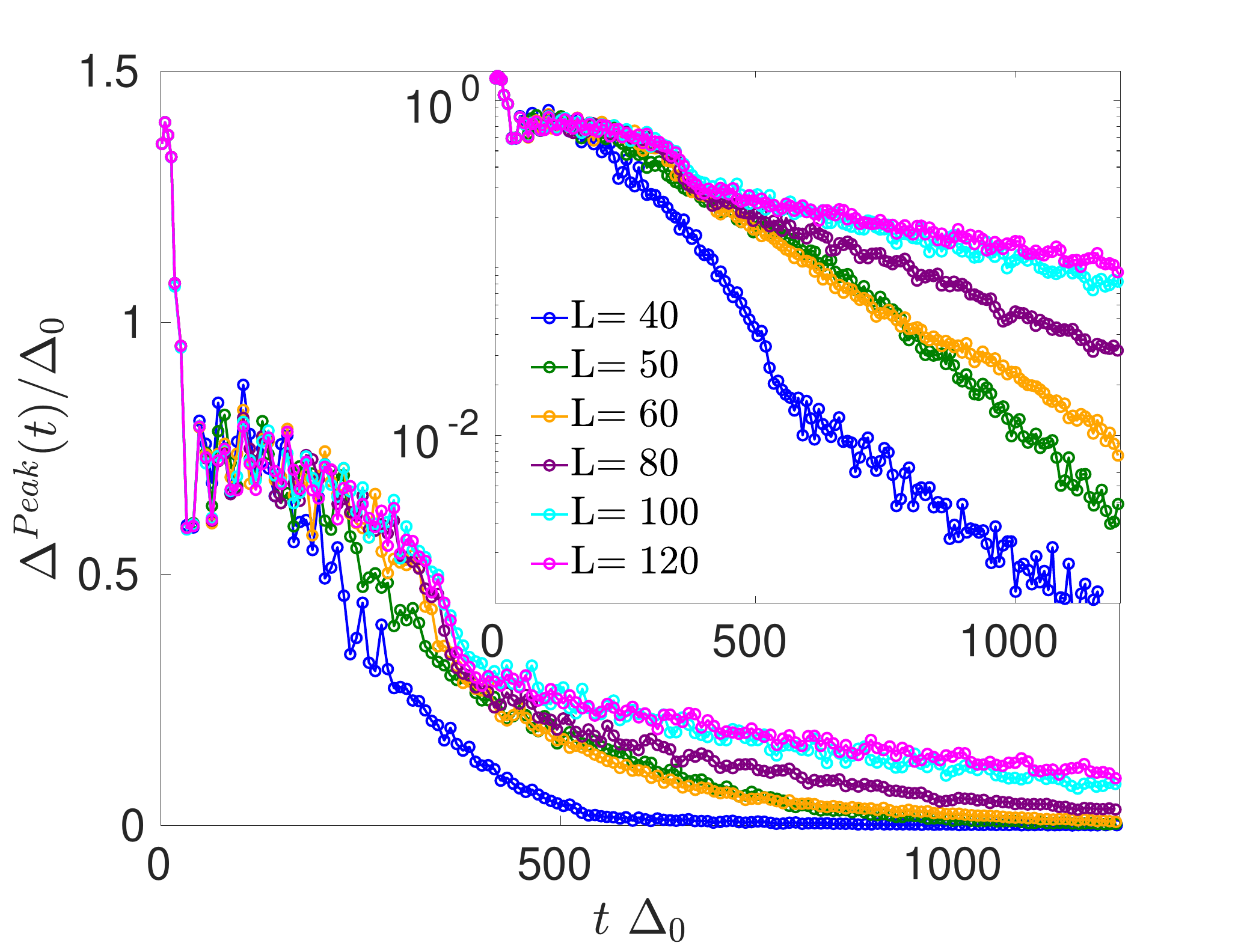}  }
		\subfigure[]{ \label{fig.BdG_V0p1_peak_decay}
			\includegraphics[width=7cm]{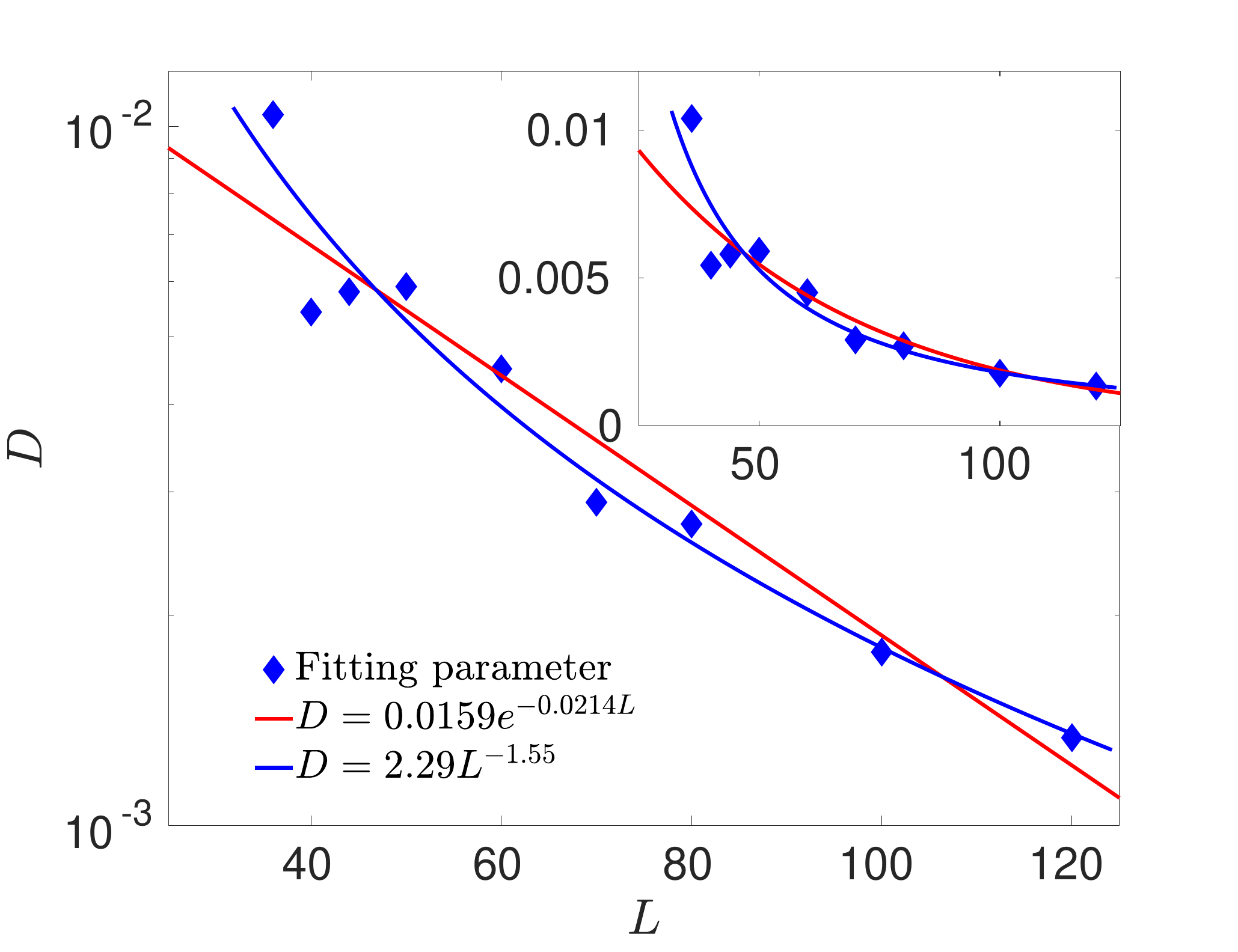}  }
		\subfigure[]{ \label{fig.BdG_V0p1_V0_peak_xy}
			\includegraphics[width=7cm]{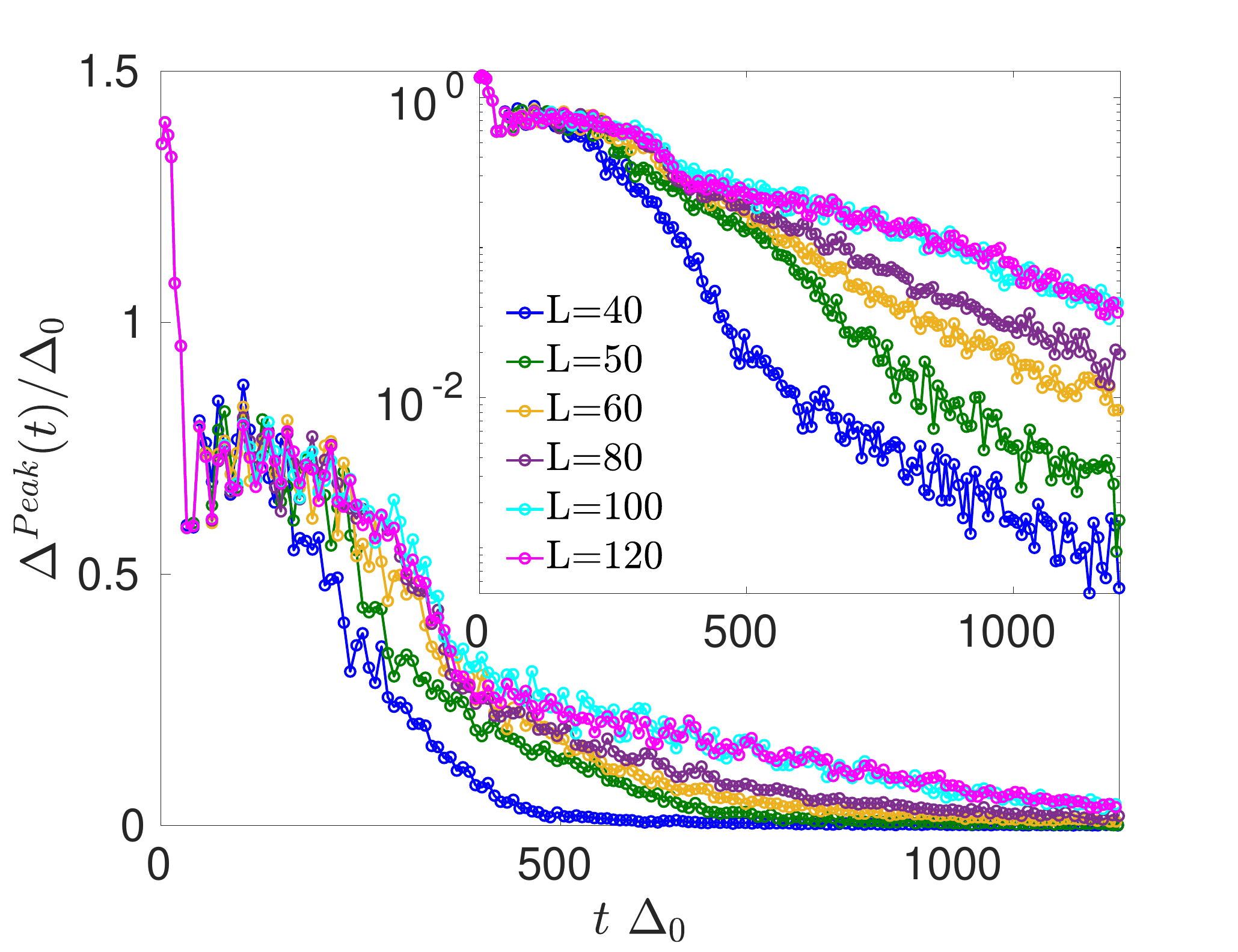}  }
		\subfigure[]{ \label{fig.BdG_V0p1_V0_peak_decay}
			\includegraphics[width=7cm]{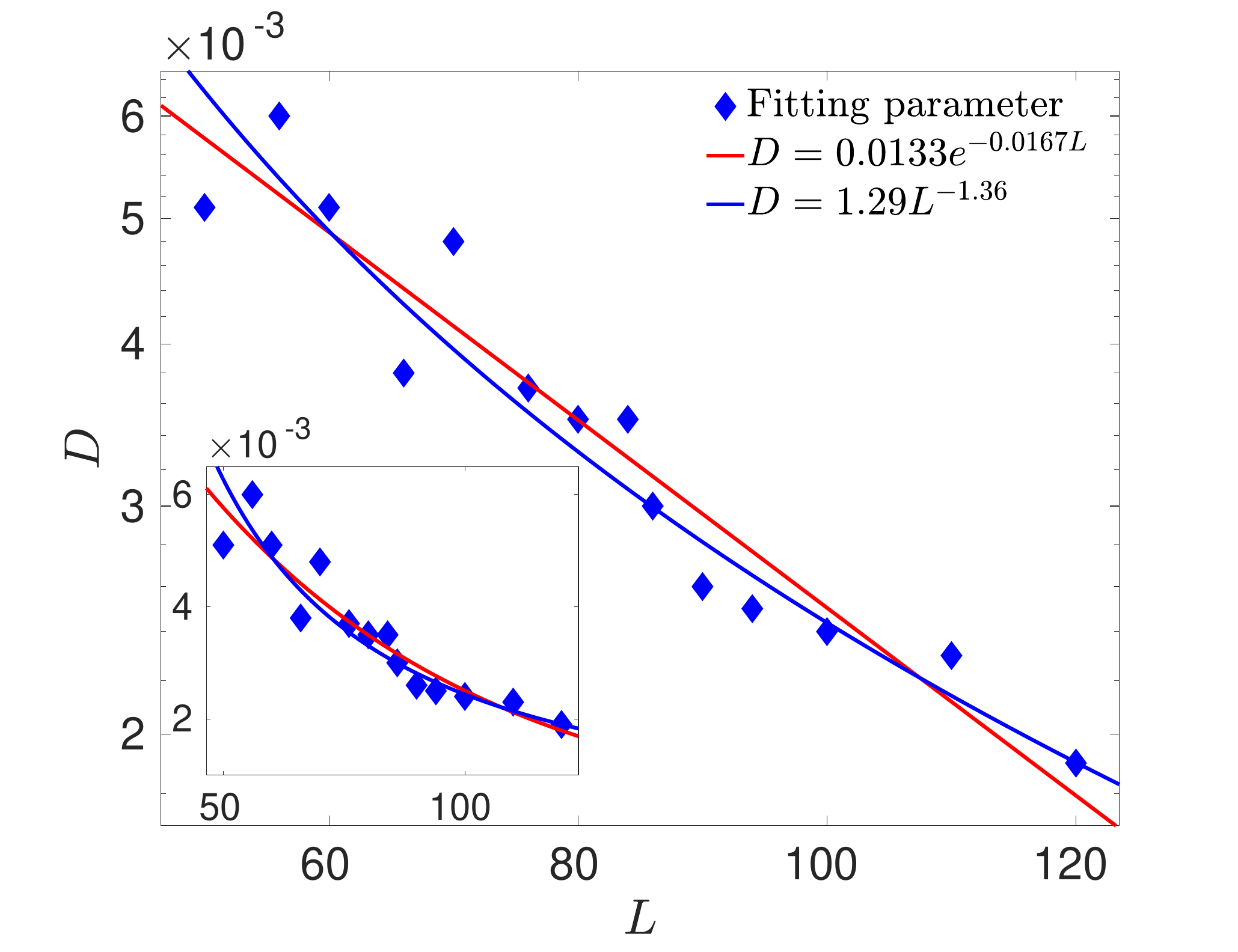}  }
		\caption{\subref{fig.BdG_V0p1_peak_xy}, \subref{fig.BdG_V0p1_V0_peak_xy}. Local maxima in time $t$ of the spatial averaged order parameter  $\Delta^{Peak}(t)$ for different system sizes $L \times L$. The right panel: the decay rate $D$ assuming an exponential decay of these maxima as a function of the size $L$. In the upper panel, disorder remains, while the disorder is removed when it is in the time crystal phase in the lower panel. The rest of parameters are system size $N = L\times L$, coupling constant $U_0 = -6$, disorder strength $V=0.1$ and chemical potential $\mu = 0$. The driving amplitude and frequency are $\alpha = 0.25$ and $\omega_d = 0.8\langle\Delta(r)\rangle$ respectively. The insets contain the same information as the main plots but in log scale.}\label{Fig:V0p1_size_effect}
	\end{center}
\end{figure}

The oscillation peaks for system sizes from $L=36$ up to $L=120$ are presented in Figure~\ref{Fig:V0p1_size_effect}, which shows a slower decay for larger sizes, strongly suggesting that the time crystal is robust in the $L \to \infty$ limit. 
When disorder increases, the time crystal oscillations are suppressed even at relatively short times. The fitting results in Figure~\ref{Fig:fit_peak_vs_L} demonstrate that the oscillation peaks decay exponentially. We extract the decay rate $D$ and depict it in Figure~\ref{fig.BdG_V0p1_peak_decay}. 
In the presence of disorder, due to the limitation of the maximum system size that we could reach, we can not rule out whether $D$ exhibits a power-law or exponential decrease with the length of samples $L$. However, the fitting result is enough to provide a robust evidence that the time crystal phase will persist in the infinite system.

To make a complete study, we only use an initial state with the same weak disorder strength $V = 0.1$. When the system enters the time crystal phase at time $t\Delta_0 \sim 19.6$, the random disordered potentials are removed, namely, the system becomes clean. The disorder in the initial state is just to introduce a perturbation to the system. We obtain similar results in this case. These findings provide compelling evidence that the time crystal phase remains robust and stable even in the presence of weak disorder. Moreover, the decrease in the decay rate with increasing system size suggests that the time crystal phase can persist in the thermodynamic limit.

\end{document}